 \documentclass[12pt,letterpaper]{article}

 \usepackage{latexsym}
 \usepackage{stmaryrd}
 \usepackage{dsfont}
 \usepackage{mathrsfs}
 \usepackage{amssymb} 
 
 \usepackage{bm}
 \usepackage{verbatim}
 \usepackage{url}
 \usepackage[svgnames]{xcolor} 
 \usepackage{soul}
 \usepackage{stackengine}
 \usepackage{stackrel} 
 \usepackage{mathtools}
 \usepackage{array}
 \usepackage{euscript}
 \usepackage{trace}
 \usepackage{etoolbox}
 \usepackage{multirow}
 \usepackage{trimclip}
 \usepackage{adjustbox}
 \usepackage{arydshln}
 \usepackage{ebproof}
 \usepackage{environ}
 \usepackage{ebprooffix}
 \usepackage{yfonts}
 \counterwithin{figure}{section}
 \counterwithin{table}{section}
  \usepackage{tikz}
 \usetikzlibrary{positioning}
 \usetikzlibrary{decorations}
 \usetikzlibrary{shapes}
 \usetikzlibrary{shapes.geometric}
 \usetikzlibrary{shapes.misc}
 \usetikzlibrary{arrows}
 \usetikzlibrary{arrows.meta}
 \usetikzlibrary{topaths}
 \usetikzlibrary{matrix}
 \usetikzlibrary{calc}
 \pgfdeclarelayer{background}
 \pgfdeclarelayer{foreground}
 \pgfsetlayers{background,main,foreground}
 \usepackage{myncbar}
 \usepackage{bibkeyhack} 

 \usepackage{myshtart2ea,myunh2ea}
 \usepackage{mathsupp2e}
 \usepackage{811paper}
 \usepackage{localadd}
 \usepackage{defs_informatica17}
 \usepackage{defs_dexa18}
 \usepackage{defs_pre_mgdb18}
 \usepackage{defs_binary_mgdb18}
 \def\citeflag{1} 
 
 \def\preformat#1{#1}
 \def\publformat#1{}
 
 \def\publcontent#1{}
 \def\twelveformat#1{\relax}
 \def\draftfooter{Version: 20220807 page \thepage}
 \threepartfoot{\draftfooter}{}{}{}
 \threepartfootone{\draftfooter}{}{}{}
\begin{document}


 \title{Inference Rules for Binary Predicates
                           \\ in a Multigranular Framework}

 \author{
         Stephen J. Hegner \\
         DBMS Research of New Hampshire \\
            PO Box, 2153, New London, NH 03257, USA \\
            {\tt dbmsnh@gmx.com}
         \vspace*{1em} \\
         M. Andrea Rodr{\'{\i}}guez\footnote{
              Author to whom all correspondence regarding this paper
              should be directed.} \\
            Millennium Institute for Foundational Research on Data \\
            Departamento Ingenier{\'i}a Inform{\'a}tica y
            Ciencias de la Computaci{\'o}n \\
            Edmundo Larenas 219,
            Universidad de Concepci{\'o}n \\
            4070409 Concepci{\'o}n, Chile \\
            {\tt andrea@udec.cl}
        }

 \date{}  

  \maketitle

 \thispagestyle{simpleheadingsone}
 \begin{abstract}
     In a multigranular framework, the two most important binary
predicates are those for subsumption and disjointness.  In the first
part of this work, a sound and complete inference system for
assertions using these predicates is developed.  It is customized for
the granular framework; particularly, it models both bottom and top
granules correctly, and it requires all granules other then the bottom
to be nonempty.  Furthermore, it is single use, in the sense that no
assertion is used more than once as an antecedent in a proof.
      \par
    In the second part of this work, a method is developed for
extending a sound and complete inference system on a framework which
admits Armstrong models to one which provides sound and complete
inference on all assertions, both positive and negative.  This method
is then applied to the binary granule predicates, to obtain a sound
and complete inference system for subsumption and disjointness, as
well as their negations.
 \end{abstract}

 \thispagestyle{empty}
 \thispagestyle{simpleheadingsone}



 \section{Introduction}\label{sec:intro}

   In a multigranular relational database system such as
\mgdbsys\ \mycite{HegnerRo17_inform} \mycite{HegnerRo18_dexa}, the
ordinary flat attributes are replaced with ones which admit a
hierarchical structure.  Specifically, a \emph{constrained granulated
attribute schema}, or \emph{CGAS} includes a set $\granules{\gsn}$,
called the \emph{granules} of $\gsn$, together with a set of
constraints $\constr{\gsn}$ which relate the granules to one another.
The knowledge model is \emph{open world} in the sense that an
assertion $\varphi$ may be true
  (if $\constr{\gsn} \sentails \varphi$),
false
  (if $\constr{\gsn} \sentails \mlnot\varphi$),
or unknown
  (if neither $\constr{\gsn} \sentails \varphi$ nor
              $\constr{\gsn} \sentails \mlnot\varphi$)
holds.
    Although it can be shown that inference is decidable, if the
nature of the constraints is not restricted carefully, it becomes
highly intractable.  To address this issue, the constraints which are
permitted are required to lie in one of two groups.  The first group,
which is the main topic of this paper, consists of the binary forms of
subsumption 
 $\subrulep{\granvar{g}_1}{\granvar{g}_2}$
 (equivalent to containment, both proper and improper, in the
underlying semantics)
 as well as of disjointness
 $\disjrulep{\granvar{g}_1}{\granvar{g}_2}$,
 together with their negations
 $\nsubrulep{\granvar{g}_1}{\granvar{g}_2}$
 and
 $\ndisjrulep{\granvar{g}_1}{\granvar{g}_2}$.
    In the context of geographic entities of Chile, simple examples of
positive binary statements include the statements that the province of
Llanquihue lies within the region of Los Lagos, and that the provinces
of Llanquihue and Osorno are disjoint, represented in
(\ref{sec:intro}-1) below.
    \[
      \tag{\ref{sec:intro}-1}
      \subrulep{\llanquihuep}{\loslagosr} \qquad
      \disjrulep{\llanquihuep}{\osornop} 
    \] 
 Similarly, simple examples of negative binary statements include that
Puyehue National Park is neither contained in nor disjoint from the
region of Los Lagos, represented in (\ref{sec:intro}-2) below.
    \[
      \tag{\ref{sec:intro}-2}
      \nsubrulep{\puyehuenpk}{\loslagosr} \qquad
      \ndisjrulep{\puyehuenpk}{\loslagosr} 
    \] 

      The second group consists of assertions which express that a
single granule is either subsumed by or else equal to the join ( or
union) of a (finite) set of other granules.
    A typical example, shown in (\ref{sec:intro}-3) below, asserts
that the region of Los Lagos is the disjoint union of its constituent
provinces.
  \[
   \tag{\ref{sec:intro}-3}
   \loslagosr =
           \biggrlatjoinddisp{}{}
        \setbr{\osornop, \llanquihuep, \chiloep, \palenap}
  \]
     Management of complexity of inference for this type of constraint
requires that the granules be grouped into \emph{granularities}, and
that such assertions be \emph{bigranular}.  For details on this,
including the meaning of bigranular, the reader is referred to
\mycite{HegnerRo18_dexa}.  Despite their importance, such assertions
will not be considered further in this paper.
   \par
      The focus of this paper is the first group of constraints; more
precisely, to provide a consistent and complete inference system for
the two forms of positive binary constraints
 $\subrulep{\granvar{g}_1}{\granvar{g}_2}$
 and
 $\disjrulep{\granvar{g}_1}{\granvar{g}_2}$,
 together with the associated negative binary constraints
 $\nsubrulep{\granvar{g}_1}{\granvar{g}_2}$
 and
 $\ndisjrulep{\granvar{g}_1}{\granvar{g}_2}$.
   The construction of this inference system is divided into two
steps.  In the first step, an inference system for positive binary
constraints only is developed, and in the second, that system is
extended to include negative binary constraints.
   \par
    For the first step, the inference rules shown in
Figs.\ \ref{fig:infrulespmain} and \ref{fig:infrulesptaut}, taken
together, are shown to be \emph{sat-conditionally complete}; that is,
they are complete when applied to a consistent set of constraints.  In
these rules, $\granvar{g}$, $\granvar{g}_1$, $\granvar{g}_2$,
$\granvar{g}_3$, $\granvar{g}_1'$, and $\granvar{g}_2'$ are variables
which may take on the value of any granule, while $\bot$ and $\top$
represent the special bottom (least) and top (greatest) granules.
  \begin{figure}[htb]
  \belowdisplayskip=0em
  \begin{align*}
    \begin{prooftreem}
       \hypoii{\subrulep{\granvar{g}_1}{\granvar{g}_2}}
              {\subrulep{\granvar{g}_2}{\granvar{g}_3}}
       \inferi{\subrulep{\granvar{g}_1}{\granvar{g}_3}}
    \end{prooftreem}
    &&
    \begin{prooftreem}
      \hypoii{\subrulep{\granvar{g}_1}{\granvar{g}_1'}}
              {\disjrulep{\granvar{g}_1'}{\granvar{g}_2}}
      \inferi{\disjrulep{\granvar{g}_1}{\granvar{g}_2}}
    \end{prooftreem}
    \end{align*}
  \caption{Main inference rules for positive binary constraints}\label{fig:infrulespmain}
  \end{figure}

  \begin{figure}[htb]
  \belowdisplayskip=0em
    \begin{align*}
    \begin{prooftreem}
      \hypoi{\phantom{X}}
      \inferi{\subrulep{\granvar{g}}{\granvar{g}}}
    \end{prooftreem}
    &&
    \begin{prooftreem}
      \hypoi{\phantom{X}}
      \inferi{\subrulep{\granvar{g}}{\top}}
    \end{prooftreem}
    &&
    \begin{prooftreem}
      \hypoi{\phantom{X}}
      \inferi{\subrulep{\bot}{\granvar{g}}}
    \end{prooftreem}
    &&
    \begin{prooftreem}
      \hypoi{\phantom{X}}
      \inferi{\disjrulep{\bot}{\granvar{g}}}
    \end{prooftreem}
   \end{align*}
  \caption{Tautological inference rules for positive binary constraints}\label{fig:infrulesptaut}
  \end{figure}

    The rules of Fig.\ \ref{fig:infrulespmain} are the main ones ---
asserting respectively that subsumption is closed under transitivity,
and that if two granules are disjoint, so too are any of their
subgranules under subsumption.
    The four rules of Fig.\ \ref{fig:infrulesptaut} represent
tautologies --- statements which are always true.  The conclusion
holds without any premises.  Although these tautological rules may
seem trivial, they are necessary to ensure complete sat-conditional
inference.
     \par
    In order to derive unsatisfiability of a set of positive
assertions, the rule of Fig.\ \ref{fig:infrulespunsat} must be
included as well.  It requires that no granule, other than $\bot$, may
have empty semantics, so such a granule may never be disjoint from
itself.
  \begin{figure}[htb]
  \belowdisplayskip=0em
    \begin{align*}
    \begin{prooftreem}
      \hypoi{\disjrulep{\granvar{g}}{\granvar{g}}}
      \inferi{\false}
    \end{prooftreem}
    {\scriptstyle\abr{|{\scriptscriptstyle(\granvar{g} \neq \bot)}}}
   \end{align*}
  \caption{Unsatisfiability inference rule on positive binary constraints}\label{fig:infrulespunsat}
  \end{figure}

    This first part of the paper is based substantially upon the
earlier work reported in \mycite{AtzeniPa88_dke}, which provides
overlapping but not identical rules in a context in which neither
$\bot$ nor $\top$ exist as special granules, and in which any granule
may have empty semantics.  Since these differences are substantial,
the full development of this inference system for positive assertions
requires considerable development.
    \par
     In the second part of this paper, it is shown that a consistent
and complete set of inference rules for both positive and negative
constraints may be obtained by augmenting the rules for positive
constraints with \emph{swapped} versions, in which a single antecedent
is swapped with the consequent, and each negated.  Shown in
Fig.\ \ref{fig:infrulespmainswap} are the main rules so obtained from
Fig.\ \ref{fig:infrulespmain}.
  \begin{figure}[htb]
   \begin{align*}
    &\begin{prooftreem}
      \hypoi{\subrulep{\granvar{g}_1}{\granvar{g}_2} \hspace*{1cm}
             \nsubrulep{\granvar{g}_1}{\granvar{g}_3}}
      \inferi{\nsubrulep{\granvar{g}_2}{\granvar{g}_3}}
      \end{prooftreem} 
   &&\begin{prooftreem}
      \hypoi{\subrulep{\granvar{g}_2}{\granvar{g}_3} \hspace*{1cm}
             \nsubrulep{\granvar{g}_1}{\granvar{g}_3}}
      \inferi{\nsubrulep{\granvar{g}_1}{\granvar{g}_2}}
    \end{prooftreem} 
   &&&\begin{prooftreem}
     \hypoii{\subrulep{\granvar{g}_1}{\granvar{g}_1'}}
             {\ndisjrulep{\granvar{g}_11}{\granvar{g}_2}}
     \inferi{\nsubrulep{\granvar{g}_1'}{\granvar{g}_2'}}
    \end{prooftreem}
   \end{align*}
  \caption{Swapped main inference rules for binary constraints}\label{fig:infrulespmainswap}
  \end{figure}
     Similarly, Fig.\ \ref{fig:infrulesntaut} shows the swapped
versions of the rules of Fig.\ \ref{fig:infrulespunsat}.  Informally
speaking, the $\false$ conclusion becomes a $\true$ premise under a
swap, which is trivial and so may be omitted.  The resulting rule thus
expresses a tautology.
     Together, the rules of
Figs.\ \ref{fig:infrulespmain}--\ref{fig:infrulesptaut}
  and
     \ref{fig:infrulespmainswap}--\ref{fig:infrulesntaut}
 provide a sat-conditionally complete inference system for all binary
constraints, positive and negative.
     (For technical reasons to be explained later, the tautological
rules of Fig.\ \ref{fig:infrulesptaut} do not need to be swapped.)
  \begin{figure}[htb]
  \belowdisplayskip=0em
    \begin{align*}
    \begin{prooftreem}
      \hypoi{\phantom{X}}
      \inferi{\ndisjrulep{\granvar{g}}{\granvar{g}}}
    \end{prooftreem}
    {\scriptstyle\abr{|{\scriptscriptstyle(\granvar{g} \neq \bot)}}}
   \end{align*}
  \caption{Tautological inference rule for negative binary constraints}\label{fig:infrulesntaut}
  \end{figure}

     \par
     In order to derive unsatisfiability for a mix of negative and
positive constraints, a simple augmentation of
Figs.\ \ref{fig:infrulespmain}--\ref{fig:infrulesntaut} suffices.
Namely, for every positive assertion $\alpha$, a rule which states
that $\alpha$ and $\mlnot\alpha$ cannot both hold is added.
     Since there are only two types of assertions in the binary logic
of granules, the two rules of Fig.\ \ref{fig:infrulesunsatpr}
suffice.

  \begin{figure}[htb]
   \begin{align*}
    \begin{prooftreem}
      \hypoi{\subrulep{\granvar{g}_1}{\granvar{g}_2} \hspace*{1cm}
             \nsubrulep{\granvar{g}_1}{\granvar{g}_2}}
      \inferi{\false}
    \end{prooftreem}
   &&
   \begin{prooftreem}
     \hypoi{\disjrulep{\granvar{g}_1}{\granvar{g}_2} \hspace*{1cm}
            \ndisjrulep{\granvar{g}_1'}{\granvar{g}_2'}}
     \inferi{\false}
    \end{prooftreem}
   \end{align*}
  \caption{Unsatisfiable-pair rules for binary constraints}\label{fig:infrulesunsatpr}
  \end{figure}

    The combined rules given in
Figs.\ \ref{fig:infrulespmain}--\ref{fig:infrulesunsatpr} are
consistent and complete for inference on all binary constraints,
positive and negative.
    \par
    An important aspect of the augmentation procedure which is
developed is that it is not limited to binary constraints of granule
logic.  Rather, it applies to any \emph{Armstrong context} in which
the positive constraints satisfy certain properties.  This opens the
door to the application of these results to other contexts.
    \par
    This report is organized as follows.  Sec.\ \ref{sec:setting}
provides a summary of granule spaces, which retain the essential
features of CGASs which are necessary for this paper, while
eliminating the unnecessary details.
     In Sec.\ \ref{sec:proofsys}, essential ideas for proof systems
are reviewed and/or developed.
     In Sec.\ \ref{sec:pinfrules}, the consistency and completeness of
the binary inference rules of
Figs.\ \ref{fig:infrulespmain}--\ref{fig:infrulespunsat} for positive
binary constraints is established.
     In Sec.\ \ref{sec:negarm}, the semantics of negation in the
setting in an Armstrong setting, which includes the logic of binary
predicates on granules, is developed.
     In Sec.\ \ref{sec:ninfrules}, the results of
Sec.\ \ref{sec:negarm} are used to develop an associated proof system,
establishing that the rules of
Figs.\ \ref{fig:infrulespmainswap}--\ref{fig:infrulesunsatpr} provide
a sufficient augmentation of those of
Figs.\ \ref{fig:infrulespmain}--\ref{fig:infrulespunsat} to obtain a
complete set of inference rules for both positive and negative granule
constraints.
    Sec.\ \ref{sec:lit} contains to information the relationship of
this work to other publications.
    Sec.\ \ref{sec:cfd} contains conclusions and further directions.


 \section{The Semantics of Granules}\label{sec:setting}

   In the framework reported \mycite{HegnerRo17_inform} and
\mycite{HegnerRo18_dexa}, a central notion is that of
\emph{granularity}, which provides a classification system for the
underlying granules.  For example, relative to the examples of Chilean
spatial entities described in Sec.\ \ref{sec:intro}, three principal
granularities are $\regiong$, $\provinceg$, and $\parkg$, with
$\loslagosr$ a granule of granularity $\regiong$, $\osornop$,
$\llanquihuep$, $\chiloep$, and $\palenap$ granules of granularity
$\provinceg$, and $\puyehuenpk$ a granule of granularity $\parkg$.
From the perspective of a multigranular DBMS, such classification of
granules into granularities is essential not only for the purposes of
organizing concepts, but also for the systematic expression and
tractable inference of constraints of the form (\ref{sec:intro}-3).
    Nevertheless, within the context of inference on binary
constraints of the forms illustrated in
 Figs.\ \ref{fig:infrulespmain}--\ref{fig:infrulesunsatpr},
 they are of secondary importance at best.
    Therefore, in this section, a simplified version of the CGAS
(constrained granulated attribute schema) of \mycite{HegnerRo18_dexa}
is developed, in which granularities are ignored, with only binary
constraints considered.  The result is termed a \emph{simple
monogranular attribute schema}, or \emph{SMAS}.
    This approach has the advantage of allowing a much simpler
development and presentation of the main results of this report.  In
addition, it allows this report to be read and understood without
first gaining a thorough understanding of the concepts developed in
\mycite{HegnerRo17_inform} and \mycite{HegnerRo18_dexa}.

 \begin{metalabpara}{definition}{}
          {Granule spaces}\envlabel{def:gransp}
   A \emph{granule space} is a triple $\granspacedef{G}$ in which
$\granulesof{G}$ is a set, called the set of \emph{granules}.  with
$\botgrsp{G},\topgrsp{G} \in \granulesof{G}$.
     $\botgrsp{G}$ is called the \emph{bottom granule} of
$\granspacename{G}$ while $\topgrsp{G}$ is called the \emph{top granule}
of $\granspacename{G}$); they are always distinct.
  The special notation $\granulesofnb{G}$ is used to denote the set
$\granulesof{G} \setminus \setbr{\botgrsp{G}}$, while
$\granulesofnbnt{G}$ is used to denote the set
 $\granulesof{G} \setminus \setbr{\botgrsp{G},\topgrsp{G}}$.
 \end{metalabpara}

 \begin{metalabpara}{mydefinition}{}
     {Granule structures and binary constraints}\envlabel{def:granstrsem}
   A \emph{granule structure} over $\granspacename{G}$ is a pair
$\gnlestrpr{\sigma}$ in which $\gnledom{\sigma}$ is a nonempty set and
  $\fn{\gnletodom{\sigma}}{\granulesof{G}}
                          {\powerset{\gnledom{\sigma}}}$
 is a function which assigns a subset of $\gnledom{\sigma}$ to each
granule, subject to the conditions that
   $\gnletodom{\sigma}(\botgrsp{G}) = \emptyset$,
   $\gnletodom{\sigma}(\topgrsp{G}) = \gnledom{\sigma}$,
 and for every $g \in \granulesofnb{G}$,
   $\gnletodom{\sigma}(g) \neq \emptyset$.
     \par
   The \emph{positive binary constraints} over $\granspacename{G}$ are
of two forms.  For $g_1, g_2 \in \granulesof{G}$, the
\emph{subsumption rule} \mbox{$\subrulep{g_1}{g_2}$} holds in the
granule structure $\sigma$ iff
 \preformat{\linebreak}
 $\gnletodom{\sigma}(g_1) \subseteq \gnletodom{\sigma}(g_2)$.
   Likewise, the \emph{disjointness rule} $\disjrulep{g_1}{g_2}$ holds
in $\sigma$ iff
 $\gnletodom{\sigma}(g_1) \intersect \gnletodom{\sigma}(g_2) = \emptyset$.
    The set of all such positive subsumption rules (resp.\ positive
disjointness rules) over $\granspacename{G}$ is denoted
$\psubconstrgr{G}$ (resp.\ $\pdisjconstrgr{G}$).
  Combining these, the set of all positive binary constraints over
$\granspacename{G}$ is
 \nlrightt
 $\pbinconstrgr{G} = \psubconstrgr{G} \union \pdisjconstrgr{G}$.
    \par
   The \emph{negative binary constraints} over $\granspacename{G}$ are the
negations of the positive constraints.  For
 $g_1, g_2 \in \granulesof{G}$, $\nsubrulep{g_1}{g_2}$ denotes that
$\subrulep{g_1}{g_2}$ does not hold; \ie, that
 $\gnletodom{\sigma}(g_1) \not\subseteq \gnletodom{\sigma}(g_2)$,
 with $\nsubconstrgr{G}$ denoting the set of all such negative
subsumption rules.
   Similarly, $\ndisjrulep{g_1}{g_2}$ denotes that
$\disjrulep{g_1}{g_2}$ does not hold; \ie, that
  $\gnletodom{\sigma}(g_1) \intersect \gnletodom{\sigma}(g_2) \neq \emptyset$.
  The set of all negative binary constraints over $\granspacename{G}$
is $\nbinconstrgr{G} = \nsubconstrgr{G} \union \ndisjconstrgr{G}$,
 with $\ndisjconstrgr{G}$ denoting the set of all such negative
subsumption rules.
   \par
   $\allbinconstrgr{G} = \pbinconstrgr{G} \union \nbinconstrgr{G}$
 is the set of \emph{all binary constraints} over $\granspacename{G}$.
   \par
   Following standard logical terminology, the granule structure
$\sigma$ is a \emph{model} of
 $\varphi \in
 \preformat{\linebreak}
 \allbinconstrgr{G}$ if $\varphi$ holds in $\sigma$, and $\sigma$ is a
\emph{model} of $\Phi \subseteq \allbinconstrgr{G}$ if it is a model
of every $\varphi \in \Phi$.  $\varphi$ (resp.\ $\Phi$) is
\emph{satisfiable} if it has at least one model.  The set of all
models of $\varphi$ (resp.\ $\Phi$) is denoted $\modelsof{\varphi}$
(resp.\ $\modelsof{\Phi}$).
   The constraint $\varphi$ (resp.\ the set $\Phi$ of constraints) is
\emph{satisfiable} if
 $\modelsof{\varphi} \neq \emptyset$
 (resp.\ $\modelsof{\Phi} \neq \emptyset$).
    \par
   For $\varphi_1, \varphi_2 \in \allbinconstrgr{G}$,
 $\varphi_1$ \emph{semantically entails} $\varphi_2$, written
 $\varphi_1 \sentails \varphi_2$, just in case
 $\modelsof{\varphi_1} \subseteq \modelsof{\varphi_2}$.
   Likewise, for $\Phi \subseteq \allbinconstrgr{G}$
and $\varphi \in \allbinconstrgrsch{G}$
   (resp.\ $\Phi' \subseteq \allbinconstrgr{G}$),
 write $\Phi \sentails \varphi$ (resp.\ $\Phi \sentails \Phi'$)
 just in case
 $\modelsof{\Phi} \subseteq \modelsof{\varphi}$.
 (resp.\  $\modelsof{\Phi} \subseteq \modelsof{\Phi'}$.
 \end{metalabpara}

 \begin{metalabpara}{mydefinition}{}
     {Positive binary tautologies}\envlabel{def:pbintaut}
    In standard logic terminology, a \emph{tautology} is a statement
which is true in every model.  Define the \emph{positive binary
tautologies} of $\granspacename{G}$ as follows.
  \begin{equation*}
  \begin{split}
     \textstyle
    \pbintautgr{G} =\;\phantom{\union}
        &\setdef{\subrulep{g}{g}}{g \in \granulesof{G}} \\
        \union\;
        &\setdef{\subrulep{\botgrsp{G}}{g}}{g \in \granulesof{G}} \\
        \union\;
        &\setdef{\subrulep{g}{\topgrsp{G}}}{g \in \granulesof{G}} \\
        \union\;
        &\setdef{\disjrulept{\botgrsp{G}}{g}}{g \in \granulesof{G}}
  \end{split}
  \end{equation*}
    That $\pbintautgr{G}$ consists of exactly those members of
$\pbinconstrgr{G}$ which are tautologies is established next.
 \end{metalabpara}

 \begin{metaemphlabpara}{proposition}{Proposition}
     {Characterization of positive binary tautologies}\envlabel{prop:pbintaut}
    Let $\granspacename{G}$ be a granule space.  Then $\pbintautgr{G}$
is precisely the subset of $\pbinconstrgr{G}$ consisting of
tautologies.  More precisely, if $\varphi \in \pbinconstrgr{G}$, then
$\varphi \in \pbintautgr{G}$ iff $\sigma \in \modelsof{\varphi}$ for
every granule structure $\sigma$ over $\granspacename{G}$.
 \begin{proof}
   First of all, it is immediate from the definition of constraint
satisfaction in \envref{def:granstrsem} that every
 $\varphi \in \pbintautgr{G}$ is a tautology; that
 $\sigma \in \modelsof{\varphi}$ for every granule structure $\sigma$
over $\granspacename{G}$.
   To show they are the only tautologies, choose any
 $\varphi \in \pbinconstrgr{G} \setminus \pbintautgr{G}$.
 If $\varphi \in \psubconstrgr{G}$, then it must be of the form
$\subrulep{g_1}{g_2}$ with
 $g_1 \neq \botgrsp{G}$, $g_2 \neq \topgrsp{G}$, and
 $g_1 \not\ideq g_2$.
 Define $\gnlestrpr{\sigma'}$ to have
 $\gnledom{\sigma'} = \setbr{x_1,x_2}$ (any two-element set) with
 $\gnletodom{\sigma'}(g_1) = \setbr{x_1,x_2}$ and
 $\gnletodom{\sigma'}(g_2) = \setbr{x_2}$.
 Clearly, $\sigma'$ is a granule structure which is not a model of
$\varphi$, so the latter cannot be a tautology.
 On the other hand, if $\varphi \in \pdisjconstrgr{G}$, then it must
be of the form $\disjrulep{g_1}{g_2}$ with
 $g_1 \neq \botgrsp{G}$ and $g_2 \neq \botgrsp{G}$.
 Then $\sigma'$, as defined above, is not a model of $\varphi$, so
again the latter is not a tautology.
 \end{proof}
 \end{metaemphlabpara}

 \begin{metalabpara}{mydefinition}{}
     {Unsatisfiable positive binary constraints}\envlabel{def:pbinunsat}
   Define the \emph{unsatisfiable positive binary constraints} of
$\granspacename{G}$ as follows.
  \begin{equation*}
  \begin{split}
     \textstyle
    \pbinunsatgr{G} =\;\phantom{\union}
        &\setdef{\subrulep{g}{\botgrsp{G}}}{g \in \granulesofnb{G}} \\
        \union\;
        &\setdef{\disjrulept{g}{g}}{g \in \granulesofnb{G}}
  \end{split}
  \end{equation*}
    That $\pbinunsatgr{G}$ consists of exactly those members of
$\pbinconstrgr{G}$ which are unsatisfiable is established next.
 \end{metalabpara}

 \begin{metaemphlabpara}{proposition}{Proposition}
     {Characterization of unsatisfiable positive binary constraints}\envlabel{prop:pbinunsat}
    Let $\granspacename{G}$ be a granule space.  Then
$\pbinunsatgr{G}$ is precisely the subset of $\pbinconstrgr{G}$
consisting of unsatisfiable assertions.  More precisely, if
 $\varphi \in \pbinconstrgr{G}$, then $\varphi \in \pbinunsatgr{G}$
iff for no granule structure $\sigma$ over $\granspacename{G}$ is it
the case that $\sigma \in \modelsof{\varphi}$.
 \begin{proof}
   First of all, it is immediate from the definition of constraint
satisfaction in \envref{def:granstrsem} that every
 $\varphi \in \pbinunsatgr{G}$ is unsatisfiable; that for no granule
structure $\sigma$ over $\granspacename{G}$ is it the case that
 $\sigma \in \modelsof{\varphi}$ .
   To show they are the only unsatisfiable members of
$\pbinconstrgr{G}$, choose any
 $\varphi \in \pbinconstrgr{G} \setminus \pbinunsatgr{G}$.
 If $\varphi \in \psubconstrgr{G}$, then it must be of the form
$\subrulep{g_1}{g_2}$ with $g_2 \neq \botgrsp{G}$.
 If $g_1 = \botgrsp{G}$, then $\varphi$ is a tautology, and so
trivially satisfiable.
 Otherwise, define $\gnlestrpr{\sigma'}$ to have
 $\gnledom{\sigma'} = \setbr{x_1,x_2}$ (any two-element set) with
 $\gnletodom{\sigma'}(g_1) = \setbr{x_1}$.
 and
 $\gnletodom{\sigma'}(g_1) = \setbr{x_1,x_2}$.
 Clearly, $\sigma'$ is a granule structure which is a model of
$\varphi$, so the latter is satisfiable.
 On the other hand, if $\varphi \in \pdisjconstrgr{G}$, then it must
be of the form $\disjrulep{g_1}{g_2}$ with either $g_1 \not\ideq g_2$
or else $g_1 = g_2 = \botgrsp{G}$.
 If $g_1 = g_2 = \botgrsp{G}$, then $\disjrulep{g_1}{g_2}$ is a
tautology and there is nothing further to prove.  Otherwise, define
$\gnlestrpr{\sigma''}$ to have
 $\gnledom{\sigma''} = \setbr{x_1,x_2}$ (any two-element set) with
 $\gnletodom{\sigma'}(g_1) = \setbr{x_1}$ and
 $\gnletodom{\sigma'}(g_2) = \setbr{x_2}$.
 Clearly, $\sigma''$ is a granule structure which is a model of
$\varphi$, rendering it satisfiable.
 \end{proof}
 \end{metaemphlabpara}

 \begin{metalabpara}{definition}{}
     {Simple monogranular attribute schemata}\envlabel{def:smas}
   A \emph{simple monogranular attribute
   \preformat{\linebreak}
   schema}, or \emph{SMAS}, is a
pair $\smasdef{G}$ in which $\granspace{G}$ is a granule space and
 $\constrgrsch{G} \subseteq  \allbinconstr{\granspace{G}}$.
   To understand the concepts of \envref{def:granstrsem} in the
context of this SMAS, it is only necessary to replace
$\granspacename{G}$ with $\granspace{G}$ everywhere.
   However, as they occur frequently, $\botgrsch{G}$
(resp.\ $\topgrsch{G}$) will often be used as synonyms for the more
cumbersome $\botgrschfull{G}$ (resp.\ $\topgrschfull{G}$).
   \par
   Define
   $\pconstrgrsch{G} = \constrgrsch{G} \intersect \pbinconstrgrsch{G}$.
 Thus, $\pconstrgrsch{G}$ is the set of all positive constraints in
$\constrgrsch{G}$.
   Similarly, define
   $\nconstrgrsch{G} = \constrgrsch{G} \intersect \nbinconstrgrsch{G}$.
    \par
   Define the \emph{closure} of $\constrgrsch{G}$ to be
  \nlrightt
    $\clconstrgrsch{G}
       = \setdef{\varphi \in \allbinconstrgrsch{G}}
                {\constrgrsch{G} \sentails \varphi}$.
    \par
   A granular structure $\sigma$ is a \emph{model} of
$\granschemaname{G}$ precisely in the case that it is a model of
$\constrgrsch{G}$.
   Call an SMAS $\granschemaname{G}$ \emph{satisfiable} if
$\constrgrsch{G}$ has that property, and \emph{unsatisfiable}
otherwise.
 \end{metalabpara}

 \begin{metalabpara}{definition}{}
     {Equivalence and identity of granules}\envlabel{def:eqidsep}
     It is important to distinguish two distinct notions of equality
between granules.
     Relative to $\granschemaname{G}$, two granules
 $g_1, g_2 \in \granulesofsch{G}$ are \emph{equivalent} if
  $\constrgrsch{G} \sentails
         \setbr{\subrulep{g_1}{g_2},\subrulep{g_2}{g_1}}$.
   Thus, two granules are equivalent if they have the same value under
every granule structure which is a model.
   On the other hand, the two granules
 $g_1, g_2 \in \granulesofsch{G}$ are \emph{identical} if they are the
same granule; that is, if they have the same name.
   Clearly, two distinct granules may be equivalent without being
identical.
   In order to distinguish these two, the expression
$\eqrulep{g_1}{g_2}$ will always mean that $g_1$ and $g_2$ are
equivalent.
   To express identity, the notation $\idrulep{g_1}{g_2}$ is always
used.
    \par
 \end{metalabpara}

 \begin{metalabpara}{convention}{}
     {Implicit symmetry of disjunction}\envlabel{conv:disjsym}
    Throughout this work, for any $g_1, g_2 \in
 \preformat{\linebreak}
 \granulesofsch{G}$, $\disjrulep{g_1}{g_2}$ and $\disjrulep{g_2}{g_1}$
will always be regarded as the same constraint.  Clearly, the
semantics does not depend upon any order on $g_1$ and $g_2$.  This
will eliminate the need for largely trivial inference rules which
would otherwise be required to assure that
 $\disjrulep{g_1}{g_2}$ and $\disjrulep{g_2}{g_1}$,
 as well as
 $\ndisjrulep{g_1}{g_2}$ and $\ndisjrulep{g_2}{g_1}$,
  are the same.
   \par
   It will also permit writing constraints of the form
$\disjrulesetp{S}$ and $\ndisjrulesetp{S}$, in which $S$ is a nonempty
set consisting of at most two granules.  If $S = \setbr{g_1,g_2}$,
then $\disjrulesetp{S}$ represents indifferently both
$\disjrulep{g_1}{g_2}$ and $\disjrulep{g_2}{g_1}$, while if
 $S = \setbr{g}$, then $\disjrulesetp{S}$ represents
$\disjrulep{g}{g}$.
   The meaning of $\ndisjrulesetp{S}$ is analogous.
 \end{metalabpara}

 \begin{metalabpara}{discussion}{}
     {Relationship of SMASs to CGASs}\envlabel{disc:cmascgas}
     This section provides a brief comparison of the SMAS, as defined
in \envref{def:smas} and the CGAS, as defined in
\mycite[Sec.\ 2]{HegnerRo18_dexa} and the earlier \emph{granularity
schema}, as defined in \mycite[Sec.\ 3]{HegnerRo17_inform}.  The
reader who is not familiar with these latter concepts, or who is not
interested in the comparison, may safely skip this discussion.
     \par
    A CGAS $\cgaschdefii{\gsn}$ consists of three components,
   a \emph{granularity poset} $\granposet{\gsn}$,
   a \emph{granule assignment} $\grasgn{\gsn}$,
   and a set $\allconstr{\gsn}$ of constraints.
   Relative to such a CGAS, an SMAS $\smasdef{G}$ differs in three
fundamental ways.
    \begin{axiomsd}{1em}{1em}{0em}
     \axitem{(a)} In an SMAS, there is no granularity poset, since
there is no classification of granules into granularities.
     \axitem{(b)} The granule space $\granspace{G}$ of the SMAS
corresponds roughly to the granule assignment $\grasgn{\gsn}$ of the
CGAS.  However, since there is no granularity hierarchy in an SMAS,
the classification of granules into granularities, which is embedded
in $\grasgn{\gsn}$ in the CGAS, is absent in $\granspace{G}$.
     \axitem{(c)} In a CGAS, some binary constraints, are already
embedded in $\grasgn{\gsn}$, including in particular those which arise
from the classification of granules into granularities.  More
specifically, the granule preorder $\gnleleqb{\gsn}$, as well as the
constraints arising from the requirement that granules of distinct
granularities be disjoint, are already embedded in the granule
assignment.  On the other hand, in a SMAS, no constraints are embedded
in the granule space; constraints are only embedded in
$\constrgrsch{G}$.
     \end{axiomsd}
 \end{metalabpara}


 \section{Background on Proof Systems}\label{sec:proofsys}

   The results of this paper are based upon simple notions of
mathematical logic in general, and of propositional logic in
particular.  There are many excellent presentations of those subjects,
at various levels and from various perspectives, including, but not
limited to, \mycite{Monk76},
\mycite{Mendelson97_book}, \mycite{Enderton01_book},
\mycite{BenAri12_book}, and \mycite{Gallier15}.
  It is assumed that the reader is familiar with the basic ideas
presented in the appropriate chapters of those books.
   The purpose of this section is to establish notation and
terminology for proof systems.

 \begin{metalabpara}{notation}{}
     {Notation}\envlabel{not:not3}
    $\natnum$ denotes the set $\setbr{0,1,2,\ldots}$ of natural
numbers.
  Unless stated specifically to the contrary, an expression of the
form $\ccinterval{n_1}{n_2}$, with $n_1,n_2 \in \natnum$, denotes the
interval $\setdef{n \in \natnum}{n_1 \leq n \leq n_2}$.
    \par
  $\cardof{S}$ denotes the cardinality of the set $S$.
 \end{metalabpara}

 \begin{metalabpara}{context}{}
     {Context --- SMAS}\envlabel{ctxt:gransch}
   Unless stated specifically to the contrary, for the rest of this
paper, $\smasdef{G}$ will denote an arbitrary SMAS. 
 \end{metalabpara}

 \begin{metalabpara}{mydefinition}{}
     {WFFs for granule logic}\envlabel{def:granwff}
     At first glance, it might seem that $\allbinconstrgrsch{G}$ forms
a suitable set over which to express inference rules for binary
granule constraints.  However, that choice is too limiting.  For
example, to express that the subsumption relation $\gnleleq{}$ is
transitive, it would be necessary to have a separate rule of the form
   \[
    \tag{\ref{def:granwff}-1}
    \begin{prooftree}
       \hypoii{\subrulep{g_1}{g_2}}
              {\subrulep{g_2}{g_3}}
       \inferi{\subrulep{g_1}{g_3}}
    \end{prooftree}
   \]
 for each triple $\abr{g_1,g_2,g_3}$ of granules.  It would be far
better to be able to assert, with a single rule, that that such a
relationship holds for any triple of granules.  To this end, it is
appropriate to introduce a set
   $\setbr{\granvar{g}_1,\granvar{g}_2,\ldots,\granvar{g}_i,\ldots}$
 of granule variables, with the understanding that each such variable,
by default, may be bound to any granule.  The general transitivity
rule then becomes
   \[
    \tag{\ref{def:granwff}-2}
    \begin{prooftree}
       \hypoii{\subrulep{\granvar{g}_1}{\granvar{g}_2}}
              {\subrulep{\granvar{g}_2}{\granvar{g}_3}}
       \inferi{\subrulep{\granvar{g}_1}{\granvar{g}_3}}
    \end{prooftree}
   \]
 with $\abr{\granvar{g}_1,\granvar{g}_2,\granvar{g}_3}$
 representing any triple of granules, as already illustrated in
Fig.\ \ref{fig:infrulespmain}.
    In the case that variable may only be bound to certain granules,
an explicit qualifier may be used, as already illustrated in the rules
of Figs.\ \ref{fig:infrulespunsat} and \ref{fig:infrulesntaut} by
the use of  $|{\scriptscriptstyle(\granvar{g} \neq \bot)}$.
     \par
    To formalize this, begin by associating with $\granschemaname{G}$
a countable set
   $\varsetgrsch{G} =
 \preformat{\linebreak}
     \setbr{\granvar{g}_1,\granvar{g}_2,\ldots,\granvar{g}_i,\ldots}$
 of \emph{granule variables}, with
  $\varsetgrsch{G} \intersect \granulesofsch{G} = \emptyset$.
    Relative to this set, the \emph{positive binary well-formed
formulas} (or \emph{positive binary WFFs}) of $\granschemaname{G}$,
denoted $\pbinwffgrsch{G}$, is the set consisting of all expressions
of the forms $\subrulep{\gamma_1}{\gamma_2}$ and
$\disjrulep{\gamma_1}{\gamma_2}$ with
  $\gamma_1, \gamma_2 \in \granulesofsch{G} \union \varsetgrsch{G}$.
    In other words, $\pbinwffgrsch{G}$ is similar to
$\pbinconstrgrsch{G}$, except that both granule names and granule
variables may be used in the expressions.  Thus, in addition to the
members of $\pbinconstrgrsch{G}$, $\pbinwffgrsch{G}$ includes all
expressions of the forms
   $\subrulep{\granvar{g}_1}{g_2}$,
   $\subrulep{g_1}{\granvar{g}_2}$,
   $\subrulep{\granvar{g}_1}{\granvar{g}_2}$,
   $\disjrulep{\granvar{g}_1}{g_2}$,
   $\disjrulep{g_1}{\granvar{g}_2}$,
  and
   $\disjrulep{\granvar{g}_1}{\granvar{g}_2}$,
  for $\granvar{g}_1, \granvar{g}_2 \in \varsetgrsch{G}$
  and $g_1, g_2 \in \granulesofsch{G}$.
    Similarly, the \emph{negative binary WFFs} of $\granschemaname{G}$,
denoted $\nbinwffgrsch{G}$, is the set consisting of all expressions
of the forms $\nsubrulep{\gamma_1}{\gamma_2}$ and
$\ndisjrulep{\gamma_1}{\gamma_2}$ with
  $\gamma_1, \gamma_2 \in \granulesofsch{G} \union \varsetgrsch{G}$.
  Finally, the set of \emph{all binary WFFs} of $\granschemaname{G}$,
denoted $\allbinwffgrsch{G}$, is
  $\pbinwffgrsch{G} \union \nbinwffgrsch{G}$.
    \par
  A more formal and complete development of inference rules is found
in \envref{def:ruleproofnot} and that which follows.
 \end{metalabpara}

 \begin{metalabpara}{context}{}
     {Context --- Granule variables}\envlabel{ctxt:granvarset3}
   Unless stated specifically to the contrary, for the rest of this
paper,
   $\varsetgrsch{G} =
     \setbr{\granvar{g}_1,\granvar{g}_2,\ldots,\granvar{g}_i,\ldots}$
 will denote a countable set of granule variables for the SMAS
$\smasname{G}$ identified in \envref{ctxt:gransch}.
 \end{metalabpara}

 \begin{metalabpara}{mydefinition}{}
     {Semantics of WFFs}\envlabel{disc:semwff}
   To extend the notions of semantic entailment and model of
\envref{def:granstrsem} to WFFs involving variables, it is necessary
to bind those variables to granules.
 More precisely, define a \emph{substitution} for $\granschemaname{G}$
to be a possibly partial function
 $\fn{s}{\varsetgrsch{G}}{\varsetgrsch{G}\union\granulesofsch{G}}$.
  Define $\substgr{\varphi}{s}$ to be the WFF in which each variable
$\granvar{g}$ in the domain of $s$ is replaced by $s(\granvar{g})$.
  For example, if
   $s_0 = \setbr{\elfd{\granvar{g}_1}{g_3},
              \elfd{\granvar{g}_2}{g_4},
              \elfd{\granvar{g}_3}{\granvar{g}_3}}$,
  then
    $\substgr{\subrulep{\granvar{g}_1}{\granvar{g}_2}}{s_0}
        = \subrulep{g_3}{g_4}$.
  A \emph{ground} substitution $s$ is one for which
     $s(\varsetgrsch{G}) \subseteq \granulesofsch{G}$;
 in other words, for every $\granvar{g}$ for which s(\granvar{g}) is
defined, $s(\granvar{g})$ is a granule, never a variable.
   A substitution \emph{complete} for
 $\varphi \in \allbinwffgrsch{G}$
 if it is defined for every $\granvar{g} \in \varsetgrsch{G}$ which
occurs in $\varphi$.
  If $s$ is ground and complete for $\varphi$, say that $\varphi$
  \emph{holds for $s$} in a granule structure $\sigma$ 
 if $\substgr{\varphi}{s}$ holds in $\sigma$.
  Thus, a WFF is assigned a truth value once all of its variables have
been bound to granules.
   Now, for $\varphi_1, \varphi_2 \in \allbinwffgrsch{G}$,
  $\varphi_1 \sentails \varphi_2$, precisely in the case that
  $\substgr{\varphi_1}{s} \sentails \substgr{\varphi_2}{s}$ for every
ground substitution $s$ which is complete for both $\varphi_1$ and
$\varphi_2$.
   \par
  These ideas extend easily to $\Phi \subseteq \allbinwffgrsch{G}$.
  Define
   $\substgr{\Phi}{s}
      = \setdef{\substgr{\psi}{s}}{\psi \in \Phi}$,
 say that $s$ is \emph{complete} for $\Phi$ if it is complete for each
$\psi \in \Phi$, and for $\varphi \in \allbinwffgrsch{G}$, define
  $\Phi \sentails \varphi$ to hold iff
  $\substgr{\Phi}{s} \sentails \substgr{\varphi}{s}$ for every ground
substitution $s$ which is complete for both $\Phi$ and $\psi$,
  Similarly, for $\Phi' \subseteq \allbinwffgrsch{G}$,
    $\Phi \sentails \Phi'$ holds iff
  $\substgr{\Phi}{s} \sentails \substgr{\Phi'}{s}$ for every ground
substitution $s$ which is complete for $\Phi \union \Phi'$.
    \par
   Finally, satisfiability is always defined with respect to ground
substitutions.  More precisely,
 $\varphi \in \allbinwffgrsch{G}$
 (resp.\ $\Phi \subseteq \allbinwffgrsch{G}$)
 is \emph{satisfiable} if there is a ground substitution $s$ which is
complete for $\varphi$ (resp.\ for $\Phi$) with
n \mbox{$\modelsof{\substgr{\varphi}{s}} \neq \emptyset$}
 (resp.\
 \preformat{\linebreak}
 $\modelsof{\substgr{\Phi}{s}} \neq \emptyset$).
     \par
    It is easy to see that these notions for $\allbinwffgrsch{G}$
generalize those of \envref{def:granstrsem} for
$\allbinconstrgrsch{G}$.
 \end{metalabpara}

 \begin{metalabpara}{context}{}
     {General logical systems and negation}\envlabel{disc:logicsys}
    Many of the concepts introduced in the remainder of this section,
as well as in Secs.\ \ref{sec:negarm} and \ref{sec:ninfrules}, apply
in a more general setting than just $\allbinwffgrsch{G}$.  In this
spirit, the framework of a general logical system $\logicsys{L}$ will
be assumed as a context, as needed.  This context includes notions of
WFFs (denoted $\wffof{L}$), structure, model, and semantic entailment
(denoted $\sentails$).  It is assumed that $\logicsys{L}$ is
\emph{nontrivial} in that there is at least one
 $\varphi \in \wffof{L}$ which is satisfiable, and that it contains at
least one tautology $\true$ and one unsatisfiable assertion $\false$.
It is furthermore assumed that $\logicsys{L}$ has a negation operator,
with $\mlnot\varphi$ denoting the negation of $\varphi \in \wffof{L}$,
$\mlnot\mlnot\varphi$ equivalent to $\varphi$, and excluded middle
($\varphi$ and $\mlnot\varphi$ cannot both be satisfiable).  Further
detailed properties of $\logicsys{L}$ will not be developed here,
except as necessary in for a particular issue.  However, it is
worthwhile noting that there is no assumption that $\logicsys{L}$ be
closed under conjunction or disjunction.  In fact, the main example of
$\logicsys{L}$ in this paper, $\allbinwffgrsch{G}$, is closed under
neither.
     \par
     For readers not comfortable with this generalization, it suffices
to fix $\wffof{L}$ to be $\allbinwffgrsch{G}$ and $\sentails$ to be
the semantic entailment relation for that context.  The assertion
$\true$ (resp.\ $\false$) may be taken to be (a synonym for) any
element of $\pbintautgrsch{G}$ (resp.\ $\pbinunsatgrsch{G}$).
     In $\allbinwffgrsch{G}$, to avoid unnecessary pedantry,
$\nsubrulep{g_1}{g_2}$ (resp. $\ndisjrulep{g_1}{g_2}$), will be
regarded as the same WFF as $\mlnot\subrulep{g_1}{g_2}$
(resp. $\mlnot\disjrulep{g_1}{g_2}$).
   \end{metalabpara}

 \begin{metalabpara}{discussion}{}
     {Proof systems}\envlabel{disc:proofsys}
     Semantic entailment is a purely model-based means of relating
WFFs.  In order to deduce that a set of formulas implies
another; that is, to determine whether $\Phi \sentails \varphi$ holds,
a computational procedure is necessary.  This is the r{\^o}le of a
proof system.
     \par
     More precisely, a {proof system} $\proofsys{A}$ for a set
$\wffset{A} \subseteq \wffof{L}$ consists of a set of inference rules,
denoted
 $\infrulesof{A}$, by which new WFFs may be obtained from others. Let
$\Phi \subseteq \wffset{A}$ and let
 $\varphi \in \wffset{A}$.
  Write
 $\Phi \syntailss[\proofsys{A}] \varphi$ just in case
 $\varphi$ may be obtained from $\Phi$ by repeated application of the
inference rules of $\proofsys{A}$.
  Write
 $\Phi \syntailssi[\proofsys{A}] \varphi$ just in case
 $\varphi$ may be obtained from $\Phi$ by a single application of a
single inference rule of $\proofsys{A}$.
     \par
     The proof system $\proofsys{A}$ is \emph{sound} if whenever $\Phi
\syntailss[\proofsys{A}] \varphi$, then
 $\Phi \sentails \varphi$.  In words, only conclusions which are true
(in the sense of $\sentails$) may be proven.  Soundness is always an
essential property of a proof system.
     \par
     The proof system $\proofsys{A}$ is \emph{complete} if
whenever $\Phi \sentails \varphi$, then
 $\Phi \syntailss[\proofsys{A}] \varphi$.  In words, all inferences
which are true may be proven using $\proofsys{A}$.
 While not essential, completeness is nevertheless a highly desirable
property.
     \par
     The proof system $\proofsys{A}$ is \emph{decidable} if there is
an algorithm (which always terminates) for determining whether or not
$\Phi \syntailss[\proofsys{A}] \varphi$ holds.  All proof systems
considered in this report are decidable, since the set $\wffset{A}$
will always be finite.
 \end{metalabpara}

 \begin{metalabpara}{mydefinition}{}
       {Sat-conditional completeness}\envlabel{def:satcompl}
    A proof system $\proofsys{A}$ for $\wffset{A}$ is said to be
\emph{sat-complete} if for any satisfiable
  $\Phi \subseteq \wffset{A}$ and any $\varphi \in \wffset{A}$,
 $\Phi \sentails \varphi$ implies
  $\Phi \syntailss[\proofsys{A}] \varphi$.
   \par
    Thus, $\proofsys{A}$ is complete as long as the input set $\Phi$
is satisfiable, but it need not be complete in the case that $\Phi$ is
not satisfiable.
    (Of course, if $\Phi$ is not satisfiable, then
  $\Phi \sentails \varphi$ holds trivially, so sat-completeness is not
a serious limitation.  It does, require, however, that an alternate
means of detecting unsatisfiability be used.)
 \end{metalabpara}

 \begin{metalabpara}{mydefinition}{}
        {Notation for rules and proofs}\envlabel{def:ruleproofnot}
    The representation of proof rules employed in this paper is widely
used; the purpose here is to clarify terminology and notation.  A
generic rule is illustrated in (\ref{def:ruleproofnot}-1) below.
     \[
       \tag{\ref{def:ruleproofnot}-1}
       \begin{prooftree}
         \hypoi{\alpha_1~\alpha_2~\ldots~\alpha_k}
         \inferi[(r)]{\beta}
       \end{prooftree}
     \]
   The set $\setdef{\alpha_i}{i \in \ccinterval{1}{k}}
          \subseteq \wffof{L}$
consists of the \emph{antecedents} of the rule
 and is denoted $\antecedentsof{r}$.
   Similarly, $\beta \in \wffof{L}$ is the \emph{consequent},
 and is denoted $\consequentof{r}$.
    This rule denotes exactly that
  $\setdef{\alpha_i}{i \in \ccinterval{1}{k}}
                      \syntailss[\proofsys{A}] \beta$
  using a single application of the rule named $r$.
    If the rule name is clear from context, it may be omitted.
  If $k=0$; \ie, if
   $\setdef{\alpha_i}{i \in \ccinterval{1}{k}} = \emptyset$,
 then the rule expresses that $\beta$ is a tautology.  If $\beta$ is
$\false$, the rule expresses that
  $\alpha_1 \mland \alpha_2 \mland \ldots \mland \alpha_k$
 is unsatisfiable.
     \par
   As already suggested in \envref{def:granwff}, substitutions may be
applied to rules (as well as formulas).  For example,
(\ref{def:granwff}-1) is the result of applying the substitution
   $s_1 = \setbr{\elfd{\granvar{g}_1}{g_1},
              \elfd{\granvar{g}_2}{g_2},
              \elfd{\granvar{g}_3}{g_3}}$,
to (\ref{def:granwff}-2).
   If the set of substitutions is restricted, this is expressed via an
appropriate expression in angle brackets, as illustrated in
Figs. \ref{fig:infrulesptaut} and \ref{fig:infrulesntaut}.
     \par
   The result of applying a substitution to a rule is called an
\emph{instance} of that rule.  It will sometimes be useful to label a
rule instance in a proof.  For example, (\ref{def:ruleproofnot}-2) below
shows a single instance labelled with $d$.  In order to distinguish
rule names from labels, the former will always be written in
parentheses (as illustrated in (\ref{def:ruleproofnot}-1), while labels
will never be enclosed in parentheses.  Note that while a rule will in
general have only one name, there may be several instances of it (and
hence labels for those instances) in a single proof.
     \[
       \tag{\ref{def:ruleproofnot}-2}
       \begin{prooftree}
         \hypoi{\alpha_1~\alpha_2~\ldots~\alpha_k}
         \inferi[d]{\beta}
       \end{prooftree}
     \]
   As examples,
Figs.\ \ref{fig:infrulespmain}--\ref{fig:infrulesunsatpr} provide the
rules on granules which will be studied in this paper.  Later in the
paper, \envref{def:bininfpos}, \envref{def:swapclpbininf}, and
\envref{def:compprrules} provide the same rules, with names.
   \par
    A great advantage of the notation of \envref{def:ruleproofnot} is
that proofs involving the application of several rules may be
presented in a clear fashion visually.  For example, formulas
(\ref{def:ruleproofnot}-3) and (\ref{def:ruleproofnot}-4) below show
two ways of proving $\subrulep{g_1}{g_5}$ from the set
 \nlrightt
   $S_{\scriptsize\mbox{\ref{def:ruleproofnot}}}
      = \setbr{\subrulep{g_1}{g_2},
           \subrulep{g_2}{g_3},
           \subrulep{g_3}{g_4},
           \subrulep{g_4}{g_5}}$,
    \newline
 using the rule to the left in Fig.\ \ref{fig:infrulespmain}.
    The various deduction instances have been labelled, with
$d_1$--$d_3$ in (\ref{def:ruleproofnot}-3) and with $d_1'$--$d_3$ in
(\ref{def:ruleproofnot}-4).  All are instances of the rule
(\ref{def:granwff}-2).
 \begin{gather*}
    \tag{\ref{def:ruleproofnot}-3}
    \begin{prooftreem}
     \hypoii{\subrulep{g_1}{g_2}}
            {\subrulep{g_2}{g_3}}
     \inferi[d_1]{\subrulep{g_1}{g_3}}
     \hypoi{\subrulep{g_3}{g_4}}
     \inferii[d_2]{\subrulep{g_1}{g_4}}
     \hypoi{\subrulep{g_4}{g_5}}
     \inferii[d_3]{\subrulep{g_1}{g_5}}
    \end{prooftreem}
     \\[0.9em]
    \tag{\ref{def:ruleproofnot}-4}
    \begin{prooftreem}
     \hypoii{\subrulep{g_1}{g_2}}
            {\subrulep{g_2}{g_3}}
     \inferi[d_1']{\subrulep{g_1}{g_3}}
     \hypoii{\subrulep{g_3}{g_4}}
            {\subrulep{g_4}{g_5}}
     \inferi[d_2']{\subrulep{g_3}{g_5}}
     \inferii[d_3']{\subrulep{g_1}{g_5}}
    \end{prooftreem}
    \end{gather*}
    Diagrams such as (\ref{def:ruleproofnot}-3) and
(\ref{def:ruleproofnot}-3) are called \emph{proof trees} (for the
inference system under consideration).
     \par
    The \emph{antecedents} of a combined rule are the formulas which
are at the top level.  In both (\ref{def:ruleproofnot}-3) and
(\ref{def:ruleproofnot}-4), the set of
antecedents is $S_{\scriptsize\mbox{\ref{def:ruleproofnot}}}$.
    Each combined rule has a single \emph{consequent}, which is at the
bottom.  In both examples, it is $\subrulep{g_1}{g_5}$.
     \par
     A formula which occurs in a proof tree, other than as an
antecedent or consequent, is called an \emph{internal formula}.
     In (\ref{def:ruleproofnot}-3), the set
of internal formulas is
 $\setbr{\subrulep{g_1}{g_3}, \subrulep{g_1}{g_4}}$,
 while in (\ref{def:ruleproofnot}-4), it is
 $\setbr{\subrulep{g_1}{g_3}, \subrulep{g_3}{g_5}}$,
    \par
   Finally, a formula $\varphi$ which is regarded as an axiom will
always be regarded as a proof of itself.
 \end{metalabpara}

 \begin{metalabpara}{mydefinition}{}
        {Soundness and minimality of rules
                    and proof systems}\envlabel{def:minrule}
    Let $\proofsys{A}$ be a proof system for $\wffset{A}$, and let $r$
be a rule, as depicted in (\ref{def:minrule}-1) below.
     \[
       \tag{\ref{def:minrule}-1}
       \begin{prooftree}
         \hypoi{\alpha_1~\alpha_2~\ldots~\alpha_k}
         \inferi[(r)]{\beta}
       \end{prooftree}
     \]
  Call $r$ \emph{sound} if it produces consistent inference by
itself; \ie, if
     $\setbr{\alpha_1,\alpha_2,\ldots,\alpha_k} \sentails \beta$.
   \par
  Call a sound $r$ \emph{minimal} if whenever any $\alpha_i$ is
removed from the antecedents, the resulting rule $r'$, as shown in
(\ref{def:minrule}-2) below, is no longer sound.
     \[
       \tag{\ref{def:minrule}-2}
       \begin{prooftree}
         \hypoi{\alpha_1~\alpha_2~\ldots~\alpha_{i-1}~~
                     \alpha_{i+1}~\ldots~ \alpha_k}
         \inferi[(r')]{\beta}
       \end{prooftree}
     \]
  Put another way, the rule $r$ is minimal if for any
 $j \in \ccinterval{1}{k}$,
  $\setdef{\alpha_i}{i \in \ccinterval{1}{k}\setminus\setbr{i}}
    \not\sentails \beta$.
    \par
   Say that $\proofsys{A}$ is \emph{sound}
(resp.\ \emph{minimal}) if all members of $\infrulesof{A}$ have that
property.
 \end{metalabpara}

 \begin{metalabpara}{mydefinition}{}
        {Single-use vs.\ multiple-use proofs}\envlabel{def:single}
    It may be the case that an antecedent of a proof is used several
times.  For example, given the set
 \nlrightt
   $S_{\scriptsize\mbox{\ref{def:single}}}
      = \setbr{\subrulep{g_1}{g_2},
           \subrulep{g_2}{g_3},
           \subrulep{g_3}{g_2},
           \subrulep{g_2}{g_3},
           \subrulep{g_3}{g_4}}$,
 \newline
 of constraints, in the proof shown in
 (\ref{def:single}-1) below, each of $\subrulep{g_1}{g_2}$,
$\subrulep{g_1}{g_3}$, and $\subrulep{g_1}{g_3}$ is used more than once.
 \[
    \tag{\ref{def:single}-1}
    \begin{prooftreem}
     \hypoii{\subrulep{g_1}{g_2}}
            {\subrulep{g_2}{g_3}}
     \inferi[d_1]{\subrulep{g_1}{g_3}}
     \hypoi{\subrulep{g_3}{g_2}}
     \inferii[d_2]{\subrulep{g_1}{g_2}}
     \hypoi{\subrulep{g_2}{g_3}}
     \inferii[d_3]{\subrulep{g_1}{g_3}}
     \hypoi{\subrulep{g_3}{g_4}}
     \inferii[d_4]{\subrulep{g_1}{g_4}}
    \end{prooftreem}
 \]
  Call such a proof \emph{multiple use}.  On the other hand, call
proofs in which no formula occurs as an antecedent more than once
\emph{single use}.  The proofs of (\ref{def:ruleproofnot}-1) and
(\ref{def:ruleproofnot}-2) are single use.
     \par
   Call a proof system \emph{single use} if for any set
 $\setbr{\alpha_1,\alpha_2,\ldots,\alpha_k}$ of formulas, and any
single formula $\beta$, if there is a proof of $\beta$ from 
 $\setbr{\alpha_1,\alpha_2,\ldots,\alpha_k}$, then there is a
single-use proof.
     \par
   It is worth noting now that even though the proof of
(\ref{def:single}-1) is multiple use, a single-use proof is also
possible, as shown in (\ref{def:single}-2) below.
 \[
    \tag{\ref{def:single}-2}
    \begin{prooftreem}
     \hypoii{\subrulep{g_1}{g_2}}
            {\subrulep{g_2}{g_3}}
     \inferi[d_1]{\subrulep{g_1}{g_3}}
     \hypoi{\subrulep{g_3}{g_4}}
     \inferii[d_4]{\subrulep{g_1}{g_4}}
    \end{prooftreem}
 \]
  The proof systems developed in this paper for granule rules
are all single use, as will be established later.
     \par
   The idea of ``using up'' formulas, without permitting their re-use,
was introduced by Girard in the context of \emph{linear logic}
\mycite{Girard95_all}.
   Since it is not central to the results of this paper, linear logic
will not be considered further.
 \end{metalabpara}

 \begin{metalabpara}{summary}{}
         {Armstrong models}\envlabel{summ:armstrong}
      Let $\armctxt{C}$ be any set of sentences in a framework for
logic.  Given $\Phi \subseteq \cal{C}$, an \emph{Armstrong model} $M$
of $\Phi$, relative to $\armctxt{C}$, has the property that for every
 $\varphi \in \cal{C}$, it is the case that $M \in \modelsof{\varphi}$
iff $\Phi \sentails \varphi$.
    In other words, an Armstrong model $M$ is minimal in the sense that it
satisfies only those constraints of $\cal{C}$ which must be
satisfied in order for $M$ to be a model of $\Phi$.
     Say that $\cal{C}$ \emph{admits Armstrong models}, or that it is
an \emph{Armstrong context}, if every satisfiable subset of
$\armctxt{C}$ has such a model.
    See \mycite{Fagin82} for a comprehensive discussion of Armstrong
models, including a succinct characterization of the conditions under
which they exist \mycite[Thm.\ 3.1]{Fagin82}.
    \par
    In the context of this paper, the important case is
  $\cal{C} = \pbinconstrgrsch{G}$.  While it follows easily from the
more general result of \mycite[3.18]{HegnerRo17_inform} that
$\pbinconstrgrsch{G}$ admits Armstrong models, a much stronger result
is established in this section.  Specifically, it is shown how to
construct an Armstrong model, called the canonical model, for
$\pconstrgrsch{G}$.  Since $\granschemaname{G}$ is arbitrary, this is
essentially equivalent to showing how to construct a canonical model
for any
 $\Phi \subseteq \pbinconstrgrsch{G}$.  Using that model, it is shown
that the positive inference rules identified in Sec.\ \ref{sec:intro}
are both sound and complete.
 \end{metalabpara}


 \section{Inference for Positive Binary Granular Constraints}\label{sec:pinfrules}

    In this section, a complete inference system is developed for
 $\pbinconstrgrsch{G}$; \ie, for binary rules of the form
\mbox{$\subrulep{g_1}{g_2}$} and \mbox{$\disjrulep{g_1}{g_2}$}.  It is
based substantially upon \mycite{AtzeniPa88_dke}, although significant
differences in the two frameworks (particularly, the absence of the
granules $\botgrsch{G}$ and $\topgrsch{G}$ in
\mycite{AtzeniPa88_dke}), as well as the need to ensure that proofs
are single use (a property not considered in \mycite{AtzeniPa88_dke}),
render the task substantial.
     \par
    While the presentation here is designed to be independent of
\mycite{AtzeniPa88_dke}, referring to the latter only for proofs,
readers may nevertheless be interested in consulting that work as a
reference point.  To that end, a brief summary of the terminology and
notation of \mycite{AtzeniPa88_dke}\footnote{
    \mycite{AtzeniPa86_icde} is a preliminary conference version of
the journal article \mycite{AtzeniPa88_dke}.  Readers who have
difficulty accessing \mycite{AtzeniPa88_dke} may find
\mycite{AtzeniPa86_icde} a useful alternative.}
 is provided below.

 \begin{metalabpara}{summary}{}
         {Summary of terminology, notation, and context of 
                \protect\mycite{AtzeniPa88_dke}}\envlabel{sum:AP}
    The work of \mycite{AtzeniPa88_dke} has its roots in general
knowledge bases, and so uses different terminology and notation than
does this report, whose roots are in the spatial domain.
     In \mycite{AtzeniPa88_dke}, the equivalent of an SMAS is called a
\emph{scheme}, with \emph{types} corresponding to the granules of an
SMAS, and \emph{objects} corresponding to the elements of the domain
$\gnledom{\sigma}$ of a granule structure $\sigma$.  A \emph{potential
state} assigns objects to types, much as the function
$\gnletodom{\sigma}$ assigns domain elements to granules.
    The subsumption constraint $\subrulep{g_1}{g_2}$ of this paper is
written $\isa{g_1}{g_2}$, while the disjointness constraint
$\disjrulep{g_1}{g_2}$ is written $\disj{g_1}{g_2}$.
     \par
    From the perspective of semantics, there are two significant
points of difference with the SMAS framework of this paper.  First,
there are no specially identified types corresponding to
$\botgrsch{G}$ and $\topgrsch{G}$, although it is possible to require
that a type $g$ be assigned the empty set of objects via the
constraint $\disj{g}{g}$.  Such a type is termed \emph{inconsistent},
with a scheme \emph{inconsistent} if it contains at least one
inconsistent type.  Second, the set of objects associated with a type
is not required to be nonempty.  This is in sharp contrast to the SMAS
framework, in which a granule structure assigns to each granule, other
than $\botgrsch{G}$, a \emph{nonempty} set of domain elements.
 \end{metalabpara}

 \begin{metalabpara}{mydefinition}{}
  {Granule quasi-structures and quasi-models}\envlabel{def:quasi}
     In \mycite{AtzeniPa88_dke}, a more relaxed version of (the notion
corresponding to) a granule structure $\sigma$ than employed in an
SMAS (see \envref{def:granstrsem}) is used, in which there are no
restrictions on the mapping $\gnletodom{\sigma}$.
     To keep the distinction clear, a structure in the sense of
\mycite{AtzeniPa88_dke} will be termed a quasi-structure.
     Formally, a \emph{granule quasi-structure} over
$\granschemaname{G}$ is a pair $\gnlestrpr{\sigma}$ in which
$\gnledom{\sigma}$ is a nonempty set and
  $\fn{\gnletodom{\sigma}}{\granulesofsch{G}}
                          {\powerset{\gnledom{\sigma}}}$
 is a function which assigns a subset of $\gnledom{\sigma}$ to each
granule, subject to the further property that
   $\bigunion\setdef{\gnletodom{\sigma}(g)}{g \in \granulesofsch{G}}
                        = \gnledom{\sigma}$.
     Note in particular that there is no requirement that granules
other than $\botgrsch{G}$ map to nonempty sets, and there are no
restrictions on the images of $\botgrsch{G}$ or $\topgrsch{G}$.
     There is, however, the weaker property that every domain element
occur in the image of some granule under $\gnletodom{\sigma}$.
     \par
   A granule quasi-structure $\sigma$ is a \emph{quasi-model} of the
subsumption constraint $\subrulep{g_1}{g_2}$ if
 $\gnletodom{\sigma}(g_1) \subseteq \gnletodom{\sigma}(g_2)$, and it
is a quasi-model of the disjointness constraint $\disjrulep{g_1}{g_2}$
if 
 $\gnletodom{\sigma}(g_1) \intersect \gnletodom{\sigma}(g_2)
                                    = \emptyset$.
   For $\Phi \subseteq \pbinconstrgrsch{G}$, $\sigma$ is a quasi-model
of $\Phi$ if it is a quasi-model of each $\varphi \in \Phi$.
   Finally, it is a quasi-model for $\granschemaname{G}$ if it is a
quasi-model of $\constrgrsch{G}$ (but not necessarily of
$\pbintautgrsch{G}$).
     \par
   The \emph{empty quasi-structure} over $\granschemaname{G}$, denoted
$\emptyqs{G}$, has $\gnledom{\emptyqs{G}} = \emptyset$ with
 $\gnletodom{\emptyqs{G}}$ the function which maps every granule to
$\emptyset$.  This quasi-structure is interesting primarily because it
is always a quasi-model, thus showing that any SMAS is satisfiable in
the context of quasi-models.
     \par
     To avoid confusion with quasi-models, the granule structures
(resp.\  models) of this
paper will sometimes be called \emph{full structures}
(resp.\ \emph{full models}).
 \end{metalabpara}

\begin{metalabpara}{summary}{}
  {Inference rules on quasi-models}\envlabel{sum:APinf}
    The inference rules of \mycite[Lem.\ 3]{AtzeniPa88_dke} are shown
in Fig.\ \ref{fig:qbininfpos}.
    Define the inference system $\qbininfpos{G}$ to have
  \nlrightt
  $\infrulesof{\qbininfpos{G}} = 
     \setbr{\text{(I1)}, \text{(I2)}, \text{(D1)},
            \text{(M1)}, \text{(M2)} }$.
   \newline
   \begin{figure}[htb]
   \def\isepi{0.5em}
   \begin{align*}
   &\mbox{(I2)}\hspace*{\isepi}
    \begin{prooftreem}
       \hypoii{\subrulep{\granvar{g}_1}{\granvar{g}_2}}
              {\subrulep{\granvar{g}_2}{\granvar{g}_3}}
       \inferi{\subrulep{\granvar{g}_1}{\granvar{g}_3}}
    \end{prooftreem}
   &&\mbox{(M1)}\hspace*{\isepi}
    \begin{prooftreem}
      \hypoii{\subrulep{\granvar{g}_1}{\granvar{g}_1'}}
              {\disjrulep{\granvar{g}_1'}{\granvar{g}_2}}
      \inferi{\disjrulep{\granvar{g}_1}{\granvar{g}_2}}
    \end{prooftreem}
    \end{align*}
    \begin{align*}
   &\mbox{(I1)}\hspace*{\isepi}
    \begin{prooftreem}
      \hypoi{\phantom{X}}
      \inferi{\subrulep{\granvar{g}}{\granvar{g}}}
    \end{prooftreem}
   &&\mbox{(D1)}\hspace*{\isepi}
    \begin{prooftreem}
      \hypoi{\disjrulep{\granvar{g}'}{\granvar{g}'}}
      \inferi{\disjrulep{\granvar{g}'}{\granvar{g}}}
    \end{prooftreem}
   &&&\mbox{(M2)}\hspace*{\isepi}
    \begin{prooftreem}
      \hypoi{\disjrulep{\granvar{g}'}{\granvar{g}'}}
      \inferi{\subrulep{\granvar{g'}}{\granvar{g}}}
    \end{prooftreem}
   \end{align*}
   \caption{Inference rules of $\qbininfpos{G}$}\label{fig:qbininfpos}
   \end{figure}
     The main rules are (I2) and (M1), which assert that subsumption
is transitive, and that disjointness is preserved under reduction of
granule size via subsumption, respectively.
     The rule (I1) asserts the tautology that every granule subsumes
itself.  These three rules make sense equally in the contexts of full
models and quasi-models.
     \par
     The remaining two rules, (D1) and (M2), assert that if a granule
is disjoint from itself, then it is also disjoint from every other
granule as well, and is also subsumed by every granule, respectively.
This makes sense in the context of quasi-models, since any granule in
a quasi-model may be required to be empty.  However, in the context of
full models, they rules apply only to the case that $\granvar{g}'$ is
bound to $\botgrsch{G}$.  This is explained in more detail in
\envref{def:bininfpos} and \envref{prop:bininfpos}.
 \end{metalabpara}

 \begin{metaemphlabpara}{proposition}{Proposition}
   {Soundness and completeness of $\qbininfpos{G}$}\envlabel{prop:qbininfpos}
   The inference rules of $\qbininfpos{G}$, shown in
Fig.\ \ref{fig:qbininfpos} are sound and complete in the context of
quasi-models.
    More precisely, they are complete when the concept of model used
is that of quasi-model.
 \begin{proof}
  See \mycite[Thm.\ 1]{AtzeniPa88_dke}.
 \end{proof}
 \end{metaemphlabpara}
    \parvert

   The strategy of augmenting $\qbininfpos{G}$ to obtain an inference
system which is complete for full models is a straightforward, and is
executed in two steps.  In the first, additional rules which enforce
all members of $\pbintautgrsch{G}$ as constraints are added, and in
the second, additional rules which require that the members of
$\pbinunsatgrsch{G}$ can never be satisfied are added.  The details
follow.

 \begin{metalabpara}{mydefinition}{}
  {Strong quasi-structures and quasi-models}\envlabel{def:squasistr}
     Call quasi-structure $\gnlestrprii{\sigma}$ over
$\granschemaname{G}$ \emph{strong} if
$\gnletodom{\sigma}(\botgrsch{G}) = \emptyset$ and
 $\gnletodom{\sigma}(\topgrsch{G}) = \gnledom{\sigma}$.
    A quasi-model of $\granschemaname{G}$ which is also a strong
quasi-structure is called a \emph{strong quasi-model} (of
$\granschemaname{G}$).
    \par
    Clearly, every full structure (resp.\ full model) is also a strong
quasi-structure (resp.\ strong quasi-model).  The only difference is
that in a strong quasi-structure $\sigma$, there is no requirement
that $\gnletodom{\sigma}(g) \neq \emptyset$ for $g \in
\granulesofschnb{G}$.
    \par
    Also, note that $\emptyqs{G}$ is always a strong quasi-model,
whatever be $\constrgrsch{G}$.  Thus, just as is the case for ordinary
quasi-structures, any SMAS is satisfiable in the context of strong
quasi-models.
 \end{metalabpara}

 \begin{metaemphlabpara}{proposition}{Proposition}
   {Tautologies and strong quasi-structures}\envlabel{prop:strongqstr}
    If $\sigma$ is a quasi-
 \preformat{\linebreak}
 structure over $\granschemaname{G}$ which
furthermore satisfies $\pbintautgrsch{G}$, then it is a strong
quasi-structure over $\granschemaname{G}$.
    Consequently, if $\sigma$ is a quasi-model of $\granschemaname{G}$
which satisfies $\pbintautgrsch{G}$, then it is a strong quasi-model.
 \begin{proof}
    The two differences between $\gnlestrpr{\sigma}$ being an ordinary
quasi-structure and a strong quasi-structure are that, in the latter,
it is additionally required that
 \preformat{\linebreak}
    (a) $\gnletodom{\sigma}(\botgrsch{G})=\emptyset$; and
    (b) $\gnletodom{\sigma}(\topgrsch{G})=\gnledom{\sigma}$.
  In a quasi-structure, neither of these conditions is required to hold.
  To prove the assertion, it suffices to show that requiring that all
constraints in $\pbintautgrsch{G}$ hold implies that (a) and (b) hold.
This is easily verified, as follows.
     \par
   For (a), letting $g \ideq \botgrsch{G}$ in
 $\disjrulep{\botgrsch{G}}{g} \in \pbintautgrsch{G}$,
 it follows that
 \preformat{\linebreak}
 $\gnletodom{\sigma}(\botgrsch{G})
     \intersect \gnletodom{\sigma}(\botgrsch{G}) = \emptyset$,
 which can only hold if
 $\gnletodom{\sigma}(\botgrsch{G}) = \emptyset$.
     \par
   For (b), the tautology
  $\subrulep{g}{\topgrsch{G}} \in \pbintautgrsch{G}$ guarantees that
 $\gnletodom{\sigma}(g) \subseteq \gnletodom{\sigma}(\topgrsch{G})$
 for every $g \in \granulesofsch{G}$, which with the additional
condition
 \preformat{\linebreak}
   $\bigunion\setdef{\gnletodom{\sigma}(g)}{g \in \granulesofsch{G}}
                        = \gnledom{\sigma}$
  (see \envref{def:quasi})
 guarantees that
     $\gnletodom{\sigma}(\topgrsch{G}) = \gnledom{\sigma}$,
 completing the proof.
 \end{proof}
 \end{metaemphlabpara}

 \begin{metalabpara}{mydefinition}{}
  {Extending $\qbininfpos{G}$
           to strong quasi-models}\envlabel{def:sqbininfpos}
    Define the inference system
 \preformat{\linebreak}
 $\sqbininfpos{G}$ to have
  \nlrightt
  $\infrulesof{\sqbininfpos{G}} =
        \infrulesof{\qbininfpos{G}} \union
     \setbr{\text{(T1)}, \text{(T2)}, \text{(T3)}}$,
   \newline
  with the rules (T1), (T2), and (T3) as defined in
Fig.\ \ref{fig:tautbininfpos}.
   \begin{figure}[htb]
   \def\isepi{0.5em}
   \begin{align*}
   &\mbox{(T1)}\hspace*{\isepi}
    \begin{prooftreem}
      \hypoi{\phantom{X}}
      \inferi{\disjrulep{\botgrsch{G}}{\granvar{g}}}
    \end{prooftreem} 
   &&\mbox{(T2)}\hspace*{\isepi}
    \begin{prooftreem}
      \hypoi{\phantom{X}}
      \inferi{\subrulep{\botgrsch{G}}{\granvar{g}}}
    \end{prooftreem}
   &&&\mbox{(T3)}\hspace*{\isepi}
    \begin{prooftreem}
      \hypoi{\phantom{X}}
      \inferi{\subrulep{\granvar{g}}{\topgrsch{G}}}
    \end{prooftreem}
    \end{align*}
   \caption{Additional tautological rules $\sqbininfpos{G}$}\label{fig:tautbininfpos}
   \end{figure}
 \end{metalabpara}

 \begin{metaemphlabpara}{proposition}{Proposition}
   {Soundness and completeness of $\sqbininfpos{G}$
     in the context of strong quasi-models}\envlabel{prop:sqbininfpos}
   The inference rules of $\sqbininfpos{G}$ are sound and complete in
the context of strong quasi-models.
    More precisely, they are complete when the concept of model used
is that of strong quasi-model.
 \begin{proof}
    The result follows from \envref{prop:strongqstr}, together with
the fact that the added rules (T1), (T2), and (T3) of
$\sqbininfpos{G}$, together with (I1) of $\qbininfpos{G}$, force all
constraints in $\pbintautgrsch{G}$ to hold.
 \end{proof}
 \end{metaemphlabpara}

 \begin{metaemphlabpara}{proposition}{Proposition}
   {Strong quasi-structures
        and $\pbinunsatgrsch{G}$}\envlabel{prop:strongqstrunsat}
    If $\sigma$ is a strong quasi-structure over $\granschemaname{G}$
which has the further property that for no
 $\varphi \in \pbinunsatgrsch{G}$
 is it the case that
 $\constrgrsch{G} \union \pbintautgrsch{G} \sentails \varphi$,
 then $\sigma$ is a (full) structure over $\granschemaname{G}$.
    \par
    Consequently, if $\sigma$ is a strong quasi-model of
$\granschemaname{G}$ which does not satisfy any member of
$\pbinunsatgrsch{G}$, then it is a (full) model of
$\granschemaname{G}$.
 \begin{proof}
    The difference between $\gnlestrpr{\sigma}$ being a strong
quasi-structure and a full structure is that, in the latter, it is
additionally required that, for $g \in \granulesofschnb{G}$,
         $\gnletodom{\sigma}(g)\neq\emptyset$.
   Now, for any $g \in \granulesofschnb{G}$, it is the case that
   $\disjrulep{g}{g} \in \pbinunsatgrsch{G}$,
 which forbids
   $\gnletodom{\sigma}(g)
       =\gnletodom{\sigma}(g) \intersect \gnletodom{\sigma}(g)
          = \emptyset$.
 \end{proof}
 \end{metaemphlabpara}

 \begin{metaemphlabpara}{theorem}{Theorem}
   {Sat-completeness of $\sqbininfpos{G}$}\envlabel{thm:satcomplsbin}
   The inference rules of
  \preformat{\linebreak}
 $\sqbininfpos{G}$ are sound and sat-complete in the context of full
models.
 \begin{proof}
    The result follows from \envref{prop:sqbininfpos} and
\envref{prop:strongqstrunsat}, since for sat-completeness, the
possibility of a structure satisfying any member of
$\pbinunsatgrsch{G}$ is ruled out.
 \end{proof}
 \end{metaemphlabpara}

 \begin{metalabpara}{mydefinition}{}
  {Extending $\sqbininfpos{G}$ to models}\envlabel{def:qbininfposext}
    Define the inference system
 \preformat{\linebreak}
 $\sqbininfposext{G}$ to have
  \nlrightt
  $\infrulesof{\sqbininfposext{G}} =
        \infrulesof{\sqbininfpos{G}} \union
            \setbr{\text{(U1)}, \text{(U2)} }$,
   \newline
  with the rules (U1) and (U2) as defined in
Fig.\ \ref{fig:sqbininfposadd}.
   \begin{figure}[htb]
   \def\isepi{0.5em}
   \begin{align*}
    &\mbox{(U1)}\hspace*{\isepi}
    \begin{prooftreem}
      \hypoi{\disjrulep{\granvar{g}}{\granvar{g}}}
      \inferi{\false}
    \end{prooftreem}
    {\scriptstyle\abr{|{\scriptscriptstyle(\granvar{g} \neq \botgrsch{G})}}}
   &&\mbox{(U2)}\hspace*{\isepi}
    \begin{prooftreem}
      \hypoi{\subrulep{g}{\botgrsch{G}}}
      \inferi{\false}
    \end{prooftreem}
    {\scriptstyle\abr{|{\scriptscriptstyle(\granvar{g} \neq \botgrsch{G})}}}
    \end{align*}
   \caption{Additional rules of $\sqbininfposext{G}$}\label{fig:sqbininfposadd}
   \end{figure}
 \end{metalabpara}

 \begin{metaemphlabpara}{proposition}{Proposition}
   {Soundness and completeness of $\qbininfposext{G}$
     in the context of full models}\envlabel{prop:qbininfposext}
   The inference rules of $\sqbininfposext{G}$ are sound and complete in
the context of full models.
 \begin{proof}
   Soundness and sat-completeness follow from
\envref{thm:satcomplsbin}, while (U1) and (U2) ensure that
unsatisfiability will be detected.
 \end{proof}
 \end{metaemphlabpara}

 \begin{metalabpara}{mydefinition}{}
  {Unnecessary rules of $\bininfpossat{G}$}\envlabel{def:bininfpos}
    While the inference system
 \preformat{\linebreak}
 $\qbininfposext{G}$ is sound and complete for full models, it
contains some rules which are not necessary.  Specifically, the rules
(D1), (M2), and (U2) may be removed.  Define the inference system
$\bininfpos{G}$ to have
  \nlrightt
  $\infrulesof{\bininfpos{G}}
      = (\infrulesof{\sqbininfposext{G}}
              \setminus \setbr{\text{(D1)},\text{(M2)},\text{(U2)}}$.
   \newline
   Due to its importance, this inference system is shown in full in
Fig.\ \ref{fig:bininfpos}.
   \begin{figure}[htb]
   \def\isepi{0.5em}
   \begin{align*}
   &\mbox{(I2)}\hspace*{\isepi}
    \begin{prooftreem}
       \hypoii{\subrulep{\granvar{g}_1}{\granvar{g}_2}}
              {\subrulep{\granvar{g}_2}{\granvar{g}_3}}
       \inferi{\subrulep{\granvar{g}_1}{\granvar{g}_3}}
    \end{prooftreem}
   &&\mbox{(M1)}\hspace*{\isepi}
    \begin{prooftreem}
      \hypoii{\subrulep{\granvar{g}_1}{\granvar{g}_1'}}
              {\disjrulep{\granvar{g}_1'}{\granvar{g}_2}}
      \inferi{\disjrulep{\granvar{g}_1}{\granvar{g}_2}}
    \end{prooftreem}
    \end{align*}
    \begin{align*}
   &\mbox{(I1)}\hspace*{\isepi}
    \begin{prooftreem}
      \hypoi{\phantom{X}}
      \inferi{\subrulep{\granvar{g}}{\granvar{g}}}
    \end{prooftreem}
   &&\mbox{(T1)}\hspace*{\isepi}
    \begin{prooftreem}
      \hypoi{\phantom{X}}
      \inferi{\disjrulep{\botgrsch{G}}{\granvar{g}}}
    \end{prooftreem} 
   &&&\mbox{(T2)}\hspace*{\isepi}
    \begin{prooftreem}
      \hypoi{\phantom{X}}
      \inferi{\subrulep{\botgrsch{G}}{\granvar{g}}}
    \end{prooftreem}
   &&&&\mbox{(T3)}\hspace*{\isepi}
    \begin{prooftreem}
      \hypoi{\phantom{X}}
      \inferi{\subrulep{\granvar{g}}{\topgrsch{G}}}
    \end{prooftreem}
    \end{align*}
    \begin{align*}
    &\mbox{(U1)}\hspace*{\isepi}
    \begin{prooftreem}
      \hypoi{\disjrulep{\granvar{g}}{\granvar{g}}}
      \inferi{\false}
    \end{prooftreem}
    {\scriptstyle\abr{|{\scriptscriptstyle(\granvar{g} \neq \botgrsch{G})}}}
   \end{align*}
   \caption{Inference rules of $\bininfpos{G}$}\label{fig:bininfpos}
   \end{figure}
    That it is sound and complete for full model is established in
\envref{prop:bininfpos} below.
 \end{metalabpara}

 \begin{metaemphlabpara}{proposition}{Proposition}
   {Soundness and completeness of $\bininfpos{G}$}\envlabel{prop:bininfpos}
   The inference rules of $\bininfpos{G}$ are sound and complete in
the context of full models.
 \begin{proof}
   First of all, the premise $\disjrulep{\granvar{g}'}{\granvar{g}'}$
in rules (D1) and (M2) can only hold in the case that $\granvar{g}'$
is bound to $\botgrsch{G}$, in view of rule (U1).  Since
$\disjrulep{\botgrsch{G}}{\botgrsch{G}}$ is a tautology by (T1), the
premise for $\granvar{g}$ bound to $\botgrsch{G}$ becomes $\true$ in
both (D1) and (M2).  The resulting rules are precisely (T1) and (T2),
respectively, rendering the original (D1) and (M2) unnecessary.
    \par
   To justify removing (U2), it suffices that to note that the
unsatisfiability of $\subrulep{g}{\botgrsch{G}}$ may be deduced from
the unsatisfiability of $\disjrulep{g}{g}$, using additionally (T2)
and (M1), as shown in Fig.\ \ref{fig:dedunsat}.
   \begin{figure}[htb]
   \[
     \begin{prooftreem}
      \hypoi{\subrulep{\granvar{g}}{\botgrsch{G}}}
      \hypoi{\phantom{X}}
      \inferi{\disjrulep{\botgrsch{G}}{\granvar{g}}}
      \inferii{\disjrulep{\granvar{q}}{\granvar{q}}}
      \inferi{\false}
     \end{prooftreem}
    {\scriptstyle\abr{|{\scriptscriptstyle(\granvar{g} \neq \botgrsch{G})}}}
   \]
   \caption{Deducing the unsatisfiability of
                $\subrulep{\granvar{g}}{\botgrsch{G}}$
           from that of 
                $\disjrulep{\granvar{g}}{\granvar{g}}$}\label{fig:dedunsat}
   \end{figure}
 \end{proof}
 \end{metaemphlabpara}

 \begin{metalabpara}{discussion}{}
   {Armstrong context strategy}\envlabel{disc:barmstrong}
     The approach to extending inference systems such as
$\bininfpos{G}$ to include negative constraints as well requires that
the context (in this case $\pbinconstrgrsch{G}$) be Armstrong (see
\envref{summ:armstrong}).
     A straightforward way to show that a context is Armstrong is to
use the characterization given in \mycite[Thm.\ 3.1]{Fagin82}.  The
proof has already been developed in
\mycite[3.15-3.18]{HegnerRo17_inform} for a much more general class of
granular constraints than $\pbinconstrgrsch{G}$, and so will not be
repeated here.  The reader is referred to that article for details.
For completeness, the formal result is recorded below.
 \end{metalabpara}

 \begin{metaemphlabpara}{theorem}{Theorem}
   {Armstrong context}\envlabel{thm:barmstrong}
     $\pbinconstrgrsch{G}$ is an Armstrong context.
 More precisely, if $\granschemaname{G}$ is satisfiable, then
 $\pconstrgrsch{G} \union \pbintautgrsch{G}$ admits an Armstrong
model.
 \begin{proof}
    See \mycite[3.18]{HegnerRo17_inform}.
 \end{proof}
 \end{metaemphlabpara}
    \parvert

   Attention is now turned to the task of showing that proofs in
$\bininfpos{G}$ are single use.  For this task, it is useful to begin
by characterizing properties of $\granschemaname{G}$ via an associated
graph.

 \begin{metalabpara}{mydefinition}{}
         {The graph of an SMAS}\envlabel{def:graphsmas}
     The \emph{graph} of $\granschemaname{G}$, denoted
$\graphofgrsch{G}$, has as vertices the elements of
$\granulesofsch{G}$.  It has two types of edges, one directed and one
undirected, defined as follows.
    \baxblkc
      \axitem{(s-edge)} For $g_1, g_2 \in \granulesofsch{G}$, there is
a directed \emph{subsumption edge} from $g_1$ to $g_2$ just in case
     $\subrulep{g_1}{g_2} \in
          \pconstrgrsch{G} \union \pbintautgrsch{G}$.
   Formally, this edge is regarded as the ordered pair
$\grpr{g_1}{g_2}$.
   The set of all subsumption edges of $\graphofgrsch{G}$ is denoted
$\sedgesofgrsch{G}$.
      \axitem{(d-edge)} For $g_1, g_2 \in \granulesofsch{G}$, there is
an undirected \emph{disjointness edge} between $g_1$ and $g_2$ just in
case
     $\disjrulep{g_1}{g_2} \in \pconstrgrsch{G} \union
\pbintautgrsch{G}$.
   Formally, if $g_1 \not\ideq g_2$, this edge is regarded as the
unordered pair $\setbr{g_1,g_2}$.  In the case that $g_1 \ideq g_2$, 
it is regarded as the singleton
  $\setbr{g_1} = \setbr{g_2} = \setbr{g_1,g_2}$.
   The set of all disjointness edges of $\graphofgrsch{G}$ is denoted
$\dedgesofgrsch{G}$.
    \eaxblk
   The set of \emph{all edges} of $\graphofgrsch{G}$ is
  \nlrightt
   $\edgesofgrsch{G} = \sedgesofgrsch{G} \union \dedgesofgrsch{G}$.
  \newline
   In the equivalent of this graph in \mycite{AtzeniPa88_dke},
subsumption edges are called \emph{black edges} and disjointness edges
are called \emph{red edges}.
     \par
    A \emph{subsumption path} in $\graphofgrsch{G}$ is a sequence
 $p = \abr{g_1,g_2,\ldots,g_k}$, with $k \geq 2$,
 of elements of $\granulesofsch{G}$ with the property that for
each $i \in \ccinterval{1}{k-1}$,
 $\grpr{g_i}{g_{i+1}} \in \sedgesofgrsch{G}$.
   Such a path $p$ is said to be \emph{from} $g_1$ \emph{to} $g_k$,
and to \emph{underlie} $\grpr{g_1}{g_k}$.
   The set of all subsumption paths of $\graphofgrsch{G}$ is denoted
$\spathsofgrsch{G}$.
   The length $k$ of $p$ is denoted $\lengthof{p}$.
    \par
    The \emph{underlying constraint sequence} of $p$ is
 $\constrseq{p} =
   \abr{\subrulep{g_1'}{g_2'},\subrulep{g_2'}{g_3'},\ldots,
      \subrulep{g_i'}{g_{i+1}'},\ldots,\subrulep{g_{k-1}'}{g_k'}}$,
 with $\subrulep{g_i'}{g_{i+1}'}$ the \emph{$i^{th}$-constraint} of
$p$.
   The \emph{underlying constraint set} of $p$ is the underlying
constraint sequence with the order removed; more precisely,
 $\constrset{p} =
   \setbr{\subrulep{g_1'}{g_2'},\subrulep{g_2'}{g_3'},\ldots,
      \subrulep{g_i'}{g_{i+1}'},\ldots,\subrulep{g_{k-1}'}{g_k'}}$,
    \par
   The relation $\sstaredgesofgrsch{G}$ on $\granulesofsch{G}$ is
defined by
  $\grpr{g_1}{g_2} \in
 \preformat{\linebreak}
 \sstaredgesofgrsch{G}$
 iff there is a
 $\abr{g_1',g_2',\ldots,g_k'} \in \spathsofgrsch{G}$
 with $g_1' \ideq g_1$ and $g_k' \ideq g_2$.
  In this case, the path $p = \abr{g_1',g_2',\ldots,g_k'}$ is said to
\emph{underlie} $\grpr{g_1}{g_2}$.
 \end{metalabpara}

 \begin{metaemphlabpara}{proposition}{Proposition}
    {Existence of underlying paths}\envlabel{prop:underpaths}
    Let $g_1, g_2 \in \granulesofsch{G}$, and assume further that 
$\granschemaname{G}$ is satisfiable.
    Then the entailment
        $\pconstrgrsch{G} \union \pbintautgrsch{G}
                      \sentails \subrulep{g_1}{g_2}$
   holds iff there is a subsumption path $p$ which underlies
$\grpr{g_1}{g_2}$.
 \begin{proof}
  That the result then holds in the context of quasi-models is shown
in \mycite[Thm.\ 2]{AtzeniPa88_dke}.  Upon adding the tautologies and
taking into account the results of \envref{prop:strongqstr}, it is
easy to see that it holds in the context of strong quasi-models as
well.
    Finally, with the requirement that $\granschemaname{G}$ be
satisfiable, in view of \envref{prop:strongqstrunsat}, the result
applies to the context of full models also.
 \end{proof}
 \end{metaemphlabpara}
    \parvert

    Since a subsumption constraint holds in $\granschemaname{G}$ iff
it has an underlying path, it also has a simple proof based upon that
path.  The details follow.

 \begin{metalabpara}{}{}
    {Linear proofs for subsumption constraints}\envlabel{def:linearproof}
     Let $p' = \abr{g_1',g_2',\ldots,g_k'} \in
 \preformat{\linebreak}
 \spathsofgrsch{G}$.
     The \emph{left-linear proof tree} (in $\bininfpos{G}$) associated
with $p$, denoted
 $\llinear{p}$, is defined as follows.
  \baxblkc
   \axitem{(a)} If $k=2$ and
 $\subrulep{g_1'}{g_2'} \in \constrgrsch{G}$, then $\llinear{p}$ is
simply $\subrulep{g_1'}{g_2'}$.
   In this case, the proof uses no rules at all, since the desired
consequence is in the set of antecedents.
   \axitem{(b)} If $k=2$ and
 $\subrulep{g_1'}{g_2'} \in \pbintautgrsch{G}\setminus\constrgrsch{G}$,
 then $\llinear{p}$ is the tree $\subrulepbar{g_1'}{g_2'}{0}$.
     In this case, the proof uses at most the rules in
 $\setbr{\text{(I1)},\text{(T1)},\text{(T2)},\text{(T3)}}$.
   \axitem{(c)} If $k>2$ and for every $i \in \ccinterval{1}{k-1}$ it
is the
 case that $\subrulep{g_i'}{g_{i+1}'} \in \constrgrsch{G}$, then
$\llinear{p}$ is defined as shown in Fig.\ \ref{fig:llinearp} .
   \begin{figure}[htb]
   \[
     \begin{prooftreem}
      \hypoii{\subrulep{g_1'}{g_2'}}
             {\subrulep{g_2'}{g_3'}}
      \inferi{\subrulep{g_1'}{g_3'}}
      \hypoi{\subrulep{g_3'}{g_4'}}
      \inferii{\subrulep{g_1'}{g_4'}}
      \ellipsis{}{\subrulep{g_1'}{g_{i}'}}
      \hypoi{\subrulep{g_i'}{g_{i+1}'}}
      \inferii{\subrulep{g_1'}{g_{i+1}'}}
      \ellipsis{}{\subrulep{g_1'}{g_{k-1}'}}
      \hypoi{\subrulep{g_k'}{g_{k-1}'}}
      \inferii{\subrulep{g_{1}'}{g_{k}'}}
     \end{prooftreem}
   \]
   \caption{A left-linear proof tree for subsumption}\label{fig:llinearp}
   \end{figure}
   In this case, only rule (I2) is used.
   \axitem{(d)} If $k>2$ and for one or more
 $i \in \ccinterval{1}{k-1}$ it is the case that
 $\subrulep{g_i'}{g_{i+1}'}
                     \in \pbintautgrsch{G}\setminus\constrgrsch{G}$,
 then
 $\llinear{p}$ is as shown in Fig.\ \ref{fig:llinearp}, but with
  $\subrulep{g_i'}{g_{i+1}'}$ replaced by the tree
$\subrulepbar{g_i'}{g_{i+1}'}{1.5}$.
  In this case, at most the rules in 
 $\setbr{\text{(I1)},\text{(I1)},\text{(T1)},\text{(T2)},\text{(T3)}}$
 are used.
  \eaxblk     
 \end{metalabpara}

 \begin{metaemphlabpara}{lemma}{Lemma}
    {Existence of left-linear proofs}\envlabel{lem:llinear}
    Let $g_1, g_2 \in \granulesofsch{G}$.  If $\granschemaname{G}$ is
satisfiable and
 $\pbinconstrgrsch{G} \union \pbintautgrsch{G}
         \sentails \subrulep{g_1}{g_2}$,
 then there is a left-linear proof of this in $\bininfpos{G}$.
 \begin{proof}
   Use \envref{prop:underpaths} to establish the existence of an
underlying subsumption path, and then \envref{def:linearproof} to
show that a proof of the given form is possible.
 \end{proof}
 \end{metaemphlabpara}

 \begin{metalabpara}{mydefinition}{}
         {Reduced subsumption paths}\envlabel{def:redsubpath}
    The subsumption path
 $p = \setbr{g_1',g_2',\ldots,g_k'} \in \spathsofgrsch{G}$
 is \emph{reduced} if either (a) $\lengthof{p}=2$, or else (b) no
granule in
 \preformat{\linebreak}
 $\granulesofschnb{G}$ occurs more than once in $p$, while
$\botgrsch{G}$ does not occur at all.
 \end{metalabpara}

 \begin{metaemphlabpara}{lemma}{Lemma}
   {Existence of reduced subsumption paths}\envlabel{lem:redsubpath}
     Let 
     $\grpr{g_1}{g_2} \in
 \preformat{\linebreak}
 \granulesofsch{G}\times\granulesofsch{G}$.
  If there is a subsumption path in $\graphofgrsch{G}$ which underlies
$\grpr{g_1}{g_2}$, then there is such a path which is reduced.
  \begin{proof}
  Let $p = \abr{g_1',g_2',\ldots,g_k'}$ be a subsumption path which
underlies $\grpr{g_1}{g_2}$.  
  If $\lengthof{p}=2$, then there is nothing further to prove, since
all such paths are reduced.
  So, assume that $\lengthof{p}>2$ and that some
 $g \in \granulesofsch{G}$ occurs more than once in $p$.  If these
multiple occurrences are the endpoints of $p$; that is, if 
 $\ideq g_1' \ideq g_k'$, then $\grpr{g_1}{g_2}$ is of the form
$\grpr{g}{g}$ for some $g \in \granulesofsch{G}$.  In that case,
replace $p$ with $\abr{g,g}$, which also underlies $\grpr{g_1}{g_2}$
and is furthermore reduced since it is of length two.
    \par
  Otherwise, writing $p = \abr{g_1',g_2',\ldots,g_k'}$, suppose that
 for $j_1,j_2 \in \ccinterval{1}{k}$ with $j_1<j_2$ and either $j_1>1$
or $j_2<k$, it is the case that $g_{j_1}' \ideq g_{j_2}'$.  Remove the
elements in the subsequence $\abr{g_{{j_1}+1},\ldots,g_{j_2}}$ from $p$,
with the resulting sequence being
 $p' = \abr{g_1',g_2',\ldots,g_{j_1}',g_{{j_2}+1}',\ldots,g_k'}$.
 It is easy to see that $p' \in \spathsofgrsch{G}$, and that is also
underlies $\grpr{g_1}{g_2}$, but with $\lengthof{p'}<\lengthof{p}$.
If another inessential cycle remains in $p'$, this process may be
repeated.  Since each iteration reduces the length of the path, the
process must terminate eventually, resulting in a path which is free
of internal cycles.
     \par
  The only satisfiable subsumption constraints which involve
$\botgrsch{G}$, are of the form $\subrulep{\botgrsch{G}}{g}$ for some
$g \in \granulesofsch{G}$, and these are already in
$\pbintautgrsch{G}$.  Thus, there is no need to use $\botgrsch{G}$ in
any path of length greater than two.
 \end{proof}
 \end{metaemphlabpara}

 \begin{metaemphlabpara}{lemma}{Lemma}
   {Single-use left-linear proofs}\envlabel{lem:singusellinear}
     If $p$ is a reduced subsumption path in $\graphofgrsch{G}$, then
the proof\/ $\llinear{p}$ (see \envref{def:linearproof}) is single
use.
  \begin{proof}
 This is immediate, since in any reduced path of length greater than
two, no granule can occur more than once.
 \end{proof}
 \end{metaemphlabpara}

 \begin{metaemphlabpara}{proposition}{Proposition}
    {Deducing satisfiability of subsumption
                                  constraints}\envlabel{prop:infsub}
     If
 \preformat{\linebreak}
 $\pconstrgrsch{G}$ is satisfiable, then for any
 $g_1, g_2 \in \granulesofsch{G}$ with
 $\pconstrgrsch{G} \sentails \subrulep{g_1}{g_2}$,
 there is a single-use left-linear proof in $\bininfpos{G}$ of
$\subrulep{g_1}{g_2}$, using only subsumption constraints in
$\pconstrgrsch{G}$ as premises, and using at most the rules which
lie in
 \preformat{\linebreak}
  $\setbr{\textup{(I2)},\textup{(I1)},
             \textup{(T1)},\textup{(T2)},\textup{(T3)}}$.
 If $g_1, g_2 \in \granulesofschnbnt{G}$ with $g_1 \not\ideq g_2$,
 then at most rules \textup{(I2)} and \textup{(T3)} need be used.
 \begin{proof}
   It suffices to combine \envref{lem:llinear} and
\envref{lem:singusellinear}.
 \end{proof}
 \end{metaemphlabpara}
     \parvert

    It is also the case that proofs of disjointness have a special
form, but in the general case, based upon two subsumption paths rather
than one.  The details follow.

 \begin{metalabpara}{mydefinition}{}
         {Protected pairs}\envlabel{def:protected}
     Define the set of \emph{doublets} of $\granschemaname{G}$ to be
  \nlrightt
   $\doubletsofgrsch{G} =
      \setdef{\setbr{g_1,g_2}}
    {g_1, g_2 \in \granulesofsch{G} ~\text{and}~ g_1 \not\ideq g_2}$.
   \par
   Say that a pair $S = \setbr{g_1,g_2} \in \doubletsofgrsch{G}$ is
\emph{protected for $\granschemaname{G}$} if there is a pair
    $S' = \setbr{g_1',g_2'} \in \granulesofsch{G}$ for which
 $\grpr{g_1}{g_1'}, \grpr{g_2}{g_2'} \in \sstaredgesofgrsch{G}$ and
  \preformat{\linebreak}
 $S' \in \dedgesofgrsch{G}$.
  See Fig.\ \ref{fig:protected}.
   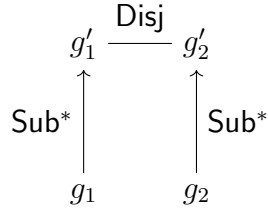
\begin{figure}[htbp]
    \begin{center}
    \begin{tikzpicture}
      \node (g1)  at (0,0) {$g_1$};
      \node (g2)  at (1.5,0) {$g_2$};
      \node (g1p) at (0,2) {$g_1'$};
      \node (g2p) at (1.5,2) {$g_2'$};
      \draw (g1p) to node[above] {$\mathsf{Disj}$} (g2p);
     \draw [->,decoration=,decorate]
            (g1) to node[left] {$\mathsf{Sub^{*}}$} (g1p);
            \draw [->,decoration=,decorate]
            (g2) to node[right] {$\mathsf{Sub^{*}}$} (g2p);
    \end{tikzpicture}
   \end{center}
   \caption{Visualization of the protected pair $\setbr{g_1,g_2}$}\label{fig:protected}
   \end{figure}
  Thus, $S$ is protected if its two granules are (not necessarily
proper) subgranules of corresponding ones in a pair $S'$ which is
declared explicitly to be disjoint, either as a tautology or else as a
constraint of $\granschemaname{G}$.
 In this case, $S'$ is called a \emph{protector} of $S$.
   A pair $S \in \doubletsofgrsch{G}$ is \emph{unprotected (for
$\granschemaname{G}$)} if it is not protected.
    The set of all protected pairs for $\granschemaname{G}$ is denoted
  $\protectedprgrsch{G}$, while
  $\unprotectedprgrsch{G}
       = \doubletsofgrsch{G}  \setminus \protectedprgrsch{G}$
 is the set of unprotected pairs for $\granschemaname{G}$.
 \end{metalabpara}

 \begin{metaemphlabpara}{lemma}{Lemma}
         {Protection implies disjointness}\envlabel{lem:protected}
   For any $S \in \doubletsofgrsch{G}$,
 \nlrightt
  $S \in \protectedprgrsch{G}$ iff 
 $\pconstrgrsch{G} \sentails \disjrulesetp{S}$.
 \begin{proof}
     See \mycite[Thm.\ 3]{AtzeniPa88_dke}.
 \end{proof}
 \end{metaemphlabpara}

 \begin{metalabpara}{}{}
    {One- and two-path left-linear proofs
     for disjointness constraints}\envlabel{def:2pathlinear}
     Suppose that $\disjrulep{g_1}{g_2}$ is the constraint to prove.
There are four cases to consider.
     First of all, if $\disjrulep{g_1}{g_2} \in \pconstrgrsch{G}$,
then the constraint itself forms its proof tree.
     Second, if $\disjrulep{g_1}{g_2}$ is a tautology; that is, if
 $g_1 \ideq \botgrsch{G}$, then the proof tree is
 $\disjrulepbar{\botgrsch{G}}{g_2}{1.5}$
 (and analogously if $g_2 \ideq \botgrsch{G}$).
     Third, neither of these hold, but for some
  $g_1' \in \granulesofsch{G}$,
 $\disjrulep{g_1'}{g_2} \in \pconstrgrsch{G}$, then the proof is of
the form shown in Fig.\ \ref{fig:1pathlinear}.  An analogous result
holds $\disjrulep{g_1}{g_2'} \in \pconstrgrsch{G}$ for some
  $g_2' \in \granulesofsch{G}$.
   \begin{figure}[htb]
   \[
     \begin{prooftreem}
      \hypoi{\Sigma_1}
      \ellipsis{}{\mbox{left-linear proof}}
      \ellipsis{}{\subrulep{g_1}{g_1'}}
      \hypoi{\disjrulep{g_1'}{g_2}}
      \inferii{\disjrulep{g_1}{g_2}}
     \end{prooftreem}
   \]
   \caption{A one-path linear proof of disjointness}\label{fig:1pathlinear}
   \end{figure}
    Finally, if none of these holds, then the proof is as shown in
Fig.\ \ref{fig:2pathlinear}.
   \begin{figure}[htb]
   \[
     \begin{prooftreem}
      \hypoi{\Sigma_1}
      \ellipsis{}{\mbox{left-linear proof}}
      \ellipsis{}{\subrulep{g_1}{g_1'}}
      \hypoi{\Sigma_2}
      \ellipsis{}{\mbox{left-linear proof}}
      \ellipsis{}{\subrulep{g_2}{g_2'}}
      \hypoi{\disjrulep{g_1'}{g_2'}}
      \inferii{\disjrulep{g_1'}{g_2}}
      \inferii{\disjrulep{g_1}{g_2}}
     \end{prooftreem}
   \]
   \caption{A two-path linear proof of disjointness}\label{fig:2pathlinear}
   \end{figure}
 \end{metalabpara}

 \begin{metaemphlabpara}{proposition}{Proposition}
    {Deducing satisfiability of disjointness
                                  constraints}\envlabel{prop:infdisj}
     If
 \preformat{\linebreak}
 $\pconstrgrsch{G}$ is satisfiable, then for any
 $g_1, g_2 \in \granulesofsch{G}$ with
 $\pconstrgrsch{G} \sentails \disjrulep{g_1}{g_2}$,
 there is a single-use left-linear proof in $\bininfpos{G}$ of at most
two paths of this and using at most those rules which lie in
     $\setbr{\textup{(I2)},\textup{(M1)},\textup{(I1)},\textup{(T1)},
            \textup{(T2)},\textup{(T3)}}$.
 \begin{proof}
   The proof follows directly from the constructions of
\envref{def:2pathlinear}.  It only remains to show single usage.  In
all but the last case, it is trivial that the proof is single use.
For the last case, it suffice to note that since $g_1'$ and $g_2'$ are
disjoint, every granule in the path underlying $\grpr{g_1}{g_1'}$ must
be disjoint from every granule in the path underlying
$\grpr{g_2}{g_2'}$, so there is no possibility of a constraint being
used in both chains.
 \end{proof}
 \end{metaemphlabpara}

 \begin{metalabpara}{myexample}{Example}{}\envlabel{ex:pdeductsat}
   For a simple example of the proof technique of \envref{prop:infdisj},
   let
  $\granschemaname[\ref{ex:pdeductsat}]{G}$ be an SMAS with
 $\setdef{g_i}{i \in \ccinterval{1}{8}} \subseteq
           \granulesofsch[\ref{ex:pdeductsat}]{G}$
 and
   \newline
      $\setbr{\subrulep{g_1}{g_2}, \subrulep{g_2}{g_3}, \subrulep{g_3}{g_4},
             \subrulep{g_5}{g_6}, \subrulep{g_6}{g_7}, \subrulep{g_7}{g_8},
             \disjrulep{g_4}{g_8}}
    \nlrightt
       \subseteq
    \pbinconstrgrsch[\ref{ex:pdeductsat}]{G}$.
    \linebreak
   Assume further that $\pbinconstrgrsch[\ref{ex:pdeductsat}]{G}$ is
satisfiable.  In Fig.\ \ref{fig:pdeductsat} is shown a two-path
left-linear proof that
 $\pbinconstrgrsch[\ref{ex:pdeductsat}]{G}
           \sentails \disjrulep{g_1}{g_5}$
 using the proof rules of
 \preformat{\linebreak}
 $\bininfpos[\ref{ex:pdeductsat}]{G}$.
   \begin{figure}[htb]
   \[
     \begin{prooftreem}
      \hypoii{\subrulep{g_1}{g_2}}
             {\subrulep{g_2}{g_3}}
      \inferi{\subrulep{g_1}{g_3}}
      \hypoi{\subrulep{g_3}{g_4}}
      \inferii{\subrulep{g_1}{g_4}}
      \hypoii{\subrulep{g_5}{g_6}}
             {\subrulep{g_6}{g_7}}
      \inferi{\subrulep{g_5}{g_7}}
      \hypoi{\subrulep{g_7}{g_8}}
      \inferii{\subrulep{g_5}{g_8}}
      \hypoi{\disjrulep{g_4}{g_8}}
      \inferiii{\disjrulep{g_1}{g_5}}
     \end{prooftreem}
   \]
   \caption{Example proof of disjointness}\label{fig:pdeductsat}
   \end{figure}
 \end{metalabpara}

 \begin{metaemphlabpara}{proposition}{Proposition}
    {Deducing unsatisfiability
               using $\bininfpos{G}$}\envlabel{prop:bininfposfalse}
     If $\granschemaname{G}$ is not satisfiable, then there is a
single-use proof in $\bininfpos{G}$ whose last step is the application
of rule (U1).
  \begin{proof}
     The completeness of $\bininfpos{G}$ guarantees that there is a
proof of the unsatisfiability of $\granschemaname{G}$ whose last step
is an instantiation of (U1), since (U1) is the only rule which allows
the deduction of $\false$.  However, such an argument alone does not
guarantee the the proof will be single use.  To ensure that condition,
additional work is required.
    \par
    For the most general case, the idea of the proof is to build upon
previous results which ensure single-use proofs, specifically
\envref{prop:infdisj}.
    The strategy is to identify a granule
 $g \in \granulesofschnb{G}$ for which $\disjrulep{g}{g}$ is provable
from $\pconstrgrsch{G}$, in a form similar to that illustrated in
Fig.\ \ref{fig:2pathlinear}, and then append to the end of it an
application of (U1) to deduce $\false$, as illustrated in
Fig.\ \ref{fig:unsatproof}.
   \begin{figure}[htb]
   \[
     \begin{prooftreem}
      \hypoi{\Sigma_1}
      \ellipsis{}{\mbox{left-linear proof}}
      \ellipsis{}{\subrulep{g}{g_1'}}
      \hypoi{\Sigma_2}
      \ellipsis{}{\mbox{left-linear proof}}
      \ellipsis{}{\subrulep{g}{g_2'}}
      \hypoi{\disjrulep{g_1'}{g_2'}}
      \inferii{\disjrulep{g_1'}{g}}
      \inferii{\disjrulep{g}{g}}
      \inferi{\false}
     \end{prooftreem}
   \]
   \caption{A two-path linear proof of unsatisfiability}\label{fig:unsatproof}
   \end{figure}
    There are, however, two details which require attention.  First of
all, since $\pconstrgrsch{G}$ is unsatisfiable, care must be taken in
identifying what a proof is, since $\false$ entails anything and
everything.  Second, since $g$ is used in both linear paths, a
situation which does not occur in Fig.\ \ref{fig:2pathlinear}, it must
be ensured via other means that no rule is used more than once.
     \par
     The strategy is to switch to the proof system $\sqbininfpos{G}$
for all deductions of Fig.\ \ref{fig:unsatproof} except the last one
which deduces $\false$.  As established in \envref{prop:sqbininfpos},
any set of constraints is satisfiable in the context of strong
quasi-models.  Furthermore, every rule of $\bininfpos{G}$, save for
(U1), is also a rule of $\sqbininfpos{G}$.  Therefore, all of the tree
of Fig.\ \ref{fig:unsatproof} except the lowest level, in which (U1)
is applied, is also a proof tree in $\sqbininfpos{G}$.  Hence, the
entire tree is a proof tree in $\bininfpos{G}$.
     \par
    Now, let $p_1 = \abr{g_{11},g_{12},\ldots,g_{1k_1}}$
   (resp.\ $p_2 = \abr{g_{21},g_{22},\ldots,g_{2k_2}}$)
 be the path which underlies the proof of
  $\subrulep{g}{g_1'}$ from $\Sigma_1$
(resp.\ of $\subrulep{g}{g_2'}$ from resp.\ $\Sigma_2$).
 As in \envref{def:2pathlinear}, it is assumed that both $p_1$ and
$p_2$ are reduced.
   If $\Sigma_1 \intersect \Sigma_2 \neq \emptyset$, then some granule
other than $g$ must occur in both $p_1$ and $p_2$.  In that case, let
$g' \ideq g_{1i_1} \ideq g_{2i_2}$
 be the granule with the highest index in $p_1$ which also occurs in
 $p_2$, and replace
  $p_1$ with $p_1' = \abr{g_{1i_1},g_{1i_2},\ldots,g_{1k_1}}$,
  $p_2$ with $p_2' = \abr{g_{2i_1},g_{2i_2},\ldots,g_{2k_1}}$,
 and $g$ with $g'$ throughout the proof.
   The result is clearly single use, since for $\Sigma_1$ and
$\Sigma_2$ to have a constraint in common would require that $p_1$ and
$p_2$ have at least two granules in common.
     \par
   In the less general case, in which a final proof of false is
appended to a tree of the form of Fig.\ \ref{fig:1pathlinear}, the
proof similar but easier, since there is only one left-linear proof of
subsumption.  For the remaining possibilities (corresponding to the
first two cases of \envref{def:2pathlinear}), the proof is trivial.
  \end{proof}
 \end{metaemphlabpara}
   \parvert

    Finally, the main result of this section may be established.

 \begin{metaemphlabpara}{theorem}{Theorem}
    {$\bininfpos{G}$ and single use}\envlabel{thm:bininfpossingleuse}
    The proof system $\bininfpos{G}$ is sound, complete, and single
use.
 \begin{proof}
   This follows from \envref{prop:infsub}, \envref{prop:infdisj}, and
\envref{prop:bininfposfalse}.
 \end{proof}
 \end{metaemphlabpara}

 \begin{metalabpara}{myexample}{Example}{}\envlabel{ex:pdeductunsat}
   For a simple example of the proof technique of
\envref{prop:bininfposfalse}(b), let
  $\granschemaname[\ref{ex:pdeductunsat}]{G}$ be an SMAS with
 $\setdef{g_i}{i \in \ccinterval{1}{7}} \subseteq
           \granulesofsch[\ref{ex:pdeductunsat}]{G}$
 and
   \newline
      $\setbr{\subrulep{g_1}{g_2}, \subrulep{g_2}{g_3},
             \subrulep{g_3}{g_4},
             \subrulep{g_1}{g_5}, \subrulep{g_5}{g_6},
             \subrulep{g_6}{g_7},
             \disjrulep{g_4}{g_7}}
    \nlrightt
       \subseteq
    \pbinconstrgrsch[\ref{ex:pdeductunsat}]{G}$.
    \linebreak
   Assume further that
      $g_1 \not\ideq \botgrsch[\ref{ex:pdeductunsat}]{G}$.
   In Fig.\ \ref{fig:pdeductunsat} is shown a proof that 
 $\pbinconstrgrsch[\ref{ex:pdeductunsat}]{G}$ is unsatisfiable,
 using the proof rules of $\bininfpos[\ref{ex:pdeductunsat}]{G}$.
   \begin{figure}[htb]
    \[
     \begin{prooftreem}
      \hypoii{\subrulep{g_1}{g_2}}
             {\subrulep{g_2}{g_3}}
      \inferi{\subrulep{g_1}{g_3}}
      \hypoi{\subrulep{g_3}{g_4}}
      \inferii{\subrulep{g_1}{g_4}}
      \hypoii{\subrulep{g_1}{g_5}}
             {\subrulep{g_5}{g_6}}
      \inferi{\subrulep{g_1}{g_6}}
      \hypoi{\subrulep{g_6}{g_7}}
      \inferii{\subrulep{g_1}{g_7}}
      \hypoi{\disjrulep{g_4}{g_7}}
      \inferii{\disjrulep{g_1}{g_4}}
      \inferiisp{\disjrulesetp{\setbr{g_1}}}{separation=-1em}
      \inferi{\false}
     \end{prooftreem}
   \]
   \caption{Example proof of unsatisfiability}\label{fig:pdeductunsat}
   \end{figure}
 \end{metalabpara}


 \section{Negation in the Armstrong Context}\label{sec:negarm}


     In the remainder of this paper, the goal is to show how to
augment the inference system $\bininfpos{G}$ of \envref{def:bininfpos}
to include all assertions in $\allbinconstrgrsch{G}$; that is, to
include those in $\nbinconstrgrsch{G}$ as well as those of
$\pbinconstrgrsch{G}$ which are already covered by $\bininfpos{G}$.
     The approach taken is a general one, not limited to
$\pbinconstrgrsch{G}$.  Rather, it applies to any Armstrong context
$\armctxt{C}$.  There is absolutely no added complexity in taking this
more general approach.  On the contrary, it is easier to develop the
ideas independently of a specific context.  This approach not only
avoids involving irrelevant properties specific to
$\allbinconstrgrsch{G}$, but also opens the door for these results to
be applied in other contexts.  Nevertheless, explicit guidance on how
the general results apply to $\allbinconstrgrsch{G}$ will always be
kept at the forefront.
      \par
     In this section, some essential properties of negation in an
Armstrong Context are developed.  In the next section,
Sec.\ \ref{sec:ninfrules}, the construction of a set of inference
rules which are complete for both positive and negative assertions is
provided.

 \begin{metalabpara}{context}{}
     {Context}\envlabel{ctxt:armctxt}
   Unless stated specifically to the contrary, for the rest of this
paper, $\logicsys{L}$ will denote a general logical system (see
\envref{disc:logicsys}), with $\armctxt{C} \subseteq \wffof{L}$ an
Armstrong context (see \envref{summ:armstrong}).  In other words,
every satisfiable subset of $\armctxt{C}$ will admit an Armstrong
model.
  For readers not interested in any generalization, it suffices to
take $\armctxt{C} = \pbinconstrgrsch{G}$, with $\logicsys{L}$ the
logic which includes all assertions in $\allbinconstrgrsch{G}$.
 \end{metalabpara}

 \begin{metalabpara}{mydefinition}{}
         {Armstrong constraint sets}\envlabel{def:armconstrset}
    It is useful to have an explicit notation for the negations of the
assertions in $\armctxt{C}$.  To this end, for any
 $\Phi \subseteq \armctxt{C}$, define
 $\notset{\Phi} = \setdef{\mlnot\varphi}{\varphi \in \Phi}$,
 and define
  $\allarm{\armctxt{C}} = \armctxt{C} \union \notset{\armctxt{C}}$.
  For the specific case of $\armctxt{C}=\pbinconstrgrsch{G}$, it is
immediate that
 $\notset{\pbinconstrgrsch{G}} = \nbinconstrgrsch{G}$
 and
 $\allarm{\pbinconstrgrsch{G}} = \allbinconstrgrsch{G}$.
   \par
    An \emph{extended Armstrong constraint set} over $\armctxt{C}$ is
is any satisfiable subset of $\allarm{\armctxt{C}}$.  In
particular,for $\armctxt{C}=\pbinconstrgrsch{G}$, an extended
Armstrong constraint set is a subset of $\allbinconstrgrsch{G}$.
    \par
    For $\Phi \subseteq \allarm{\armctxt{C}}$, define
  $\posof{\Phi} = \Phi \intersect \armctxt{C}$, and define
  $\negof{\Phi} = \Phi \intersect \notset{\armctxt{C}}$.
  Thus, if $\armctxt{C}=\pbinconstrgrsch{G}$, then
 $\posof{\Phi} = \Phi \intersect \pbinconstrgrsch{G}$
 and
 $\negof{\Phi} = \Phi \intersect \nbinconstrgrsch{G}$.
 \end{metalabpara}

 \begin{metalabpara}{mydefinition}{}
         {Equivalent and complementary pairs}\envlabel{def:cfree}
    Two WFFs may be equivalent without being identical.  For example,
in $\pbinconstrgrsch{G}$, $\subrulep{\botgrsch{G}}{g_1}$ and
$\subrulep{g_2}{\topgrsch{G}}$ are equivalent for any
 $g_1, g_2 \in \granulesofsch{G}$, since both are tautologies and so
true in any granular structure.
    More generally, $\varphi_1, \varphi_2 \in \wffof{L}$ are
\emph{equivalent} if they have the same models.  In this case, write
$\equivwff{\varphi_1}{\varphi_2}$,
 and say that $\setbr{\varphi_1,\varphi_2}$ forms an \emph{equivalent
pair}.
    \par
   A two-element set $\setbr{\varphi_1,\varphi_2} \subseteq \wffof{L}$
forms a \emph{complementary pair} if
$\setbr{\varphi_1,\mlnot\varphi_2}$ forms an equivalent pair, with
$\varphi_1$ and $\varphi_2$ called \emph{complements} of one another.
 \end{metalabpara}

 \begin{metalabpara}{myexample}{Example}
         {Complementary pairs in $\pbinconstrgrsch{G}$}\envlabel{ex:cpair}
    For any $g_1, g_2 \in \granulesofschnb{G}$,
 $\setbr{\subrulep{\botgrsch{G}}{g_1},\disjrulep{g_2}{g_2}}$
 forms a complementary pair of elements in
 $\armctxt{C}=\pbinconstrgrsch{G}$,
 with $\subrulep{\botgrsch{G}}{g_1}$ a tautology and
 $\disjrulep{g_2}{g_2}$ unsatisfiable.
 \end{metalabpara}
    \parvert

    It holds generally that a complementary pair in $\armctxt{C}$ must
consist of a tautology and an an unsatisfiable assertion.  Central to
proving this is to note that the empty set admits an Armstrong model,
as established below.

 \begin{metaemphlabpara}{lemma}{Lemma}
       {Armstrong model for $\emptyset$}\envlabel{lem:armempty}
  In the context identified in \envref{ctxt:armctxt}, the empty set
$\emptyset \subseteq \armctxt{C}$ admits an Armstrong model.
   \begin{proof}
   $\logicsys{L}$ is assumed to be nontrivial in the sense that there
is a $\varphi \in \wffof{L}$ which admits a model (see
\envref{disc:logicsys}).
   Since any model of $\varphi$ is also a model of $\emptyset$, it
follows that the empty set is satisfiable.  Since $\armctxt{C}$ is an
Armstrong context and $\emptyset \subseteq \armctxt{C}$, $\emptyset$
must have an Armstrong model, as required.
   \end{proof}
 \end{metaemphlabpara}

 \begin{metaemphlabpara}{lemma}{Lemma}
   {Complementary pairs in the Armstrong context}\envlabel{lem:cparm}
  Let $\setbr{\varphi_1,\varphi_2} \subseteq \armctxt{C}$.  If
$\setbr{\varphi_1,\varphi_2}$ is a complementary pair, then one must
be a tautology, with the other unsatisfiable.
   \begin{proof}
     Let $M_{\emptyset}$ be an Armstrong model for $\emptyset$, which
must exist in view of \envref{lem:armempty}, and let
 $\setbr{\varphi_1,\varphi_2} \subseteq \armctxt{C}$ be a
complementary pair.  $M_{\emptyset}$ is a model of $\varphi_1$ iff it
is a tautology (just by definition of tautology; see,
\mycite[Def.\ 8.22]{Monk76} or \mycite[p.\ 23]{Enderton01_book}), and
similarly for $\varphi_2$.  Thus, if neither is a tautology, then
$M_{\emptyset}$ is a model of neither, and so it must be a model of
both $\mlnot\varphi_1$ and $\mlnot\varphi_2$, which is impossible,
since they are complements of one another.  Thus, one of $\varphi_1$
or $\varphi_2$ must be a tautology, completing the proof.
   \end{proof}
 \end{metaemphlabpara}

  \parvert
    The characterization of overlap of $\armctxt{C}$ and
$\notset{\armctxt{C}}$ now follows easily.

 \begin{metaemphlabpara}{proposition}{Proposition}
   {Overlap of $\armctxt{C}$ and $\notset{\armctxt{C}}$ in the
Armstrong context}\envlabel{prop:ovlparm}
  If $\varphi_1 \in \armctxt{C}$ and
  $\varphi_2 \in \notset{\armctxt{C}}$ with
$\equivwff{\varphi_1}{\varphi_2}$, then $\varphi_1$ must be either a
tautology or else unsatisfiable.
   \begin{proof}
 Assume that $\varphi_1 \in \armctxt{C}$,
  $\varphi_2 \in \notset{\armctxt{C}}$ with
$\equivwff{\varphi_1}{\varphi_2}$.  Then $\varphi_2$ is of the form
$\mlnot\varphi_3$ for some $\varphi_3 \in \armctxt{C}$, whence
 $\equivwff{\varphi_1}{\mlnot\varphi_3}$.
 It then follows that $\setbr{\varphi_1,\varphi_3}$ is a complementary
pair, whence by \envref{lem:cparm}, one of $\varphi_1$, $\varphi_3$
must be a tautology, with the other unsatisfiable.
   \end{proof}
 \end{metaemphlabpara}

  \parvert
   With the preceding results in hand, the focus is now turned to the
more general topic of incorporating negation in inference.  Although
the idea of contraposition is well-known in logic, it is worthwhile to
recall it precisely here, since the idea is central to results of this
section.

 \begin{metaemphlabpara}{lemma}{Lemma}
       {General swapping of premise and conclusion}\envlabel{lem:genswap}
     Let $\wffset{A} \subseteq \wffof{L}$ be closed under logical
negation; \ie, $\notset{\wffset{A}} \subseteq \wffset{A}$.
(Note that this holds in particular for
  $\wffset{A} = \allarm{\armctxt{C}}$.)
 In addition, let $\Phi \subseteq \wffset{A}$ be finite with
 $\varphi, \psi \in \wffset{A}$,
  \baxblkc
    \axitem{(a)} 
 $\Phi \union \setbr{\varphi} \sentails \psi$
     ~~iff~~
 $\Phi \union \setbr{\mlnot\psi} \sentails \mlnot\varphi$.
    \axitem{(b)} If $\Phi \union \setbr{\varphi} \sentails \psi$
 and $\Phi\union\setbr{\varphi}$ is satisfiable, with
$\varphi$ is essential in the sense that
   $\Phi \not\sentails \psi$,
 then $\Phi \union \setbr{\mlnot\varphi}$ is also satisfiable.
  \eaxblk
   In particular, taking $\psi$ to be unsatisfiable, the following
special cases are obtained.
   \baxblkc
     \axitem{(a$'$)} 
 $\Phi \union \setbr{\varphi}$ is unsatisfiable
     ~~iff~~
 $\Phi \sentails \mlnot\varphi$,
     \axitem{(b$'$)} If  $\Phi \union \setbr{\varphi}$ is unsatisfiable
and $\varphi$ is essential to that unsatisfiability in the sense that 
   $\Phi$ is satisfiable, then $\Phi \union \setbr{\mlnot\varphi}$ is
also satisfiable.
  \eaxblk
   \begin{proof}
   Part (a) is just \emph{contraposition}
\mycite[Cor.\ 24D]{Enderton01_book}.
  For part (b), since $\Phi \not\sentails \psi$, there is a model $M$
of $\Phi \union \setbr{\mlnot\psi}$, which must also be a model of
$\mlnot\varphi$ by (a).  Hence $M$ is also a model of
 $\Phi \union \setbr{\mlnot\varphi}$.
   Parts (a$'$) and (b$'$) follow by taking $\psi$ to be the identically
false assertion $\false$ in (a) and (b).
   \end{proof}
 \end{metaemphlabpara}

 \begin{metalabpara}{remark}{Remark}
     {Swapping in the presence of unsatisfiability}\envlabel{rmk:ruleswap}
    In the first part of \envref{lem:genswap}, it is important to note
that if $\Phi \sentails \psi$; that is, if $\varphi$ is superfluous in
$\Phi \union \setbr{\varphi} \sentails \psi$, then
 $\Phi \union \setbr{\mlnot\psi}$ must be unsatisfiable, so that
 $\Phi \union \setbr{\mlnot\psi} \sentails \mlnot\varphi$
 holds in a rather trivial way, as $\false \sentails \mlnot\varphi$.
 Just take $\varphi$ to be the identically true assertion $\true$ to
see this.
    \par
    Similarly, in the second part, if $\Phi$ is unsatisfiable, so that
 $\Phi \union \setbr{\varphi}$ is unsatisfiable regardless of $\varphi$,
 then $\Phi \sentails \mlnot\varphi$ holds again in a rather trivial
way, since false implies anything.
    \par
    The point is that while the result holds in general, it is, for
the most part, interesting only in the case that $\varphi$ is
essential on the left-hand side for the entailment to hold.
 \end{metalabpara}

   \parvert
     When applied within an Armstrong context, the general idea of
swapping premise and conclusion leads to remarkable results.
Specifically, negative premises are of no consequence in deriving
positive conclusions, and in the derivation of a negative conclusion,
at most one negative premise is necessary.  These ideas are formalized
in the next two results.

 \begin{metaemphlabpara}{lemma}{Lemma}
    {Satisfiability and unsatisfiability
             in extended Armstrong constraint sets}\envlabel{lem:suarmstrong}
     Let $\Phi$ be an extended Armstrong constraint set over
$\armctxt{C}$,
 \baxblkc
  \axitem{(a)} If $\Phi$ satisfiable, then any Armstrong model of
$\posof{\Phi}$ relative to $\armctxt{C}$ is also a model of all of
$\Phi$.
  \axitem{(b)} If $\Phi$ is not satisfiable, then either
$\posof{\Phi}$ is unsatisfiable or else there is some
 $\mlnot\varphi \in \negof{\Phi}$ with
 $\posof{\Phi} \union \setbr{\mlnot\varphi}$ unsatisfiable.
 \eaxblk
 \begin{proof}
     For part (a), if $\Phi = \posof{\Phi}$, then there is nothing
further to prove.  Otherwise, let $M$ be an Armstrong model of $\Phi$
relative to $\armctxt{C}$, and let $\mlnot\varphi \in \negof{\Phi}$.
Since $\Phi \union \setbr{\mlnot\varphi}$ is satisfiable, it follows
from \envref{lem:genswap}(a) that $\Phi \not\sentails \varphi$.  Thus,
since $M$ is an Armstrong model of $\Phi$, it cannot be a model of
$\varphi$, whence it is a model of $\mlnot\varphi$.  Since
$\mlnot\varphi$ was chosen arbitrarily from $\negof{\Phi}$, it follows
that $M$ is a model of all of $\Phi$, as required.
   \par
   For part (b), assume that $\Phi$ is not satisfiable.  If
$\posof{\Phi}$ is not satisfiable, there is nothing further to prove.
So, assume that $\posof{\Phi}$ is satisfiable, and let $M$ be an
Armstrong model of $\posof{\Phi}$ relative to $\armctxt{C}$.  Since
$\Phi$ is not satisfiable, there must be some
 $\mlnot\varphi \in \negof{\Phi}$ with
 $M \not\in \modelsof{\mlnot\varphi}$,
 whence $M \in \modelsof{\varphi}$.
 Thus, by the definition of Armstrong model,
 $\posof{\Phi} \sentails \varphi$.
 Finally, applying \envref{lem:genswap}(a$'$),
 $\posof{\Phi} \union \setbr{\mlnot\varphi}$
 is unsatisfiable, completing the proof.
 \end{proof}
 \end{metaemphlabpara}

 \begin{metaemphlabpara}{proposition}{Proposition}
    {Entailment in extended Armstrong constraint sets}\envlabel{prop:entarm}
     Let $\Phi$ be an extended Armstrong constraint set over
$\armctxt{C}$,
 and let $\beta \in \armctxt{C}$.
   \baxblkc
     \axitem{(a)} If $\Phi \sentails \beta$,
  then it must be the case that $\posof{\Phi} \sentails \beta$.
     \axitem{(b)} If $\Phi \sentails \mlnot\beta$,
 then either $\posof{\Phi} \sentails \mlnot\beta$, or else there is
 a $\mlnot\varphi \in \negof{\Phi}$ with the property that
    $\posof{\Phi} \union \setbr{\mlnot\varphi} \sentails \mlnot\beta$
 (and hence
    $\posof{\Phi} \union \setbr{\beta} \sentails \varphi$)
 holds,  with both $\posof{\Phi} \union \setbr{\mlnot\varphi}$ and
   $\posof{\Phi} \union \setbr{\beta}$ satisfiable.
  \eaxblk
  \begin{proof}
    Part (a): If $\Phi$ is not satisfiable, the result holds
trivially.  So, assume that $\Phi$ is satisfiable with
 $\Phi \sentails \beta$, and let $M$ be an Armstrong model for
$\posof{\Phi}$.  By \envref{lem:suarmstrong}(a), $M$ is a model of all
of $\Phi$, and hence a model of $\beta$ as well.  Finally, by the
definition of Armstrong model, the only constraints in $\armctxt{C}$
which hold in $M$ are those implied by $\posof{\Phi}$, whence
 $\posof{\Phi} \sentails \beta$.
     \par
    Part (b): If $\Phi$ is not satisfiable, then the result is
immediate.  So, assume that $\Phi$ is satisfiable with
 $\Phi \sentails \mlnot\beta$.  In view of \envref{lem:genswap}(a$'$),
$\Phi \union \setbr{\beta}$ must be unsatisfiable.  Then, applying
\envref{lem:suarmstrong}(b), it must be the case that either
 $\posof{\Phi}\union\setbr{\beta}$ is unsatisfiable or else there is
$\mlnot\varphi \in \negof{\Phi}$ with
 $\posof{\Phi} \union \setbr{\beta} \union \setbr{\mlnot\varphi}$
 unsatisfiable.
 If $\posof{\Phi}\union\setbr{\beta}$ is unsatisfiable, a second
application of \envref{lem:genswap}(a$'$) yields $\posof{\Phi}
\sentails \mlnot\beta$.
 If $\posof{\Phi}\union\setbr{\beta}$ is satisfiable but there is
$\mlnot\varphi \in \negof{\Phi}$ with
 $\posof{\Phi} \union \setbr{\beta} \union \setbr{\mlnot\varphi}$
 unsatisfiable, an application of \envref{lem:genswap}(a) yields 
 $\posof{\Phi} \union \setbr{\mlnot\varphi} \sentails \mlnot\beta$,
 completing the proof.
  \end{proof}
 \end{metaemphlabpara}


 \section{Inference on Binary Rules Involving Negation}\label{sec:ninfrules}

     In this section, it is shown how to extend a sound and complete
proof system $\proofsys{A}$ for an Armstrong context $\armctxt{C}$ to
one which applies to all of $\allarm{\armctxt{C}}$.  It builds upon
the theoretical ideas developed in Sec.\ \ref{sec:negarm}, together
with the fundamental operation of \emph{swapping} to each rule in
$\proofsys{A}$, in which the conclusion of a rule is swapped with a
premise, and each negated.  It is shown that augmenting $\proofsys{A}$
with the set of all such swapped rules results in a sound and complete
inference system for $\allarm{\armctxt{C}}$.

 \begin{metalabpara}{context}{}
     {Context}\envlabel{ctxt:armctxtii}
 Throughout this section, unless stated specifically to the contrary,
$\proofsys{A}$\/ will be taken to be a sound but not necessarily
complete inference system for $\armctxt{C}$.  Note that the context
introduced in \envref{ctxt:armctxt} continues to hold.
    \par
    As in the previous section, the reader who is not interested in
the general case may fix $\armctxt{C}$ to be $\pbinconstrgrsch{G}$.
 \end{metalabpara}
    \parvert

    While the following result is straightforward, it is recorded
formally because it provides the key connection between the results of
Sec.\ \ref{sec:negarm} and the type of proof system developed in this
section.

 \begin{metaemphlabpara}{proposition}{Proposition}
    {Satisfiable inference in the Armstrong setting}\envlabel{prop:infarm}
     Let $\Phi$ be a satisfiable subset of $\armctxt{C}$, and let
 $\beta \in \armctxt{C}$ also be satisfiable.
    Assume furthermore that $\proofsys{A}$ complete.
   \baxblke
    \axitem{(a)} $\Phi \sentails \beta$ holds iff
   $\posof{\Phi} \syntailss[\proofsys{A}] \beta$.
    \axitem{(b)} $\Phi \sentails \mlnot\beta$ holds
  iff either
    $\posof{\Phi} \union \setbr{\beta} \syntailss[\proofsys{A}]
                                             \false$
 or else there is a
  $\mlnot\varphi \in \negof{\Phi}$ for which
    $\posof{\Phi} \union \setbr{\beta}
           \syntailss[\proofsys{A}] \varphi$.
   \eaxblk
 \begin{proof}
   The proof follows immediately from \envref{prop:entarm} and the
fact that $\proofsys{A}$\/ is sound and complete for $\armctxt{C}$.
 \end{proof}
 \end{metaemphlabpara}

 \begin{metalabpara}{mydefinition}{}
     {Swapping and swap closure}\envlabel{def:ruleswap}
     Let $\proofsys{A}$\, be an inference system for $\armctxt{C}$.
     Say that a rule $r \in \infrulesof{\proofsys{A}}$
\emph{has nonempty antecedent set} if
 $\antecedentsof{r} \neq \emptyset$.
     \par
     Let $r \in \infrulesof{\proofsys{A}}$ have nonempty
antecedent set
 $\antecedentsof{r} = \setdef{\alpha_i}{i \in \ccinterval{1}{k}}$
 and $\consequentof{r} = \setbr{\beta}$.
     The \emph{$\alpha_i$-swap} of $r$, denoted $\rswap{r}{\alpha_i}$,
is the rule obtained by replacing $\alpha_i$ with $\mlnot\beta$ as an
antecedent, and replacing $\beta$ with $\mlnot\alpha_i$ as the
consequent.  Visually, the transformation is
  \def\transto#1{\hspace*{#1} &\rightsquigarrow \hspace*{#1}}
  \def\sepval{0.5cm}
    \begin{align*}
    \tag{\ref{def:ruleswap}-1}
    \begin{prooftreem}
     \hypoi{\alpha_1~\alpha_2~\ldots~\alpha_{i-1}~\alpha_i~\alpha_{i+1}~
                      \ldots~\alpha_{k-1}~\alpha_k}
     \inferi{\beta}
    \end{prooftreem}
     \transto{\sepval}
    \begin{prooftreem}
     \hypoi{\alpha_1~\alpha_2~\ldots~\alpha_{i-1}~~~\mlnot\beta~\alpha_{i+1}~
                      \ldots~\alpha_{k-1}~\alpha_k}
     \inferi{\mlnot\alpha_i}
    \end{prooftreem}
    \end{align*}

    If $\beta$ is unsatisfiable; \ie, $\beta\equiv\false$, then
$\mlnot\beta$ is simply omitted as a premise.  In this case, the
transformation is
    \begin{align*}
    \tag{\ref{def:ruleswap}-2}
    \begin{prooftreem}
     \hypoi{\alpha_1~\alpha_2~\ldots~\alpha_{i-1}~\alpha_i~\alpha_{i+1}~
                      \ldots~\alpha_{k-1}~\alpha_k}
     \inferi{\false}
    \end{prooftreem}
    \transto{\sepval}
    \begin{prooftreem}
     \hypoi{\alpha_1~\alpha_2~\ldots~\alpha_{i-1}~~~\alpha_{i+1}~~
                      \ldots~\alpha_{k-1}~\alpha_k}
     \inferi{\mlnot\alpha_i}
    \end{prooftreem}
    \end{align*}
    \par
    Define the \emph{swap closure} $\swapcl{\proofsys{A}}$ of
$\proofsys{A}$ to be the inference system on
 $\allarm{\armctxt{C}}$ given by 
  \newline
    $\infrulesof{\swapcl{\proofsys{A}}} =
     \infrulesof{\proofsys{A}} \union
  \nlrightt
      \setdef{\rswap{r}{\alpha}}
             {r \in \infrulesof{\proofsys{A}}
              \text{ and }
              \antecedentsof{r}\neq\emptyset
              \text{ and }
              \alpha \in \antecedentsof{r}}$.
 \end{metalabpara}

 \begin{metaemphlabpara}{proposition}{Proposition}
         {Soundness of swapping}\envlabel{prop:genswap}
  \baxblke    
    \axitem{(a)} If $r$ is a sound and minimal inference rule of
$\proofsys{A}$, then $\rswap{r}{\alpha_i}$ is a sound and minimal
inference rule on $\allarm{\armctxt{C}}$.
   \axitem{(b)} If every rule of $\proofsys{A}$ is sound and
minimal, then $\swapcl{\proofsys{A}}$ is sound as well.
  \eaxblk
   \begin{proof}
      The proof of (a) follows directly from \envref{lem:genswap}.
The proof of (b) then follows immediately from the definition of swap
closure.
   \end{proof}
 \end{metaemphlabpara}

 \begin{metalabpara}{myexample}{}
      {The swap closure of $\bininfpos{G}$}\envlabel{def:swapclpbininf}
   Upon applying the construction of \envref{def:ruleswap}(b) to the
rules of $\bininfpos{G}$, a new inference system, denoted
 $\bininf{G}$, is obtained.
   Its rules are defined by
 \newline
    $\infrulesof{\bininf{G}} =
      \infrulesof{\swapcl{\bininfpos{G}}}
  \nlrightt
       = \setbr{\text{(I2)},
                \text{(M1)},
                \text{(I1)},
                \text{(T1)},
                \text{(T2)},
                \text{(T3)},
                \text{(U1)},
                \text{(I2-sa)},
                \text{(I2-sb)},
                \text{(M2-sa)},
                \text{(M2-sb)},
                \text{(U1-s)}
               }$
     with five additional rules not found in $\infrulesof{\bininfpos}$
shown in Fig.\ \ref{fig:swaprulesi}, with the way in which each rule
was obtained via swapping shown to its far right.  Note that rules
(I1) and (T1)--(T3) are not swapped, since they all have empty
antecedent set.
   \begin{figure}[htb]
   \begin{center}
   \baxblke
     \axitem{(I2-sa)}
   $\begin{prooftreem}
      \hypoii{\subrulep{\granvar{g}_1}{\granvar{g}_2}}
             {\nsubrulep{\granvar{g}_1}{\granvar{g}_3}}
      \inferi{\nsubrulep{\granvar{g}_2}{\granvar{g}_3}}
      \end{prooftreem}$
      \hfill ($\rswap{\text{(I2)}}{\subrulep{\granvar{g}_2}{\granvar{g}_3}}$)
     \axitem{(I2-sb)}
   $\begin{prooftreem}
      \hypoii{\nsubrulep{\granvar{g}_1}{\granvar{g}_3}}
             {\subrulep{\granvar{g}_2}{\granvar{g}_3}}
      \inferi{\nsubrulep{\granvar{g}_1}{\granvar{g}_2}}
     \end{prooftreem}$
      \hfill ($\rswap{\text{(I2)}}{\subrulep{\granvar{g}_1}{\granvar{g}_2}}$)
     \axitem{(M1-sa)}
   $\begin{prooftreem}
     \hypoii{\subrulep{\granvar{g}_1}{\granvar{g}_1'}}
             {\ndisjrulep{\granvar{g}_1}{\granvar{g}_2}}
     \inferi{\ndisjrulep{\granvar{g}_1'}{\granvar{g}_2'}}
    \end{prooftreem}$
      \nlrightt ($\rswap{\text{(M1)}}{\disjrulep{\granvar{g}_1'}{\granvar{g}_2'}}$)
     \axitem{(M1-sb)}
   $\begin{prooftreem}
      \hypoii{\ndisjrulep{\granvar{g}_1}{\granvar{g}_2}}
             {\disjrulep{\granvar{g}_1'}{\granvar{g}_2}}
      \inferi{\nsubrulep{\granvar{g}_1}{\granvar{g}_1'}}
    \end{prooftreem}$
      \nlrightt ($\rswap{\text{(M1)}}{\subrulep{\granvar{g}_1}{\granvar{g}_1'}}$)
      \axitem{(U1-s)}
    \begin{prooftreem}
      \hypoi{\phantom{X}}
      \inferi{\ndisjrulep{\granvar{g}}{\granvar{g}}}
    \end{prooftreem}
    ${\scriptstyle\abr{|{\scriptscriptstyle(\granvar{g} \neq \bot)}}}$
     \hfill ($\rswap{\text{(U1)}}{\disjrulep{\granvar{g}}{\granvar{g}}}$
   \eaxblk
   \end{center}
   \caption{Swapped rules for $\bininfpos{G}$}\label{fig:swaprulesi}
   \end{figure}
 \end{metalabpara}
    \parvert

   The goal is to show that $\bininf{G}$ is both sound and complete.
The first part is easy.

 \begin{metaemphlabpara}{proposition}{Proposition}
    {Soundness of $\bininf{G}$}\envlabel{prop:bininfsound}
   The inference system $\bininf{G}$ is sound.
  \begin{proof}
   The proof follows immediately from \envref{prop:genswap}.
  \end{proof}
 \end{metaemphlabpara}
   \parvert

 \begin{metalabpara}{mydefinition}{}
    {Conversion to positive-only proofs}\envlabel{def:proofconv}
    Let the setting be as in \envref{prop:infarm}(b), and suppose that
$\Phi \sentails \mlnot\beta$.
    Of course, there is no proof of this using $\proofsys{A}$, since
that proof system does not include members of $\notset{\armctxt{C}}$
for either antecedents or consequents of rules.
    However, using \envref{prop:infarm}(b) and the completeness of
$\proofsys{A}$, it must be the case that either
 $\posof{\Phi} \union \setbr{\beta} \syntailss[\proofsys{A}] \false$
 or else
 $\posof{\Phi} \union \setbr{\beta} \syntailss[\proofsys{A}] \varphi$
 for some $\mlnot\varphi \in \negof{\Phi}$.
    The idea is to take a proof in $\proofsys{A}$ for 
 $\posof{\Phi} \union \setbr{\beta} \syntailss[\proofsys{A}] \false$
 or
 $\posof{\Phi} \union \setbr{\beta} \syntailss[\proofsys{A}] \varphi$,
 as the case may be, and convert it to a proof 
 $\Phi \syntailss[\swapcl{\proofsys{A}}] \beta$
 by \emph{contrapositioning} certain rules in the original proof,
which is developed in detail in \envref{def:proofflip} below.
  In support of this, it is first useful to introduce an alternative
representation for proof trees, using trees represented in vertex-edge
format..
  \end{metalabpara}

 \begin{metalabpara}{mydefinition}{}
    {The tree of an inference rule}\envlabel{def:irulegraph}
     The notation for proof rules in which the premises are separated
from the conclusion by a single horizontal line is something of a
standard in logic.  Implicitly, it represents a tree structure.
However, for the purposes of formalizing the process of
contrapositioning, it is advantageous to make this tree structure
explicit, as illustrated in (\ref{def:irulegraph}-1) and
(\ref{def:irulegraph}-2).
  \def\transto#1{\hspace*{#1} &\rightsquigarrow \hspace*{#1}}
  \def\equivto#1{\hspace*{#1} &\equiv \hspace*{#1}}
  \def\sepval{0.5cm}
    \begin{align*}
    \tag{\ref{def:irulegraph}-1}
    \begin{prooftreem}
     \hypoi{\alpha_1~\alpha_2~\ldots~\alpha_{i-1}~\alpha_i~\alpha_{i+1}~
                      \ldots~\alpha_{k-1}~\alpha_k}
     \inferi[d]{\beta}
    \end{prooftreem}
     \equivto{\sepval}
    \begin{tikzpicture} [->, baseline=(current bounding box.center) ]
    \node {$\beta$} [grow' = up, align=center,
                     sibling distance = 2em, level distance = 5em]
    child {node {$\alpha_1$} edge from parent 
            node [left=0.1, near end] {$\scriptstyle d$} }
    child {node {$\alpha_2$} edge from parent node [right] {$\ldots$}
            node [left, near end] {$\scriptstyle d$} }
    child {node {$\ldots$} \efpn}
    child {node {$\alpha_{i-1}$} edge from parent
            node [left, near end] {$\scriptstyle d$} }
    child {node {$\alpha_i$} edge from parent
            node [left, near end] {$\scriptstyle d$} }
    child {node {$\alpha_{i+1}$} edge from parent node [right] {$\ldots$}
            node [right, near end] {$\scriptstyle d$} }
    child {node {$\ldots$} \efpn}
    child {node {$\alpha_{k-1}$} edge from parent
            node [left, near end] {$\scriptstyle d$} }
    child {node {$\alpha_k$} edge from parent
            node [right, near end] {$\scriptstyle d$} };
    \end{tikzpicture}
    \\
    \tag{\ref{def:irulegraph}-2}
    \begin{prooftreem}
     \hypoi{\phantom{\true}}
     \inferi[e]{\beta}
    \end{prooftreem}
     \equivto{\sepval}
    \begin{tikzpicture} [->, baseline=(current bounding box.center) ]
    \node {$\beta$} [grow = up, align=center,
                     sibling distance = 2em, level distance = 3em]
    child {node {$\true$} edge from parent 
            node [left, midway] {$\scriptstyle e$} };
    \end{tikzpicture}
    \end{align*}
    Each antecedent, as well as the consequent, is a vertex.  The
consequent is the root of the tree, with a directed edge from it to
each antecedent, as illustrated in (\ref{def:irulegraph}-1).  If the
antecedent set is empty, an artificial antecedent vertex $\true$ is
added, as illustrated in (\ref{def:irulegraph}-2).  Formally, each edge is
labelled with the instance name for that rule.  However, those labels
may be omitted if the name of the instance is clear from context.
 \end{metalabpara}

 \begin{metalabpara}{mydefinition}{}
    {Proof trees in vertex-edge format}\envlabel{def:prooftreesve}
     Proof trees are represented in vertex-edge format in the obvious
way, by connecting the individual trees together.
Fig.\ \ref{fig:verep} shows the proof trees for the proofs of
(\ref{def:ruleproofnot}-3) and (\ref{def:ruleproofnot}-4).
   \begin{figure}[htb]
    \begin{tikzpicture} [->, baseline=(current bounding box.center)]
    \node {$\subrulep{g_1}{g_5}$} [grow' = up, align=center, anchor=north,
                     sibling distance = 6em, level distance = 4em]
    child {node {$\subrulep{g_1}{g_4}$} 
      child {node {$\subrulep{g_1}{g_3}$}
        child {node {$\subrulep{g_1}{g_2}$} edge from parent
                           node[below left=-0.4em] {$\scriptscriptstyle d_)$}
              }
        child {node {$\subrulep{g_2}{g_3}$} edge from parent
                          node[below right=-0.4em] {$\scriptscriptstyle d_1$}
              }
          edge from parent node[below left=-0.4em] {$\scriptscriptstyle d_2$}
           }
      child {node {$\subrulep{g_3}{g_4}$} edge from parent
                         node[below right=-0.4em] {$\scriptscriptstyle d_2$}}
          edge from parent node[below left=-0.4em] {$\scriptscriptstyle d_3$}
            }
    child {node {$\subrulep{g_4}{g_5}$} edge from parent 
                       node[below right=-0.4em] {$\scriptscriptstyle d_3$}
          };
    \end{tikzpicture}
    \begin{tikzpicture} [->, baseline=(current bounding box.center)]
    \node {$\subrulep{g_1}{g_5}$} [grow' = up, align=center, anchor=north,
                     sibling distance = 12em, level distance = 4em]
    child {node {$\subrulep{g_1}{g_3}$} [sibling distance=6em]
      child {node {$\subrulep{g_1}{g_2}$}
       edge from parent node[below left=-0.4em] {$\scriptscriptstyle d_1'$}
            }
      child {node {$\subrulep{g_2}{g_3}$}
       edge from parent node[below right=-0.4em] {$\scriptscriptstyle d_1'$}
            }
       edge from parent node[below left=-0.4em] {$\scriptscriptstyle d_3'$}
          }
    child {node {$\subrulep{g_3}{g_5}$} [sibling distance=6em]
      child {node {$\subrulep{g_3}{g_4}$}
       edge from parent node[below left=-0.4em] {$\scriptscriptstyle d_2'$}
            }
      child {node {$\subrulep{g_4}{g_5}$}
       edge from parent node[below right=-0.4em] {$\scriptscriptstyle d_2'$}
            }
       edge from parent node[below right=-0.4em] {$\scriptscriptstyle d_3'$}
          };
    \end{tikzpicture}
    \caption{Vertex-edge representations for (\ref{def:ruleproofnot}-3) and
                   (\ref{def:ruleproofnot}-4)}\label{fig:verep}
     \end{figure}
 \end{metalabpara}

 \begin{metalabpara}{mydefinition}{}
    {Swapped inference rules in vertex-edge format}\envlabel{def:irulegraphswap}
    Given an inference rule $r$, the representation of
$\rswap{r}{\alpha_i}$ is, in principle, straightforward, as
illustrated in (\ref{def:irulegraphswap}-1).  All that is necessary is
to swap the identities of the vertex labelled $\alpha_i$ with that of
the conclusion $\beta$, and negate each.  The edges remain fixed; thus, this representation is called \emph{fixed-edge positioning}.
   \def\transto#1{\hspace*{#1} &\rightsquigarrow \hspace*{#1}}
   \def\sepval{0.5cm}
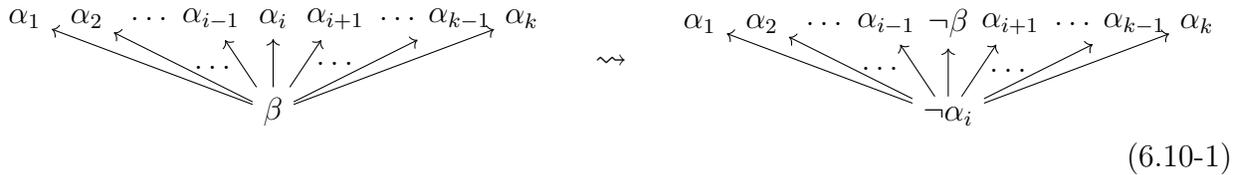
    \begin{align*}
    \tag{\ref{def:irulegraphswap}-1}
    \begin{tikzpicture} [->, baseline=(current bounding box.center) ]
    \node {$\beta$} [grow' = up, align=center,
                     sibling distance = 2em, level distance = 3em]
    child {node {$\alpha_1$}}
    child {node {$\alpha_2$} edge from parent node [right] {$\ldots$}}
    child {node {$\ldots$} \efpn}
    child {node {$\alpha_{i-1}$}}
    child {node {$\alpha_i$}}
    child {node {$\alpha_{i+1}$} edge from parent node [right] {$\ldots$}}
    child {node {$\ldots$} \efpn}
    child {node {$\alpha_{k-1}$}}
    child {node {$\alpha_k$}};
    \end{tikzpicture}
     \transto{\sepval}
    \begin{tikzpicture} [->, baseline=(current bounding box.center) ]
    \node {$\mlnot\alpha_i$} [grow' = up, align=center,
                     sibling distance = 2em, level distance = 3em]
    child {node {$\alpha_1$}}
    child {node {$\alpha_2$} edge from parent node [right] {$\ldots$}}
    child {node {$\ldots$} \efpn}
    child {node {$\alpha_{i-1}$}}
    child {node {$\mlnot\beta$}  edge from parent []}
    child {node {$\alpha_{i+1}$} edge from parent node [right] {$\ldots$}}
    child {node {$\ldots$} \efpn}
    child {node {$\alpha_{k-1}$}}
    child {node {$\alpha_k$}};
    \end{tikzpicture}
    \end{align*}
   While this works fine for an isolated inference rule, it leads to
problems when such rules are interconnected to form proofs because
vertices must be moved around, creating a representational maze.  It
is more suitable to keep the position of the vertices fixed, and move
the edges as necessary.  This \emph{fixed-vertex positioning} is shown
in (\ref{def:irulegraphswap}-2).
   \def\sepval{0.5cm}
    \begin{align*}
    \begin{tikzpicture} [->, baseline=(current bounding box.center) ]
    \node {$\beta$} [grow' = up, align=center,
                     sibling distance = 2em, level distance = 3em]
    child {node {$\alpha_1$}}
    child {node {$\alpha_2$} edge from parent node [right] {$\ldots$}}
    child {node {$\ldots$} \efpn}
    child {node {$\alpha_{i-1}$}}
    child {node {$\alpha_i$}}
    child {node {$\alpha_{i+1}$} edge from parent node [right] {$\ldots$}}
    child {node {$\ldots$} \efpn}
    child {node {$\alpha_{k-1}$}}
    child {node {$\alpha_k$}};
    \end{tikzpicture}
     \transto{\sepval}
    \tag{\ref{def:irulegraphswap}-2}
    \begin{tikzpicture} [->, baseline=(current bounding box.center),
                         inner sep=0em ]
    \node {$\mlnot\beta$} [grow' = up, align=center,
                     sibling distance = 2em, level distance = 3em]
    child {node (a1)   {$\alpha_1$} \efpn}
    child {node (a2)   {$\alpha_2$} \efpn}
    child {node (ld1)  {$\ldots$} \efpn}
    child {node (aim1) {$\alpha_{i-1}$} \efpn}
    child {node (ai)   {$\mlnot\alpha_i$} edge from parent [<-] }
    child {node (aip1) {$\alpha_{i+1}$} \efpn}
    child {node (ld2)  {$\ldots$} \efpn}
    child {node (akm1) {$\alpha_{k-1}$} \efpn}
    child {node (ak)   {$\alpha_k$} \efpn};
    \draw (ai.250) to[myncbar, arm=0.6,angle=270] (a1.south);
    \draw (ai.235) to[myncbar, arm=0.4,angle=270] (a2.south);
    \draw (ai.220) to[myncbar, arm=0.2,angle=270] (aim1.south);
    \draw (ai.320) to[ myncbar, arm=0.2,angle=270] (aip1.south);
    \draw (ai.305) to[myncbar, arm=0.4,angle=270] (akm1.south);
    \draw (ai.290) to[myncbar, arm=0.6,angle=270] (ak.south);
    \node [below=6pt of ld1] {$\ldots$};
    \node [below=6pt of ld2] {$\ldots$};
    \end{tikzpicture}
    \end{align*}
   Although the vertices $\alpha_i$ and $\beta$ have been negated, all
retain their original positions.  On the other hand, the edges must be
redrawn.  In addition to reversing the original arrow from $\beta$ to
$\alpha_i$, each edge from $\beta$ to $\alpha_j$ for $j\neq i$ is
replaced by an edge from $\alpha_i$ to $\alpha_j$.
 \end{metalabpara}

 \begin{metalabpara}{mydefinition}{}
    {Swapping of rule instances}\envlabel{def:swapinst}
     The notion of swapping introduced in \envref{def:ruleswap}
extends naturally to rule instances.  Specifically, if $d$ is an
instance of rule $r$ obtained via substitution $s$ (see
\envref{disc:semwff} and \envref{def:ruleproofnot}), then for
 $\alpha_i \in \antecedentsof{r}$, $\rswap{d}{\substgr{\alpha_i}{s}}$
is the rule instance obtained by applying $s$ to all entries of
$\rswap{r}{\alpha_i}$.
 \end{metalabpara}

 \begin{metalabpara}{notation}{}
    {Proof trees in multiple-use contexts}\envlabel{not:multuse}
     In order to describe the process of proof contrapositioning (see
\envref{def:proofflip}), it is necessary to have a means of
identifying the vertices of the proof tree unambiguously.  For the
case of single-use proofs, this is an easy task, since the WFF
associated with that vertex identifies it uniquely.  However, if the
proof is not single use, there may several vertices labelled with the
same WFF.  In that case, the solution is to identify a non-root vertex
as a pair $\abr{\alpha,d}$ in which $\alpha$ is the WFF and $d$ is the
rule for which $\alpha \in \antecedentsof{d}$.  The root vertex is
identified specially as $\abr{\beta,\textsf{root}}$.  As a concrete
example, consider the proof tree of (\ref{def:single}-1), which is
shown in Fig.\ \ref{fig:multvertex}.
    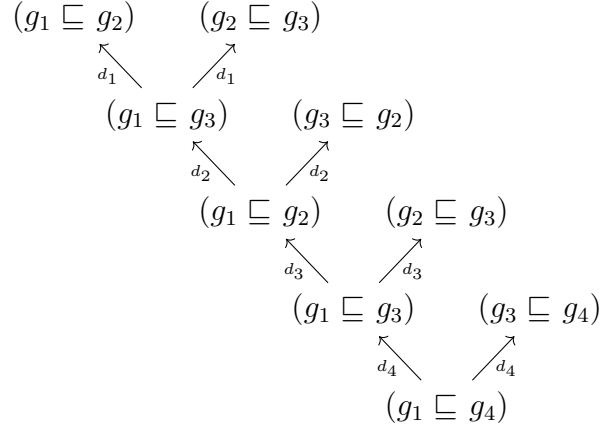
\begin{figure}[htb]
    \begin{center}
    \begin{tikzpicture} [->, baseline=(current bounding box.center)]
    \node {$\subrulep{g_1}{g_4}$} [grow' = up, align=center, anchor=north,
                     sibling distance = 6em, level distance = 4em]
    child {node {$\subrulep{g_1}{g_3}$} 
      child {node {$\subrulep{g_1}{g_2}$}
        child {node {$\subrulep{g_1}{g_3}$}
           child {node {$\subrulep{g_1}{g_2}$} edge from parent
                  node[below left=-0.4em] {$\scriptscriptstyle d_1$}
                 }
           child {node {$\subrulep{g_2}{g_3}$} edge from parent
                  node[below right=-0.4em] {$\scriptscriptstyle d_1$}
                 }
          edge from parent node[below left=-0.4em] {$\scriptscriptstyle d_2$}
              }
        child {node {$\subrulep{g_3}{g_2}$} edge from parent
                          node[below right=-0.4em] {$\scriptscriptstyle d_2$}
              }
          edge from parent node[below left=-0.4em] {$\scriptscriptstyle d_3$}
           }
      child {node {$\subrulep{g_2}{g_3}$} edge from parent
                         node[below right=-0.4em] {$\scriptscriptstyle d_3$}}
          edge from parent node[below left=-0.4em] {$\scriptscriptstyle d_4$}
            }
    child {node {$\subrulep{g_3}{g_4}$} edge from parent 
                       node[below right=-0.4em] {$\scriptscriptstyle d_4$}
          };
    \end{tikzpicture}
    \caption{Vertex-edge representation for
                             (\ref{def:single}-1)}\label{fig:multvertex}
    \end{center}
     \end{figure}
     There are two vertices labelled $\subrulep{g_1}{g_2}$; one is a
leaf, an antecedent of rule $d_1$, while the second is an interior
vertex, an antecedent of rule $d_3$.  These are represented as
$\grpr{\subrulep{g_1}{g_2}}{d_1}$ and $\grpr{\subrulep{g_1}{g_2}}{d_3}$,
respectively.  Note that the root vertex is denoted
$\grpr{\subrulep{g_1}{g_4}}{\textsf{root}}$.
     \par
     Given a proof tree $\tau$, notation such as
 $\grpr{\alpha}{d} \in \antecedentsof{\tau}$ will be used with the
obvious meaning --- that $\alpha \in \antecedentsof{\tau}$.
     \par
    It should be observed that this convention assumes that a single
rule instance does not have multiple instances of the same antecedent.
In most situations, this is a reasonable assumption.  To work with
multiple instances of the same antecedent, it is necessary to employ
trees which have ordered vertices, and to tag the instances with an
index based upon that order.  The details are straightforward, but
will not be used in this work, because they would render the
presentation much more complex.
 \end{metalabpara}

 \begin{metalabpara}{mydefinition}{}
    {Proof contrapositioning}\envlabel{def:proofflip}
  Figs.\ \ref{fig:tocontrap}, \ref{fig:contrapi}, and
\ref{fig:contrapii} illustrate the strategy of \emph{proof
contrapositioning}, in which rule swapping is applied to an entire
proof tree.  Figure \ref{fig:tocontrap} shows a \emph{source tree},
call it $\tau$, in vertex-edge format.
     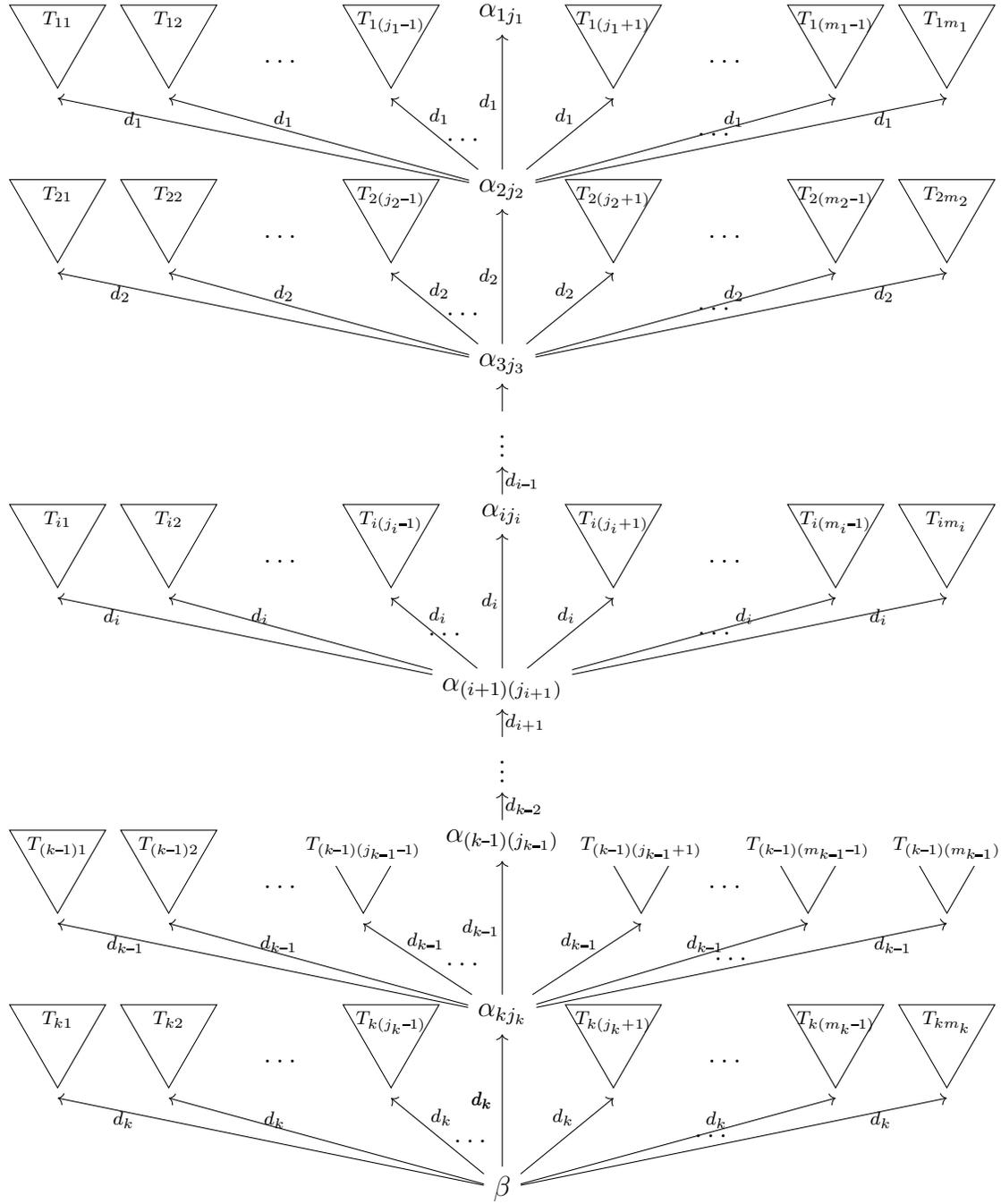
\begin{figure}[htb]
      \def\markneg#1{#1}
     \def\hsepc{\hspace*{1em}}
    \begin{tikzpicture} [->, baseline=(current bounding box.center) ]
    \node {$\beta$} [grow' = up, align=center, anchor=north,
                     sibling distance = 4em, level distance = 7em]
    child {node {$\treeupps{T_{k1}}$} \efpas
            \edgelabelds{left}{0.3}{near end}{k}
          }
    child {node {$\treeupps{T_{k2}}$} \efpas
            \edgelabelds{right}{0.1}{near end}{k} \noderldots{1.75}
          }
    child {node [below=2em] {$\ldots$} \efpn
          }
    child {node {$\treeupps{T_{k({j_k}\shortminus1)}}$} \efpas
            \edgelabelds{right}{0.1}{near end}{k}
          }
    child {node {$\alpha_{k{j_k}}$} \efpas
     child {node {$\treeupps{T_{(k\shortminus1)1}}$} \efpas
            \edgelabelds{left}{0.1}{near end}{k\shortminus1}
           }
     child {node {$\treeupps{T_{(k\shortminus1)2}}$} \efpas \noderldots{1.75}
            \edgelabelds{right}{0.1}{near end}{k\shortminus1}
           }
     child {node [below=2em] {$\ldots$} \efpn}
     child {node [xshift=-1em]
            {$\treeuppsb{T_{(k\shortminus1)({j_{k\shortminus1}\shortminus1)}}}$} \efpas
            \edgelabelds{right}{0.1}{near end}{k\shortminus1}
           }
     child {node (akm1jkm1) {$\alpha_{(k\shortminus1)(j_{k\shortminus1})}$} \efpas
            \edgelabelds{left}{-0.1}{midway}{k\shortminus1}
           }
     child {node [xshift=1em]
                    {$\treeuppsb{T_{(k\shortminus1)({j_{k\shortminus1}}+1)}}$} \efpas
                                                      \noderldots{1.75}
             node [left=0.1, near end] {$\scriptstyle d_{k\shortminus1}$}
           }
     child {node [below=2em] {$\ldots$} \efpn}
     child {node [xshift=-1em]
            {$\treeuppsb{T_{(k\shortminus1)({m_{k\shortminus1}}\shortminus1)}}$} \efpas
            \edgelabelds{left}{0.1}{near end}{k\shortminus1}
           }
     child {node {$\treeuppsb{T_{(k\shortminus1)(m_{k\shortminus1})}}$} \efpas
            \edgelabelds{right}{0.3}{near end}{k\shortminus1}
           }
            node [left=0.0, midway] {$\scriptstyle d_k$}
            \edgelabelds{left}{0.0}{midway}{k}
          }
    child {node {$\treeupps{T_{k({j_k}+1)}}$} \efpas \noderldots{1.75}
            \edgelabelds{left}{0.1}{near end}{k}
           }
    child {node [below=2em] {$\ldots$} \efpn}
    child {node {$\treeupps{T_{k({m_k}\shortminus1)}}$} \efpas
            \edgelabelds{left}{0.3}{near end}{k}
          }
    child {node {$\treeupps{T_{k{m_k}}}$} \efpas
            \edgelabelds{right}{0.3}{near end}{k}
          };
    \node (vdots1) [ above=1em of akm1jkm1] {$\vdots$};
    \draw (akm1jkm1) -- (vdots1)
            \edgelabelds{right}{-0.1}{midway}{k\shortminus2} ;
    \node (aip1jip1) [above=1em of vdots1]  {$\alpha_{(i+1)(j_{i+1})}$}
                    [grow' = up, align=center, anchor=north,
                     sibling distance = 4em, level distance = 7em]
    child {node {$\treeupps{T_{i1}}$} \efpas
            \edgelabelds{left}{0.3}{near end}{i}
          }
    child {node {$\treeupps{T_{i2}}$} \efpas
            \edgelabelds{right}{0.1}{near end}{i} \noderldots{1.75}
          }
    child {node [below=2em] {$\ldots$} \efpn}
    child {node {$\treeupps{T_{i({j_i}\shortminus1)}}$} \efpas
            \edgelabelds{right}{0.1}{near end}{i}
          }
    child {node (aiji) {$\alpha_{i{j_i}}$} \efpas
            \edgelabelds{left}{-0.1}{midway}{i}
          }
    child {node {$\treeupps{T_{i({j_i}+1)}}$} \efpas \noderldots{1.75}
            \edgelabelds{left}{0.1}{near end}{i}
          }
    child {node [below=2em] {$\ldots$} \efpn}
    child {node {$\treeupps{T_{i({m_i}\shortminus1)}}$} \efpas
            \edgelabelds{left}{0.1}{near end}{i}
          }
    child {node {$\treeupps{T_{i{m_i}}}$} \efpas
            \edgelabelds{right}{0.1}{near end}{i}
          };
    \draw (vdots1) -- (aip1jip1)
            \edgelabelds{right}{-0.1}{midway}{i+1} ;
    \node (vdots2) [ above=1em of aiji] {$\vdots$};
    \draw (aiji) -- (vdots2)
            \edgelabelds{right}{-0.1}{midway}{i\shortminus1} ;
    \node (a3j3) [above=1em of vdots2] {$\alpha_{3{j_3}}$}
                    [grow' = up, align=center, anchor=north,
                     sibling distance = 4em, level distance = 7em]
    child {node {$\treeupps{T_{21}}$} \efpas
            \edgelabelds{left}{0.3}{near end}{2}
          }
    child {node {$\treeupps{T_{22}}$} \efpas
            \edgelabelds{right}{0.3}{near end}{2} \noderldots{1.75}
          }
    child {node [below=2em] {$\ldots$} \efpn}
    child {node {$\treeupps{T_{2({j_2}\shortminus1)}}$} \efpas
            \edgelabelds{right}{0.1}{near end}{2}
          }
    child {node (a2ji) {$\alpha_{2{j_2}}$} \efpas
     child {node {$\treeupps{T_{11}}$} \efpas
            \edgelabelds{left}{0.1}{near end}{1}
           }
     child {node {$\treeupps{T_{12}}$} \efpas
            \edgelabelds{right}{0.3}{near end}{1} \noderldots{1.75}
           }
     child {node [below=2em] {$\ldots$} \efpn}
     child {node {$\treeupps{T_{1({j_1}\shortminus1)}}$} \efpas
            \edgelabelds{right}{0.1}{near end}{1}
           }
     child {node (a1j1) {$\alpha_{1{j_1}}$} \efpas
            \edgelabelds{left}{-0.1}{midway}{1}
           }
     child {node {$\treeupps{T_{1({j_1}+1)}}$} \efpas
            \edgelabelds{left}{0.1}{near end}{1} \noderldots{1.75}
           } 
     child {node [below=2em] {$\ldots$} \efpn}
     child {node {$\treeupps{T_{1({m_1}\shortminus1)}}$} \efpas
            \edgelabelds{left}{0.1}{near end}{1}
           }
     child {node {$\treeupps{T_{1m_1}}$} \efpas
            \edgelabelds{right}{0.3}{near end}{1}
           }
            \edgelabelds{left}{-0.1}{midway}{2}
          }
    child {node {$\treeupps{T_{2({j_2}+1)}}$} \efpas
            \edgelabelds{left}{0.1}{near end}{2} \noderldots{1.75}
          }
    child {node [below=2em] {$\ldots$} \efpn}
    child {node {$\treeupps{T_{2({m_2}\shortminus1)}}$} \efpas
            \edgelabelds{left}{0.1}{near end}{2}
          }
    child {node {$\treeupps{T_{2{m_2}}}$} \efpas
            \edgelabelds{right}{0.3}{near end}{2}
          };
    \draw (vdots2) to (a3j3);
    \end{tikzpicture}
    \caption{The source proof to undergo contrapositioning}\label{fig:tocontrap}
    \end{figure}
     It is an arbitrary proof tree using the rules $\proofsys{A}$; no
special properties are assumed.  Each entry of the form
{\footnotesize$\treeup{T_{ij}}$} represents a subtree of the entire
proof tree which remains unchanged during the transformation.
    The most important vertex in the tree is
$\grpr{\alpha_{1{j_1}}}{d_1}$, called the \emph{(designated) swap
leaf} of $\tau$.  It is the antecedent which will be swapped with the
consequent $\beta$ to obtain a proof of $\mlnot\alpha_{1{j_1}}$ from
 $(\antecedentsof{\tau} \setminus \setbr{\alpha_{1{j_1}}})
                               \union \setbr{\mlnot\beta}$.
   The \emph{swap path} is
 \nlrightt
    $\abr{\grpr{\alpha_{1{j_1}}}{d_1},\grpr{\alpha_{2{j_2}}}{d_2},
          \ldots,\grpr{\alpha_{i{j_i}}}{d_i},\ldots,
          \grpr{\alpha_{(k-1){j_{k-1}}}}{d_{i-1}},
          \grpr{\alpha_{k{j_k}}}{d_k},\grpr{\beta}{\rootv}}$,
  \linebreak
  the (unique) path from the swap leaf to the root.  The members of
this sequence are precisely the vertices whose assertions will be
negated in the contrapositioned proof.  In Fig.\ \ref{fig:tocontrap},
this path goes right down the middle of the diagram.
     For  $i \in \ccinterval{1}{k-1}$, 
         $\consequentof{d_{i}} = \alpha_{(i+1){j_{i+1}}}
                             \in \antecedentsof{d_{i+1}}$,
 with
 $\consequentof{d_{k}} = \beta$.
    \par
  In the result, called the \emph{contrapositioned tree}, each rule
instance $d_{i}$ in the swap-rule sequence is replaced by
 $\rswap{d_{i}}{\alpha_{i{j_i}}}$.  This tree is illustrated in
Fig.\ \ref{fig:contrapi}, using fixed-vertex positioning as described
in \envref{def:irulegraphswap}.  For compactness of notation, this
replacement rule is labelled $d_{i}'$ in Fig.\ \ref{fig:contrapi}.
     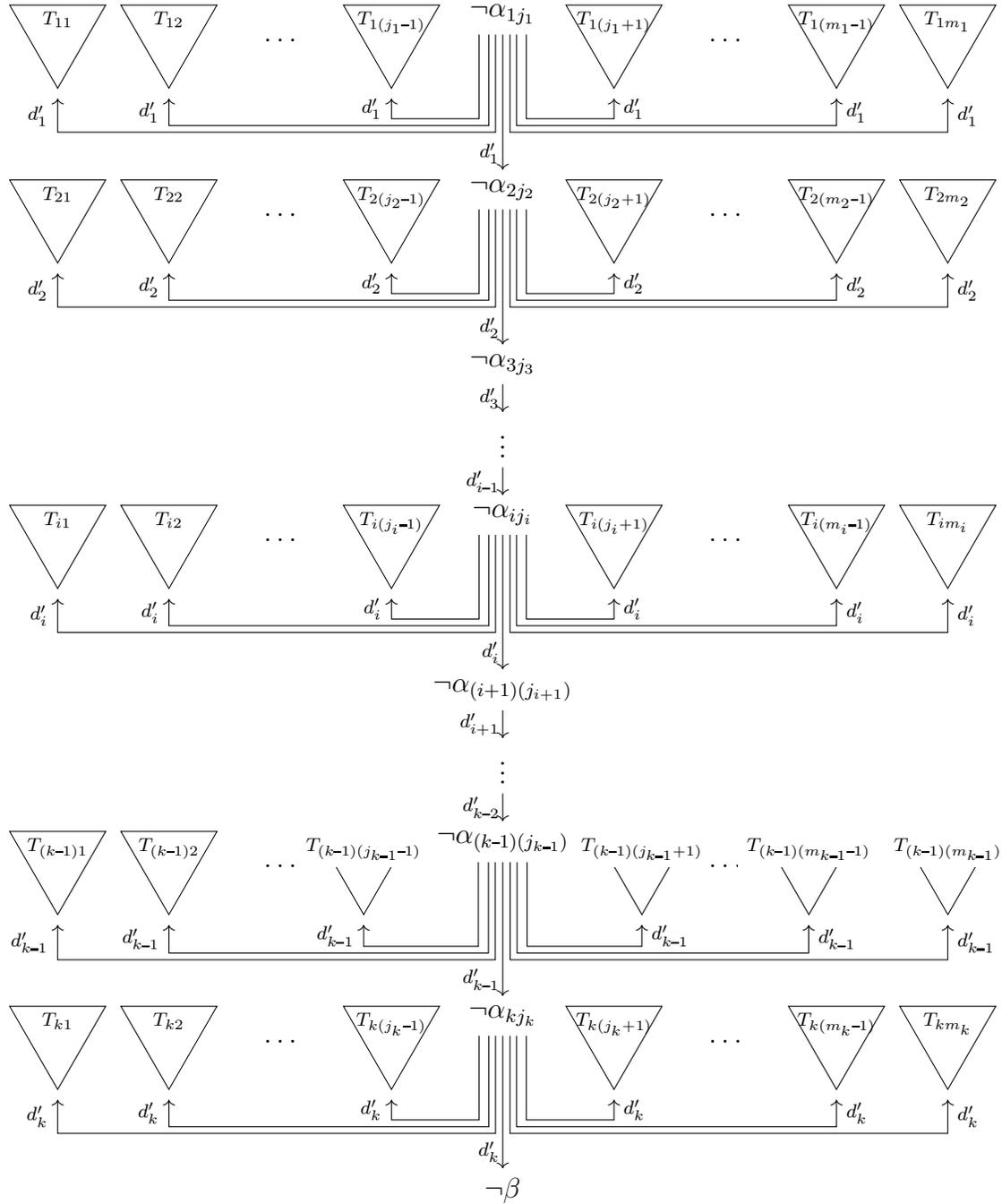
\begin{figure}[htb]
      \def\markneg#1{#1}
     \def\hsepc{\hspace*{1em}}
   \begin{tikzpicture} [->, baseline=(current bounding box.center) ]
    \node (beta) {$\mlnot\beta$} [grow' = up, align=center, anchor=north,
                     sibling distance = 4em, level distance = 7em]
    child {node (tk1) {$\treeupps{T_{k1}}$} \efpn}
    child {node (tk2) {$\treeupps{T_{k2}}$} \efpn}
    child {\nodebldots{0.5} \efpn}
    child {node (tkjkm1) {$\treeupps{T_{k({j_k}\shortminus1)}}$} \efpn}
    child {node (akjk) {$\mlnot\alpha_{k{j_k}}$} edge from parent [<-]
     child {node (tkm11) {$\treeupps{T_{(k\shortminus1)1}}$} \efpn}
     child {node (tkm12) {$\treeupps{T_{(k\shortminus1)2}}$} \efpn}
     child {\nodebldots{0.5} \efpn}
     child {node (tkm1jkm1m1) [xshift=-1em]
            {$\treeuppsb{T_{(k\shortminus1)({j_{k\shortminus1}\shortminus1)}}}$} \efpn}
     child {node (akm1jkm1) {$\mlnot\alpha_{(k\shortminus1)(j_{k\shortminus1})}$}
                                     edge from parent [<-]
           \edgelabeldsp{left}{-0.1}{very near start}{k\shortminus1}
           }
     child {node (tkm1jkm1p1) [xshift=1em]
             {$\treeuppsb{T_{(k\shortminus1)({j_{k\shortminus1}}+1)}}$} \efpn}
     child {\nodebldots{0.5} \efpn}
     child {node (tkm1mkm1m1) [xshift=-1em]
            {$\treeuppsb{T_{(k\shortminus1)({m_{k\shortminus1}}\shortminus1)}}$} \efpn}
     child {node (tkm1mkm1) {$\treeuppsb{T_{(k\shortminus1)(m_{k\shortminus1})}}$}
            \efpn
           }
           \edgelabeldsp{left}{-0.1}{very near start}{k}
          }
    child {node (tkjkp1) {$\treeupps{T_{k({j_k}+1)}}$} \efpn}
    child {\nodebldots{0.5} \efpn}
    child {node (tkmkm1) {$\treeupps{T_{k({m_k}\shortminus1)}}$} \efpn}
    child {node (tkmk) {$\treeupps{T_{k{m_k}}}$} \efpn};
    \draw (akm1jkm1.250) to[myncbar, arm=1.45,angle=270]
            node[left] {\edgelabelnames{d_{k\shortminus1}'}} (tkm11.south);
    \draw (akm1jkm1.235) to[myncbar, arm=1.35,angle=270]
            node[left] {\edgelabelnames{d_{k\shortminus1}'}} (tkm12.south);
    \draw (akm1jkm1.220) to[myncbar, arm=1.25,angle=270]
            node[left] {\edgelabelnames{d_{k\shortminus1}'}} (tkm1jkm1m1.south);
    \draw (akm1jkm1.320) to[myncbar, arm=1.25,angle=270]
            node[right] {\edgelabelnames{d_{k\shortminus1}'}} (tkm1jkm1p1.south);
    \draw (akm1jkm1.305) to[myncbar, arm=1.35,angle=270]
            node[right] {\edgelabelnames{d_{k\shortminus1}'}} (tkm1mkm1m1.south);
    \draw (akm1jkm1.290) to[myncbar, arm=1.45,angle=270]
            node[right] {\edgelabelnames{d_{k\shortminus1}'}} (tkm1mkm1.south);
    \draw (akjk.250) to[myncbar, arm=1.45,angle=270]
            node[left] {\edgelabelnames{d_{k}'}} (tk1.south);
    \draw (akjk.235) to[myncbar, arm=1.35,angle=270]
            node[left] {\edgelabelnames{d_{k}'}} (tk2.south);
    \draw (akjk.220) to[myncbar, arm=1.25,angle=270]
            node[left] {\edgelabelnames{d_{k}'}} (tkjkm1.south);
    \draw (akjk.320) to[myncbar, arm=1.25,angle=270]
            node[right] {\edgelabelnames{d_{k}'}} (tkjkp1.south);
    \draw (akjk.305) to[myncbar, arm=1.35,angle=270]
            node[right] {\edgelabelnames{d_{k}'}} (tkmkm1.south);
    \draw (akjk.290) to[myncbar, arm=1.45,angle=270]
            node[right] {\edgelabelnames{d_{k}'}} (tkmk.south);
    \node (vdots1) [ above=1em of akm1jkm1] {$\vdots$};
    \draw [<-] (akm1jkm1) -- (vdots1)
           \edgelabeldsp{left}{-0.1}{midway}{k\shortminus2};
    \node (aip1jip1) [above=1em of vdots1]  {$\mlnot\alpha_{(i+1)(j_{i+1})}$}
                    [grow' = up, align=center, anchor=north,
                     sibling distance = 4em, level distance = 7em]
    child {node (ti1) {$\treeupps{T_{i1}}$}
                    edge from parent [child anchor=south, draw=none]}
    child {node (ti2) {$\treeupps{T_{i2}}$}
                    edge from parent [child anchor=south, draw=none]}
    child {\nodebldots{0.5} \efpn}
    child {node (tijim1) {$\treeupps{T_{i({j_i}\shortminus1)}}$}
                    edge from parent [child anchor=south, draw=none]}
    child {node (aiji) {$\mlnot\alpha_{i{j_i}}$} edge from parent [<-]
           \edgelabeldsp{left}{-0.1}{very near start}{i}          }
    child {node (tijip1) {$\treeupps{T_{i({j_i}+1)}}$}
                    edge from parent [child anchor=south, draw=none]}
    child {\nodebldots{0.5} \efpn}
    child {node (timim1) {$\treeupps{T_{i({m_i}\shortminus1)}}$}
                    edge from parent [child anchor=south, draw=none]}
    child {node (timi) {$\treeupps{T_{i{m_i}}}$}
                    edge from parent [child anchor=south, draw=none]};
    \draw (aiji.250) to[myncbar, arm=1.45,angle=270]
            node[left] {\edgelabelnames{d_{i}'}} (ti1.south);
    \draw (aiji.235) to[myncbar, arm=1.35,angle=270]
            node[left] {\edgelabelnames{d_{i}'}} (ti2.south);
    \draw (aiji.220) to[myncbar, arm=1.25,angle=270]
            node[left] {\edgelabelnames{d_{i}'}} (tijim1.south);
    \draw (aiji.320) to[myncbar, arm=1.25,angle=270]
            node[right] {\edgelabelnames{d_{i}'}} (tijip1.south);
    \draw (aiji.305) to[myncbar, arm=1.35,angle=270]
            node[right] {\edgelabelnames{d_{i}'}} (timim1.south);
    \draw (aiji.290) to[myncbar, arm=1.45,angle=270]
            node[right] {\edgelabelnames{d_{i}'}} (timi.south);
    \draw [<-] (vdots1) -- (aip1jip1)
           \edgelabeldsp{left}{-0.1}{midway}{i+1};
    \node (vdots2) [ above=1em of aiji] {$\vdots$};
    \draw [<-] (aiji) -- (vdots2)
           \edgelabeldsp{left}{-0.1}{midway}{i\shortminus1};
    \node (a3j3) [above=1em of vdots2] {$\mlnot\alpha_{3{j_3}}$}
                    [grow' = up, align=center, anchor=north,
                     sibling distance = 4em, level distance = 7em]
    child {node (t21) {$\treeupps{T_{21}}$}
                edge from parent [child anchor=south,draw=none]}
    child {node (t22) {$\treeupps{T_{22}}$}
                edge from parent [child anchor=south,draw=none]}
    child {\nodebldots{0.5} \efpn}
    child {node (t2j2m1) {$\treeupps{T_{2({j_2}\shortminus1)}}$}
                edge from parent [child anchor=south,draw=none]}
    child {node (a2j2) {$\mlnot\alpha_{2{j_2}}$} edge from parent [<-] 
     child {node (t11) {$\treeupps{T_{11}}$}
                edge from parent [child anchor=south,draw=none]}
     child {node (t12) {$\treeupps{T_{12}}$}
                edge from parent [child anchor=south,draw=none]}
     child {\nodebldots{0.5} \efpn}
     child {node (t1jm1) {$\treeupps{T_{1({j_1}\shortminus1)}}$}
                edge from parent [child anchor=south,draw=none]}
     child {node (a1j1) {$\mlnot\alpha_{1{j_1}}$} edge from parent [<-]
           \edgelabeldsp{left}{-0.1}{very near start}{1}
           }
     child {node (t1j1p1) {$\treeupps{T_{1({j_1}+1)}}$}
                edge from parent [child anchor=south,draw=none]}
     child {\nodebldots{0.5} \efpn}
     child {node (t1m1m1) {$\treeupps{T_{1({m_1}\shortminus1)}}$}
                edge from parent [child anchor=south,draw=none]}
     child {node (t1m1) {$\treeupps{T_{1m_1}}$}
                edge from parent [child anchor=south,draw=none]}
           \edgelabeldsp{left}{-0.1}{very near start}{2}
          }
    child {node (t2j2p1) {$\treeupps{T_{2({j_2}+1)}}$}
                edge from parent [child anchor=south,draw=none]}
    child {\nodebldots{0.5} \efpn}
    child {node (t2m2m1) {$\treeupps{T_{2({m_2}\shortminus1)}}$}
                edge from parent [child anchor=south,draw=none]}
    child {node (t2m2) {$\treeupps{T_{2{m_2}}}$}
               edge from parent [child anchor=south,draw=none]};
    \draw [<-] (vdots2) -- (a3j3)
           \edgelabeldsp{left}{-0.1}{midway}{3};
    \draw (a1j1.250) to[myncbar, arm=1.45,angle=270]
            node[left] {\edgelabelnames{d_{1}'}} (t11.south);
    \draw (a1j1.235) to[myncbar, arm=1.35,angle=270]
            node[left] {\edgelabelnames{d_{1}'}} (t12.south);
    \draw (a1j1.220) to[myncbar, arm=1.25,angle=270]
            node[left] {\edgelabelnames{d_{1}'}} (t1jm1.south);
    \draw (a1j1.290) to[myncbar, arm=1.45,angle=270]
            node[right] {\edgelabelnames{d_{1}'}} (t1m1.south);
    \draw (a1j1.305) to[myncbar, arm=1.35,angle=270]
            node[right] {\edgelabelnames{d_{1}'}} (t1m1m1.south);
    \draw (a1j1.320) to[myncbar, arm=1.25,angle=270]
            node[right] {\edgelabelnames{d_{1}'}} (t1j1p1.south);
    \draw (a2j2.250) to[myncbar, arm=1.45,angle=270]
            node[left] {\edgelabelnames{d_{2}'}} (t21.south);
    \draw (a2j2.235) to[myncbar, arm=1.35,angle=270]
            node[left] {\edgelabelnames{d_{2}'}} (t22.south);
    \draw (a2j2.220) to[myncbar, arm=1.25,angle=270]
            node[left] {\edgelabelnames{d_{2}'}} (t2j2m1.south);
    \draw (a2j2.320) to[myncbar, arm=1.25,angle=270]
            node[right] {\edgelabelnames{d_{2}'}} (t2j2p1.south);
    \draw (a2j2.305) to[myncbar, arm=1.35,angle=270]
            node[right] {\edgelabelnames{d_{2}'}} (t2m2m1.south);
    \draw (a2j2.290) to[myncbar, arm=1.45,angle=270]
            node[right] {\edgelabelnames{d_{2}'}} (t2m2.south);
    \end{tikzpicture}
     \caption{The contrapositioned proof in fixed-position format}\label{fig:contrapi}
     \end{figure}
    Notice that the changes described in \envref{def:irulegraphswap}
are followed.  In particular, in the transition from $d_i$ to $d_i'$,
the antecedent $\alpha_{i{j_i}}$ and the consequent
$\alpha_{(i+1){j_{i+1}}}$ ($\beta$ for $d_k$) are negated, and the
arrow connecting them is reversed.  Furthermore, the source of all
arrows of $d_i$ is changed to $\mlnot\alpha_{i{j_i}}$ in $d_i'$.
     \par
     In Fig,\ \ref{fig:contrapii}, this same contrapositioned tree is
shown in fixed-edge format.  
     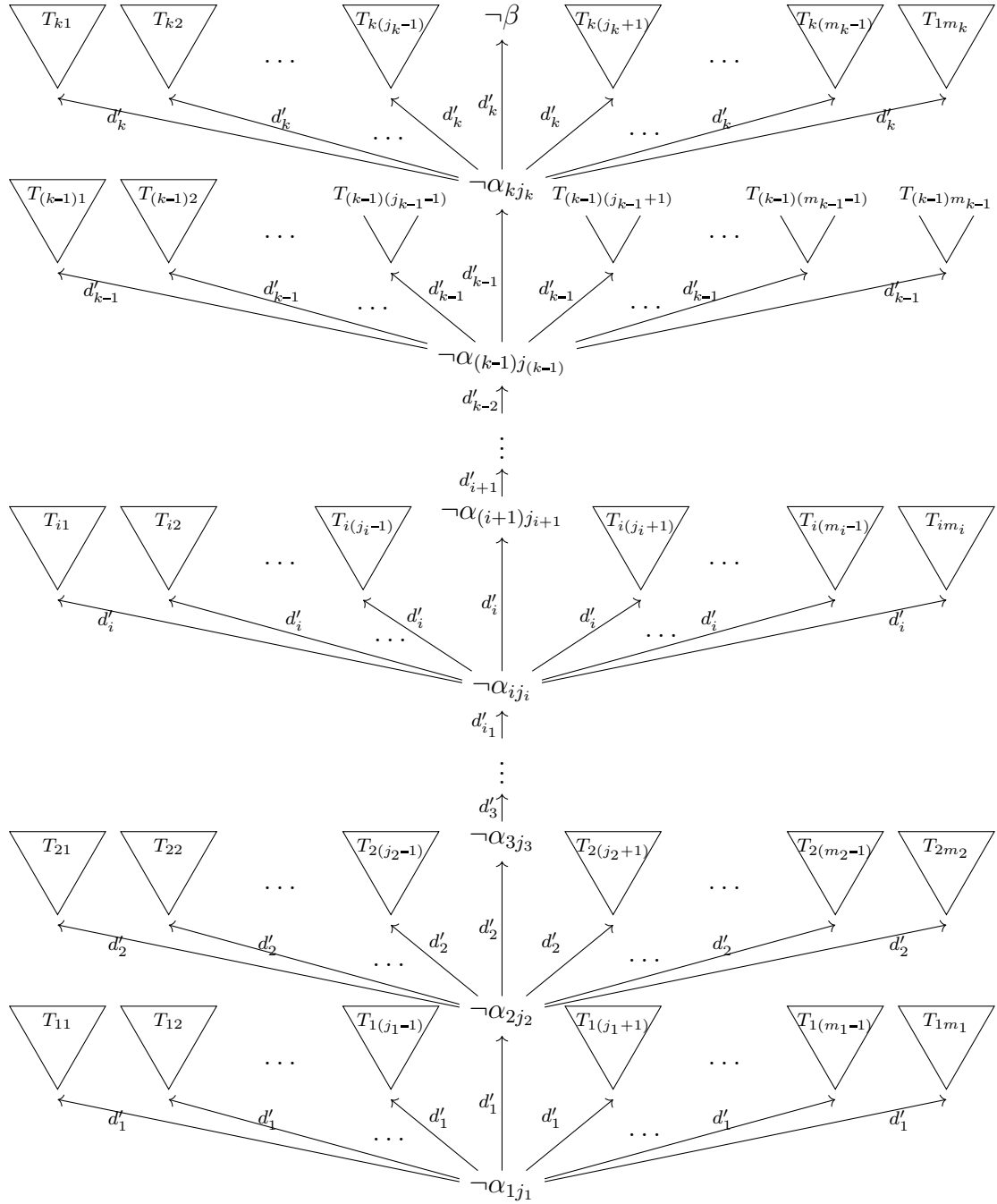
\begin{figure}[htb]
      \def\markneg#1{#1}
     \def\hsepc{\hspace*{1em}}
    \begin{tikzpicture} [->, baseline=(current bounding box.center) ]
    \node {$\mlnot\alpha_{1{j_1}}$} [grow' = up, align=center, anchor=north,
                     sibling distance = 4em, level distance = 7em]
    child {node {$\treeupps{T_{11}}$} \efpas
            \edgelabeldsp{left}{0.3}{near end}{1}
          }
    child {node {$\treeupps{T_{12}}$} \efpas
            \edgelabeldsp{right}{0.1}{near end}{1}
                                     node [right=1.75em] {$\ldots$}
          }
    child {node [below=2em] {$\ldots$} \efpn}
    child {node {$\treeupps{T_{1({j_1}\shortminus1)}}$}
                                     \efpas
            \edgelabeldsp{right}{0.1}{near end}{1}
          }
    child {node {$\mlnot\alpha_{2{j_2}}$} \efpas
     child {node {$\treeupps{T_{21}}$}
                                     \efpas
            \edgelabeldsp{left}{0.3}{near end}{2}
           }
     child {node {$\treeupps{T_{22}}$}
                                      \efpas
            \edgelabeldsp{right}{0.1}{near end}{2}
                                      node [right=1.75em] {$\ldots$}
           }
     child {node [below=2em] {$\ldots$} \efpn}
     child {node {$\treeupps{T_{2(j_2\shortminus1)}}$}
                                     \efpas
            \edgelabeldsp{right}{0.1}{near end}{2}
           }
     child {node (akm1jkm1) {$\mlnot\alpha_{3j_3}$} \efpas
            \edgelabeldsp{left}{-0.1}{midway}{2}
           }
     child {node {$\treeupps{T_{2({j_2}+1)}}$} \efpas
                                     node [right=1.75em] {$\ldots$}
            \edgelabeldsp{left}{0.3}{near end}{2}
           } 
     child {node [below=2em] {$\ldots$} \efpn}
     child {node {$\treeupps{T_{2({m_2}\shortminus1)}}$} \efpas
            \edgelabeldsp{left}{0.3}{near end}{2}
           }
     child {node {$\treeupps{T_{2m_2}}$} \efpas
            \edgelabeldsp{right}{0.5}{near end}{2}
           }
            \edgelabeldsp{left}{-0.1}{midway}{1}
          }
    child {node {$\treeupps{T_{1({j_1}+1)}}$} \efpas
            \edgelabeldsp{left}{0.3}{near end}{1}
                                     node [right=1.75em] {$\ldots$}}
    child {node [below=2em] {$\ldots$} \efpn}
    child {node {$\treeupps{T_{1({m_1}\shortminus1)}}$} \efpas
            \edgelabeldsp{left}{0.3}{near end}{1}
                }
    child {node {$\treeupps{T_{1{m_1}}}$} \efpas
            \edgelabeldsp{right}{0.5}{near end}{1}
          };
    \node (vdots1) [ above=1em of akm1jkm1] {$\vdots$};
    \draw (akm1jkm1) -- (vdots1)
            \edgelabeldsp{left}{-0.1}{midway}{3};
    \node (aip1jip1) [above=1em of vdots1]  {$\mlnot\alpha_{i{j_i}}$}
                    [grow' = up, align=center, anchor=north,
                     sibling distance = 4em, level distance = 7em]
    child {node {$\treeupps{T_{i1}}$} \efpas
            \edgelabeldsp{left}{0.5}{near end}{i}
          }
    child {node {$\treeupps{T_{i2}}$} \efpas
            \edgelabeldsp{right}{0.5}{near end}{i}
                                     node [right=1.75em] {$\ldots$}
          }
    child {node [below=2em] {$\ldots$} \efpn}
    child {node [xshift=-1em] {$\treeupps{T_{i({j_i}\shortminus1)}}$} \efpas
            \edgelabeldsp{right}{0.1}{near end}{i}
          }
    child {node (aiji) {$\mlnot\alpha_{(i+1){j_{i+1}}}$} \efpas
            \edgelabeldsp{left}{-0.1}{midway}{i}
          }
    child {node [xshift=1em] {$\treeupps{T_{i({j_i}+1)}}$} \efpas
            \edgelabeldsp{left}{0.1}{near end}{i}
                                     node [right=1.75em] {$\ldots$}
          }
    child {node [below=2em] {$\ldots$} \efpn}
    child {node {$\treeupps{T_{i({m_i}\shortminus1)}}$} \efpas
            \edgelabeldsp{left}{0.5}{near end}{i}
          }
    child {node {$\treeupps{T_{i{m_i}}}$} \efpas
            \edgelabeldsp{right}{0.5}{near end}{i}
          };
    \draw (vdots1) -- (aip1jip1)
            \edgelabeldsp{left}{-0.1}{midway}{i_1};
    \node (vdots2) [ above=1em of aiji] {$\vdots$};
    \draw (aiji) -- (vdots2)
            \edgelabeldsp{left}{-0.1}{midway}{i+1};
    \node (a3j3) [above=1em of vdots2]
                     {$\mlnot\alpha_{(k\shortminus1){j_{(k\shortminus1)}}}$}
                    [grow' = up, align=center, anchor=north,
                     sibling distance = 4em, level distance = 7em]
    child {node {$\treeupps{T_{(k\shortminus1)1}}$} \efpas
            \edgelabeldsp{left}{0.3}{near end}{k\shortminus1}
          }
    child {node {$\treeupps{T_{(k\shortminus1)2}}$} \efpas
            \edgelabeldsp{right}{0.3}{near end}{k\shortminus1}
                                     node [right=1.75em] {$\ldots$}}
    child {node [below=2em] {$\ldots$} \efpn}
    child {node {$\treeuppsb{T_{(k\shortminus1)({j_{k\shortminus1}}\shortminus1)}}$}
                \efpas
            \edgelabeldsp{right}{0.1}{near end}{k\shortminus1}
          }
    child {node (a2i2) {$\mlnot\alpha_{k{j_k}}$} \efpas
     child {node {$\treeupps{T_{k1}}$} \efpas
            \edgelabeldsp{left}{0.3}{near end}{k}
           }
     child {node {$\treeupps{T_{k2}}$} \efpas
            \edgelabeldsp{right}{0.3}{near end}{k}
                                      node [right=1.75em] {$\ldots$}}
     child {node [below=2em] {$\ldots$} \efpn}
     child {node {$\treeupps{T_{k({j_k}\shortminus1)}}$} \efpas
            \edgelabeldsp{right}{0.3}{near end}{k}
           }
     child {node (a1i1) {$\mlnot\beta$} \efpas
            \edgelabeldsp{left}{-0.1}{midway}{k}
           }
     child {node {$\treeupps{T_{k({j_k}+1)}}$}
                                     \efpas
            \edgelabeldsp{left}{0.3}{near end}{k}
                                     node [right=1.75em] {$\ldots$}} 
     child {node [below=2em] {$\ldots$} \efpn}
     child {node {$\treeupps{T_{k({m_k}\shortminus1)}}$} \efpas
            \edgelabeldsp{left}{0.3}{near end}{k}
           }
     child {node {$\treeupps{T_{1m_k}}$} \efpas
            \edgelabeldsp{right}{0.3}{near end}{k}
           }
            \edgelabeldsp{left}{-0.1}{midway}{k\shortminus1}
          }
    child {node {$\treeuppsb{T_{(k\shortminus1)({j_{k\shortminus1}}+1)}}$} \efpas
            \edgelabeldsp{left}{0.1}{near end}{k\shortminus1}
                                     node [right=1.75em] {$\ldots$}}
    child {node [below=2em] {$\ldots$} \efpn}
    child {node [xshift=-1em]
           {$\treeuppsb{T_{(k\shortminus1)({m_{k\shortminus1}}\shortminus1)}}$} \efpas
            \edgelabeldsp{left}{0.3}{near end}{k\shortminus1}
          }
    child {node {$\treeuppsb{T_{(k\shortminus1){m_{k\shortminus1}}}}$} \efpas
            \edgelabeldsp{right}{0.3}{near end}{k\shortminus1}
          };
    \draw (vdots2) -- (a3j3)
            \edgelabeldsp{left}{-0.1}{midway}{k\shortminus2};
    \end{tikzpicture}
     \caption{The contrapositioned proof in hierarchical format}\label{fig:contrapii}
     \end{figure}
Although the result may seem more aesthetically pleasing, note
that there has been substantial rearrangement --- the swapped rules
occur in the flipped tree in the opposite order from that in the
source tree, since the swapping operation exchanges an antecedent of
the rule with the consequent (and then complements both).  Thus, while
fixed-edge positioning may result in a tree which looks more natural,
it is fixed-vertex positioning which illustrates, in a far more
understandable manner, the relationship between the source tree and its
contrapositioned counterpart.
     \par
     The contrapositioning of the tree $\tau$ is denoted
$\contraposn{\tau}{\grpr{\alpha_{1{j_i}}}{d_1}}$.  The swap leaf
determines the entire swap path.
     \par
     For examples of contrapositioning, the reader is encouraged to
look ahead to \envref{ex:neginf1} and \envref{ex:neginf2}.
     The immediate topic is to establish the key completeness result
surrounding contrapositioning.
 \end{metalabpara}

 \begin{metaemphlabpara}{lemma}{Lemma}
    {Contrapositioning preserves proof}\envlabel{lem:contraptree}
 If $\tau$ is a proof tree over $\proofsys{A}$ with
 $\grpr{\alpha}{d} \in \antecedentsof{\tau}$, then
$\contraposn{\tau}{\grpr{\alpha}{d}}$ is a proof of $\mlnot\alpha$
from
  $(\antecedentsof{\tau} \setminus \setbr{\alpha})
     \union \setbr{\mlnot\consequentof{\tau}}$
 in the proof system $\swapcl{\proofsys{A}}$.
 \begin{proof}
     The result follows directly from the construction given in
\envref{def:proofflip}.  It is easy to verify that
$\contraposn{\tau}{\grpr{\alpha}{d}}$ is a proof tree constructed from
the rules of $\swapcl{\proofsys{A}}$, with the required antecedents
and consequent.
 \end{proof}
 \end{metaemphlabpara}
 
 \begin{metaemphlabpara}{theorem}{Theorem}
    {Sat-completeness of
         Armstrong inference systems}\envlabel{thm:armcomplete}
  If $\proofsys{A}$ is sound, complete and single use, then
$\swapcl{\proofsys{A}}$ is sound, sat-complete, and single use
also.
 \begin{proof}
   Consistency of $\swapcl{\proofsys{A}}$ follows from
\envref{prop:genswap}(b).
   To show sat-completeness, assume that $\proofsys{A}$ is complete,
and let $\Phi \subseteq \allarm{\armctxt{C}}$ be satisfiable, and let
 $\psi \in \armctxt{C}$.  There are two cases to consider: 
 $\Phi \sentails \psi$ and $\Phi \sentails \mlnot\psi$.
     \par
    For the first case, $\Phi \sentails \psi$, it suffices to invoke
\envref{prop:infarm}(a), which ensures that
 $\posof{\Phi} \syntailss[\proofsys{A}] \psi$,
 whence $\Phi \syntailss[\proofsys{A}] \psi$ holds \emph{a fortiori}.
     \par
   For the second case, if $\Phi \sentails \mlnot\psi$, then by
\envref{prop:infarm}(b), either
  $\posof{\Phi} \union \setbr{\psi}
                  \syntailss[\proofsys{A}] \false$,
 or else there is a single
 $\mlnot\varphi \in \negof{\Phi}$ with the property that 
   $\posof{\Phi} \union \setbr{\psi} \sentails \varphi$.
    For the first subcase,  $\posof{\Phi} \union \setbr{\psi}
                  \syntailss[\proofsys{A}] \false$,
 let $\tau$ be the proof tree in $\proofsys{A}$
modelled after that of Fig.\ \ref{fig:tocontrap}, with $\alpha_{1{j_1}}=\psi$
and $\beta=\false$.
    For the second subcase,
     $\posof{\Phi} \union \setbr{\psi} \sentails \varphi$ for some
$\mlnot\varphi \in \negof{\Phi}$, let $\tau$ be the proof tree in
$\proofsys{A}$ modelled after that of Fig.\ \ref{fig:tocontrap}, with
 $\alpha_{1{j_1}}=\psi$ and $\beta=\varphi$.
    In either subcase, the completeness of $\proofsys{A}$ guarantees
that such a tree exists.  If $\psi$ is used more than once as an
antecedent, pick any occurrence.  (It will be shown shortly that such
multiple occurrences are impossible.)
    \par
    Now, consider the tree
$\tau'=\contraposn{\tau}{\grpr{\psi}{d_1}}$, with $d_1$ the rule which
has $\psi=\alpha_{1{j_1}}$ as antecedent, as in
Fig.\ \ref{fig:tocontrap}.  In both subcases identified above, the
contrapositioned tree $\tau'$ will have $\mlnot\psi$ as the
conclusion.  In the first,
 $\mlnot\beta=\mlnot\false=\true$, so $\mlnot\beta$ is simply omitted
as an antecedent of $\tau'$.
    In the second, $\mlnot\beta=\mlnot\varphi$, which is
exactly the single negative sentence which must be added as an
antecedent of the entire proof.
    It cannot be the case that there are multiple occurrences of
$\psi$ in the original tree $\tau$.  If there were, the resulting tree
would exhibit a proof of $\mlnot\psi$ from a set of premises which
includes $\psi$.  Since all of the rules are sound, this is
impossible.
    \par
    From this, it follows that $\contraposn{\tau}{\grpr{\psi}{d_1}}$
is a proof of $\psi$ from $\Phi$ in $\swapcl{\proofsys{A}}$, so the
latter is sat-complete.  The single-use property follows from the
argument of the previous paragraph.
  \end{proof}
 \end{metaemphlabpara}

   \parvert
    To obtain a system which is complete even for unsatisfiable
formulas of $\allarm{\armctxt{C}}$, the notion of complement-pair
rules is introduced.    

 \begin{metalabpara}{mydefinition}{}
    {Complement-pair rules}\envlabel{def:compprrules}
     A \emph{complement-pair rule} is one which asserts that a WFF and
its negation cannot both be true.  A generic such rule is illustrated
in (\ref{def:compprrules}-1).
     \[
       \tag{\ref{def:compprrules}-1}
       \begin{prooftree}
         \hypoii{\alpha}
                {\mlnot\alpha}
         \inferi{\false}
       \end{prooftree}
     \]
    The \emph{complement-pair set} of $\armctxt{C}$, denoted
$\compprset{\armctxt{C}}$, includes, for each
 $\alpha \in \armctxt{C}$, a complement-pair rule of the form shown
(\ref{def:compprrules}-1).  In the case that the rules of
$\proofsys{A}$ are represented generically using variables, it is only
necessary to add rules which instantiate to all complement-pair rules.
For the main example $\pbinconstrgrsch{G}$,
$\compprset{\pbinconstrgrsch{G}}$ is given by the two rules shown in
Fig.\ \ref{fig:cprgrsch}.
  \begin{figure}[htb]
  \begin{center}
   \def\isepi{0.5em}
   \begin{align*}
   &\mbox{(C1)}\hspace*{\isepi}
    \begin{prooftreem}
       \hypoii{\subrulep{\granvar{g}_1}{\granvar{g}_2}}
              {\nsubrulep{\granvar{g}_1}{\granvar{g}_2}}
       \inferi{\false}
    \end{prooftreem}
   &&\mbox{(C2)}\hspace*{\isepi}
    \begin{prooftreem}
      \hypoii{\disjrulep{\granvar{g}_1}{\granvar{g}_2}}
              {\ndisjrulep{\granvar{g}_1}{\granvar{g}_2}}
      \inferi{\false}
    \end{prooftreem}
    \end{align*}
  \end{center}
  \caption{Complement-pair rules for $\pbinconstrgrsch{G}$}\label{fig:cprgrsch}
  \end{figure}
     \par
   The \emph{full swap closure} of $\proofsys{A}$, denoted
$\fullswapcl{\proofsys{A}}$, is
 $\swapcl{A} \union \compprset{\armctxt{C}}$.
    Thus,
     $\fullswapcl{\bininfpos{G}} = \swapcl{\bininfpos{G}} \union
                                  \compprset{\pbinconstrgrsch{G}} =
       \nlrightt
          \setbr{\text{(I2)},   \text{(M1)}, \text{(I1)},
            \text{(T1)},  \text{(T2)},  \text{(T3)},
            \text{(U1},
            \text{(I2-sa)},  \text{(I2-sb)},
                 \text{(M1-sa)}, \text{(M1-sb)},
                 \text{(U1-s)},
            \text{(C1)}, \text{(C2)}
                }$.
     \par
   The justification for this terminology is that a complement-pair
rule may be regarded as the swap of an identity rule, as illustrated
in (\ref{def:compprrules}-2).  The middle inference rule is equivalent
to the leftmost, adding only a $\true$ premise, while the rightmost
rule results from the swap of the antecedent $\true$ and the
consequent $\alpha$.

  \def\sepval{0.5cm}
     \[
       \tag{\ref{def:compprrules}-2}
       \begin{prooftree}
         \hypoi{\alpha}
         \inferi{\alpha}
       \end{prooftree}
     \equivto{\sepval}
       \begin{prooftree}
         \hypoii{\alpha}
                 {\true}
         \inferi{\alpha}
       \end{prooftree}
     \transto{\sepval}
       \begin{prooftree}
         \hypoii{\alpha}
                {\mlnot\alpha}
         \inferi{\false}
       \end{prooftree}
     \]
   Thus, if identity rules, such as the leftmost rule of
(\ref{def:compprrules}-2), were added to $\proofsys{A}$, then the
complement-pair rules would be included in the swap closure.
 \end{metalabpara}

 \begin{metaemphlabpara}{lemma}{Lemma}
    {Characterization of unsatisfiability in
          $\allarm{\armctxt{C}}$}\envlabel{lem:compprrules}
     Let $\Phi \subseteq \allarm{\armctxt{C}}$.  If $\Phi$ is
unsatisfiable, then either $\posof{\Phi}$ is unsatisfiable or else
there is a\/ $\mlnot\varphi \in \negof{\Phi}$ with the property that
$\posof{\Phi} \sentails \varphi$.
 \begin{proof}
     If $\posof{\Phi}$ is unsatisfiable, then applying
\envref{prop:entarm}(b) with
  $\beta \equiv \true$ (so $\mlnot\beta \equiv \false$),
 it follows that either $\posof{\Phi} \sentails \false$;
 \ie, $\posof{\Phi}$ is unsatisfiable, or else there is a 
 $\mlnot\varphi \in \negof{\Phi}$
 with the  property that
 $\posof{\Phi} \union \setbr{\mlnot\varphi} \sentails \false$;
 \ie, that $\posof{\Phi} \union \setbr{\mlnot\varphi}$ is
unsatisfiable.  From this, it follows from \envref{lem:genswap}(a$'$)
that $\posof{\Phi} \sentails \varphi$, as required.
 \end{proof}
 \end{metaemphlabpara}

 \begin{metaemphlabpara}{theorem}{Theorem}
    {Completeness of $\fullswapcl{\proofsys{A}}$}\envlabel{thm:fullswapcompl}
     If $\proofsys{A}$ is complete, then the proof system
$\fullswapcl{\proofsys{A}}$ is both sound and complete.
 \begin{proof}
    Assume that $\proofsys{A}$ is complete.  Soundness and
sat-completeness of $\fullswapcl{\proofsys{A}}$ follow from
\envref{thm:armcomplete}.
    Indeed, the rules of $\compprset{\proofsys{A}}$ are clearly sound,
and adding them to $\swapcl{\proofsys{A}}$  has no effect on
sat-completeness.
    \par
   To show that completeness extends to unsatisfiable inferences, let
 $\Phi \subseteq \allarm{\armctxt{C}}$ be unsatisfiable.  If
$\posof{\Phi}$ is unsatisfiable, then this is already provable using
only the rules of $\proofsys{A}$, since it is assumed to be complete.
Trivially, a proof that $\posof{\Phi}$ is unsatisfiable is also a
proof that all of $\Phi$ is unsatisfiable.
     \par
   If $\posof{\Phi}$ is satisfiable, then in view of
\envref{lem:compprrules}, there is a $\mlnot\varphi \in \negof{\Phi}$
with the property that $\posof{\Phi} \sentails \varphi$.  A proof tree
establishing that $\Phi$ is unsatisfiable is shown in
(\ref{thm:fullswapcompl}-1).
   \[
    \tag{\ref{thm:fullswapcompl}-1}
     \begin{prooftreem}
      \hypoi{\Sigma}
      \ellipsis{}{\mbox{\shortstack{proof using $\proofsys{A}$}}}
      \ellipsis{}{\varphi}
      \hypoi{\mlnot\varphi}
      \inferii{\false}
     \end{prooftreem}
   \]
  In that proof, $\Sigma$ is a minimal subset of $\posof{\Phi}$ for
which $\Sigma \sentails \varphi$.  The fact that $\proofsys{A}$ is
complete guarantees that the associated proof exists.  Thus, $\Sigma
\union \setbr{\mlnot\varphi}$ is provably unsatisfiable, whence $\Phi$
has this property as well.
 \end{proof}
 \end{metaemphlabpara}

 \parvert
   To close this section, two examples are provided which illustrate
the process of proof using contrapositioning.

 \begin{metaemphlabpara}{corollary}{Corollary}
    {Completeness of $\bininf{G}$}\envlabel{cor:bininfcomplete}
   The inference system $\bininf{G}$ is sound and complete.
  \begin{proof}
   Soundness follows from \envref{prop:bininfsound}, while
completeness follows from \envref{thm:bininfpossingleuse} and
\envref{thm:fullswapcompl}
  \end{proof}
 \end{metaemphlabpara}

 \begin{metalabpara}{myexample}{Example}
    {Inference involving contrapositioning}\envlabel{ex:neginf1}
   In this example, a proof which involves contrapositioning is
presented.
   Let $\granschemaname[\ref{ex:neginf1}]{G}$ be an SMAS with
 $\setdef{g_i}{i \in \ccinterval{1}{6}} \subseteq
           \granulesofsch[\ref{ex:neginf1}]{G}$
 and
  \nlrightt
     $\setbr{\subrulep[\gsn]{g_1}{g_2},
                      \subrulep[\gsn]{g_2}{g_3},
                      \subrulep[\gsn]{g_4}{g_5},
                      \subrulep[\gsn]{g_5}{g_6},
                      \nsubrulep[\gsn]{g_1}{g_6}}
       \subseteq \allbinconstrgrsch[\ref{ex:neginf1}]{G}$.
    \linebreak
   It will be shown that 
     $\allbinconstrgrsch[\ref{ex:neginf1}]{G}
               \sentails \nsubrulep[\gsn]{g_3}{g_4}$
 using the proof rules of $\bininf[\ref{ex:neginf1}]{G}$.
   As the first step, it is necessary to identify the proof of
positive assertions which will be contrapositioned.  Using
\envref{lem:genswap}(a) with
  $ \Phi \assign \allbinconstrgrsch[\ref{ex:neginf1}]{G}
           \setminus \setbr{\nsubrulep{g_1}{g_6}}$,
 $\varphi \assign \subrulep{g_3}{g_4}$, and
 $\psi \assign \subrulep{g_1}{g_6}$,
   it suffices to show that
  \nlrightt
   $(\allbinconstrgrsch[\ref{ex:neginf1}]{G}
      \setminus \setbr{\nsubrulep[\gsn]{g_1}{g_6}})
     \union \setbr{\subrulep{g_3}{g_4}} \sentails \subrulep{g_1}{g_6}$.
  \newline
   The various trees which illustrate the proof process are shown in
Fig.\ \ref{fig:neginf1}.
  \begin{figure}[htb]
  \def\sepval{0.5cm}
    \begin{prooftreem}
     \hypoii{\subrulep{g_1}{g_2}}
            {\subrulep{g_2}{g_3}}
     \inferi{\subrulep{g_1}{g_3}}
     \hypoi{\subrulep{g_3}{g_4}}
     \inferii{\subrulep{g_1}{g_4}}
     \hypoi{\subrulep{g_4}{g_5}}
     \inferii{\subrulep{g_1}{g_5}}
     \hypoi{\subrulep{g_5}{g_6}}
     \inferii{\subrulep{g_1}{g_6}}
    \end{prooftreem}
    \\[1em]
    \rlap{$\equivto{0em}$}
    \begin{tikzpicture} [->, baseline=(current bounding box.center)]
    \node {$\subrulep{g_1}{g_6}$} [grow' = up, align=center, anchor=north,
                     sibling distance = 6em, level distance = 4em]
    child {node {$\subrulep{g_1}{g_5}$} edge from parent [double]
      child {node {$\subrulep{g_1}{g_4}$} edge from parent [double]
        child {node {$\subrulep{g_1}{g_3}$}
          child {node {$\subrulep{g_1}{g_2}$}}
          child {node {$\subrulep{g_2}{g_3}$}}
              }
        child {node {$\subrulep{g_3}{g_4}$} edge from parent [double]
              }
            }
      child {node {$\subrulep{g_4}{g_5}$}
            }
          }
    child {node {$\subrulep{g_5}{g_6}$}};
    \end{tikzpicture}
    \hspace*{-2em}$\smash{\transto{0em}}$\hspace*{-2em}
    \begin{tikzpicture} [->, 
                         baseline=(current bounding box.center)]
    \node (g16) {$\nsubrulep{g_1}{g_6}$}
                    [grow' = up, align=center, anchor=north,
                     sibling distance = 6em, level distance = 4em]
    child {node (g15) {$\nsubrulep{g_1}{g_5}$} \efpn
      child {node (g14) {$\nsubrulep{g_1}{g_4}$} \efpn
        child {node (g13) {$\subrulep{g_1}{g_3}$} \efpn
          child {node (g12) {$\subrulep{g_1}{g_2}$}}
          child {node (g11) {$\subrulep{g_2}{g_3}$}}
              }
        child {node (g34) {$\nsubrulep{g_3}{g_4}$} \efpn}
            } 
      child {node (g45) {$\subrulep{g_4}{g_5}$} \efpn}
          }
    child {node (g56) {$\subrulep{g_5}{g_6}$} \efpn};
    \draw (g15) edge [double,->] (g16);
    \draw (g15) edge [dashed,->] (g56);
    \draw (g14) edge [double,->] (g15);
    \draw (g14) edge [dashed,->] (g45);
    \draw (g34) edge [double,->] (g14);
    \draw (g34) edge [dashed,->] (g13);
    \end{tikzpicture}
    \\[1em]
    \phantom{xxxxx}\hfill$\equivto{\sepval}$
    \begin{tikzpicture} [->, baseline=(current bounding box.center),
           level 1/.style={sibling distance=12em},
           level 2/.style={sibling distance=6em},
           level 3/.style={sibling distance=6em}
                        ]
    \node {$\nsubrulep{g_3}{g_4}$} [grow' = up, align=center, anchor=north,
                     level distance = 4em]
    child {node {$\subrulep{g_1}{g_3}$}
      child {node {$\subrulep{g_1}{g_2}$}}
      child {node {$\subrulep{g_2}{g_3}$}}
          }
    child {node {$\nsubrulep{g_1}{g_4}$}
      child {node {$\nsubrulep{g_1}{g_5}$}
        child {node {$\subrulep{g_5}{g_6}$}}
        child {node {$\nsubrulep{g_1}{g_6}$} edge from parent [double]
              }
            edge from parent [double]
            }
      child {node {$\subrulep{g_4}{g_3}$}}
          edge from parent [double]
          };
    \end{tikzpicture}
    \\[1em]
    \phantom{xxxxx}\hfill$\equivto{\sepval}$
   \begin{prooftreem}
     \hypoii{\subrulep{g_1}{g_2}}
            {\subrulep{g_2}{g_3}}
     \inferi{\subrulep{g_1}{g_3}}
     \hypoii{\subrulep{g_5}{g_6}}
            {\nsubrulep{g_1}{g_6}}
     \inferi{\nsubrulep{g_1}{g_5}}
     \hypoi{\subrulep{g_4}{g_3}}
     \inferii{\nsubrulep{g_1}{g_4}}
     \inferii{\nsubrulep{g_3}{g_4}}
    \end{prooftreem}
     \caption{Proof trees for \envref{ex:neginf1}}\label{fig:neginf1}
    \end{figure}
   The first tree, at the top of the figure, represents the proof of
  \nlrightt
   $(\allbinconstrgrsch[\ref{ex:neginf1}]{G}
     \setminus \setbr{\nsubrulep[\gsn]{g_1}{g_6}})
     \union \setbr{\subrulep{g_3}{g_4}} \sentails \subrulep{g_1}{g_6}$
  \newline
 in standard proof-tree format.  Below it, to the left, is this same
tree in vertex-edge format, with the swap path highlighted using
double-line arrows.  This tree is the instantiation, for this example,
of the generic tree of Fig.\ \ref{fig:tocontrap}.  To its right is the
contrapositioning of this tree using fixed vertex positioning,
corresponding to the generic tree of Fig.\ \ref{fig:contrapi}.  The
swap path is again highlighted using double-line arrows, while edges
whose source vertex has been altered as a result of this
transformation are show using dashed lines.  Just below is the same
tree in fixed-edge positioning, corresponding to
Fig.\ \ref{fig:contrapii}, while at the bottom of the figure is the
same result proof in standard inference format.  In all cases, the the
proof rules used have not been identified explicitly, but these are
very easy to determine.
  \end{metalabpara}

 \begin{metalabpara}{myexample}{Example}
    {Inference using a complement-pair rule}\envlabel{ex:neginf2}
   In this example, a proof of unsatisfiability which involves the use
of a complement-pair rule is developed.
   Let $\granschemaname[\ref{ex:neginf2}]{G}$ be an SMAS with
 $\setdef{g_i}{i \in \ccinterval{1}{6}} \subseteq
           \granulesofsch[\ref{ex:neginf2}]{G}$
 and
  \newline
   $\setbr{\subrulep[\gsn]{g_1}{g_2},
          \subrulep[\gsn]{g_2}{g_3},
          \subrulep[\gsn]{g_4}{g_5},
          \subrulep[\gsn]{g_5}{g_6},
          \disjrulep[\gsn]{g_3}{g_6},
          \ndisjrulep[\gsn]{g_1}{g_4}}
   \nlrightt
       \subseteq \allbinconstrgrsch[\ref{ex:neginf2}]{G}$.
    \linebreak
   It will be shown that 
     $\allbinconstrgrsch[\ref{ex:neginf2}]{G}$
 is unsatisfiable, using the proof rules of
$\bininf[\ref{ex:neginf1}]{G}$.
   \begin{figure}[htbp]
    \begin{center}
    \begin{tikzpicture} [->, baseline=(current bounding box.center),
           level 1/.style={sibling distance=12em},
           level 2/.style={sibling distance=12em},
           level 3/.style={sibling distance=6em}
                        ]
    \node {$\false$} [grow' = up, align=center, anchor=north,
                      level distance = 4em]
      child {node {$\disjrulep{g_1}{g_4}$}
        child {node {$\subrulep{g_1}{g_3}$}
          child {node {$\subrulep{g_1}{g_2}$}
                }
          child {node {$\subrulep{g_2}{g_3}$}
                }
              }
        child {node {$\subrulep{g_4}{g_6}$}
          child {node {$\subrulep{g_4}{g_5}$}
                }
          child {node {$\subrulep{g_5}{g_6}$}
                }
              }
        child {node {$\disjrulep{g_3}{g_6}$}
              }
            }
      child {node {$\ndisjrulep{g_1}{g_4}$}
            };
    \end{tikzpicture}
    \\[1em]
    \[
    \begin{prooftreem}
     \hypoii{\subrulep{g_1}{g_2}}
            {\subrulep{g_2}{g_3}}
     \inferi{\subrulep{g_1}{g_3}}
     \hypoii{\subrulep{g_4}{g_5}}
            {\subrulep{g_5}{g_6}}
     \inferi{\subrulep{g_4}{g_6}}
     \hypoi{\disjrulep{g_3}{g_6}}
     \inferiii{\disjrulep{g_1}{g_4}}
     \hypoi{\ndisjrulep{g_1}{g_4}}
     \inferii{\false}
    \end{prooftreem}
    \]
    \end{center}
    \caption{Proof trees for \envref{ex:neginf2}}\label{fig:neginf2}
    \end{figure}
   The strategy is simple.  First, develop a proof that
   $\allbinconstrgrsch[\ref{ex:neginf2}]{G}
         \setminus \setbr{\ndisjrulep{g_1}{g_4}}
         \sentails \disjrulep{g_1}{g_4}$,
  and then apply the complement-pair rule which states that both
elements of the pair
  $\setbr{\disjrulep{g_1}{g_4},\ndisjrulep{g_1}{g_4}}$
 cannot be true.  The associated proof trees, in both vertex-edge
format and traditional logic format, are shown in
Fig.\ \ref{fig:neginf2}.
 \end{metalabpara}

 \begin{metalabpara}{discussion}{}
    {Relationship to RCC5}\envlabel{disc:rcc5}
     Given two granules $g_1, g_2 \in \granulesofsch{G}$, the truth
values of the three sentences $\subrulep{g_1}{g_2}$,
$\subrulep{g_2}{g_1}$, and $\disjrulep{g_1}{g_2}$ together represent
what is known about their relationship.  To formalize this,
for $\varphi \in \pbinconstrgrsch{G}$, define
   \[
     \gstateof{G}{\varphi} =
        \begin{cases}
          \true ~\text{if}~ \constrgrsch{G} \sentails \varphi, \\
          \false ~\text{if}~ \constrgrsch{G} \sentails \mlnot\varphi, \\
          \unknown ~\text{otherwise},
        \end{cases}
   \]
 and then define the \emph{state vector of} $\grpr{g_1}{g_2}$
\emph{in} $\granschemaname{G}$ to be the triple
 \nlrightt
 $\stateof{G}{g_1}{g_2} =
  \abr{\sstateof{G}{g_1}{g_2},
       \sstateof{G}{g_2}{g_1},
       \dstateof{G}{g_1}{g_2}}$.
  \newline
 Call $\stateof{G}{g_1}{g_2}$ \emph{complete} if no entry has the
value $\unknown$.
   \par
   The \emph{region connection calculus}, or \emph{RCC}, and its
variants, have complete state vectors, rather than individual
predicates, as their starting point.  The papers
 \mycite{RandellCBLR89_kr}, \mycite{RandellCC92_ickrr},
 \mycite{CohnBGG97_geoinf}, \mycite{LiYa03_ai},
 and \mycite{Stell04_amai}
 form a small sample of the vast work in this area.
    The most commonly investigated variant is RCC8, whose associated
state vectors have four entries, since boundaries of regions are also
modelled.  For a comparison to the work of this paper, the more
appropriate variant is RCC5, which focuses on regions without
boundaries, so the state vector has only three entries.  Table
\ref{table:rcc5} shows the state vectors for the version of RCC5
described in \mycite[Table 1]{Stell04_amai}, which is particularly
appropriate for comparison to the granule logic of this paper, since
it supports empty granules, unlike most variants of RCC.\footnote{
     In \mycite[Table 1]{Stell04_amai}, rather than
$\disjrulep{g_1}{g_2}$, a predicate which is best described as
$\subrulep{g_1}{\overline{g_2}}$ is used for the third state
parameter, with $\overline{g_2}$ the complement of $g_2$.  Since
complements are not part of the framework of an SMAS,
$\disjrulep{g_1}{g_2}$ is used instead.  It is easy to verify that the
results are identical.  }
   \begin{table}[htb]
   \begin{center}
   $\begin{array}{*5{|c}|}
     \hline
        & \subrulep{g_1}{g_2}
        & \subrulep{g_2}{g_1}
        & \disjrulep{g_2}{g_1}
        & \text{Symmetric} \\
     \hline\hline
        \rccdc{g_1}{g_2}
                               & \false & \false & \true 
                               & \checkmark \\ \hline
        \rccpo{g_1}{g_2}
                               & \false & \false & \false
                               & \checkmark \\ \hline
        \rcceq{g_1}{g_2}
                               & \true & \true & \false
                               & \checkmark \\ \hline
        \rccpp{g_1}{g_2}
                               & \true & \false & \false
                               &            \\ \hline
        \rccppi{g_1}{g_2}
                               & \false & \true & \false
                               &            \\ \hline
        \rcceqe{g_1}{g_2}
                               & \true & \true & \true
                               & \checkmark \\ \hline
        \rccppe{g_1}{g_2}
                               & \true & \false & \true
                               &            \\ \hline
        \rccppie{g_1}{g_2}
                               & \false & \true & \true
                               &            \\ \hline
   \end{array}$
   \end{center}
   \caption{Predicates of RCC5+}\label{table:rcc5}
   \end{table}
    To avoid any confusion, the variant of RCC described in Table
\ref{table:rcc5} will be called \emph{RCC5+}.
    The predicate names which are used are those which have become
standard in the field.  $\rccdcname$ stands for \emph{disconnected} (\ie,
disjoint), $\rccponame$ for \emph{partial overlap}, $\rcceqname$ for
\emph{equal}, $\rccppname$ for \emph{proper part}, and $\rccppiname$
for \emph{proper part inverse}.  The relations $\rccdcname$,
$\rccponame$, and $\rcceqname$ are \emph{symmetric} in that they are
their own inverses.  On the other hand $\rccppname$ is not symmetric;
hence the need for $\rccppiname$.
    These five relations constitute ordinary RCC5, and are fine as
long as the regions are nonempty.  However, they fail on empty
regions.  The assertions
  $\rcceq{\botgrsch{G}}{\botgrsch{G}}$,
  $\rccpp{\botgrsch{G}}{g_2}$,
 and
  $\rccppi{\botgrsch{G}}{g_2}$
 all fail to hold, the latter two for any $g \in \granulesofschnb{G}$,
counter to the expected meaning of these predicates.
  The solution it to introduce three additional relation names.
  The statements $\rcceq{g_1}{g_2}$, $\rccpp{g_1}{g_2}$, and
$\rccppi{g_1}{g_2}$ cover the cases for which
 $g_1 \not\ideq \botgrsch{G}$, while $\rcceqe{g_1}{g_2}$,
$\rccppe{g_1}{g_2}$, and $\rccppie{g_1}{g_2}$ cover the cases for
which $g_1 \ideq \botgrsch{G}$.
   \par
  A key feature of this framework is that for any pair
$\grpr{g_1}{g_2}$ of granules, $R\grpr{g_1}{g_2}$ holds for exactly
one
 $R \in \setbr{\rccdcname,\rccponame,\rcceqname,\rccppname,
               \rccppiname,\rcceqename,\rccppename,\rccppiename}$.
   This can be seen from the fact that every possible state vector is
represented by one of the predicates.
 This property is sometimes called \emph{JEPD}, short for
\emph{jointly exclusive and pairwise disjoint}.
    \par
  Inference within RCC is often performed via the use of a
\emph{composition table} \mycite{LiYa03_ai}.  Such a table presents,
for
 $R_1, R_2 \in \setbr{\rccdcname,\rccponame,\rcceqname,\rccppname,
               \rccppiname,\rcceqename,\rccppename,\rccppiename}$,
 the relations 
  $R_3 \in \setbr{\rccdcname,\rccponame,\rcceqname,\rccppname,
               \rccppiname,\rcceqename,\rccppename,\rccppiename}$
 for which $R_3 \subseteq R_1 \comp R_2$, with $\comp$ denoting
ordinary relational composition.  Less formally, the table identifies
the relations for which $R_3\grpr{g_1}{g_3}$ could hold, given that
$R_1\grpr{g_1}{g_2}$ and $R_2\grpr{g_2}{g_3}$ hold.  
    \par
   Presenting the entire composition table for RCC5+ is beyond the
scope of this discussion.  However, a simple example of inference is
appropriate, in order to compare RCC5+ to the framework of this paper.
   Let $\granschemaname[\ref{disc:rcc5}]{G}$ have
  $\granulesofschnbnt[\ref{disc:rcc5}]{G} =
         \setbr{g_1,g_2,g_3}$,
 with
 \newline
 \mbox{
  $\constrgrsch{G} =
   \setbr{
    \nsubrulep{g_1}{g_2},
    \nsubrulep{g_2}{g_1},
    \ndisjrulep{g_1}{g_2},
    \subrulep{g_2}{g_3},
    \nsubrulep{g_3}{g_2},
    \ndisjrulep{g_2}{g_3}
         }$.
      }
   It is immediate that these constraints correspond exactly to the
RCC5+ conditions
 \preformat{\linebreak}
  $\setbr{\rccpo{g_1}{g_2}, \rccpp{g_2}{g_3}}$.
   Even without constructing the entire composition table for RCC5+,
it is straightforward to see that $\rccpo{g_1}{g_3}$ and
$\rccpp{g_1}{g_3}$ are the only possibilities for the relationship on
$\grpr{g_1}{g_3}$.  This can be done by looking at a composition table
for RCC8, such as \mycite[Table 2]{LiYa03_ai}, and collapsing
$\rcctppname$ and $\rccntppname$ into $\rccppname$, or just by direct
inspection, drawing a Venn diagram.
    \par
   Now, consider applying the proof rules of
$\bininf[\ref{disc:rcc5}]{G}$.  The only possible deductions are shown
in Fig. \ref{fig:proofsrcc5}.
   \begin{figure}[htb]
    \begin{align*}
     &\begin{prooftreem}
      \hypoii{\subrulep{g_2}{g_3}}
             {\nsubrulep{g_2}{g_1}}
      \inferi{\nsubrulep{g_3}{g_1}}
     \end{prooftreem}
     &&\begin{prooftreem}
      \hypoii{\subrulep{g_2}{g_3}}
             {\ndisjrulep{g_1}{g_2}}
      \inferi{\ndisjrulep{g_1}{g_3}}
     \end{prooftreem}
   \end{align*}
   \caption{Proofs for \envref{disc:rcc5}}\label{fig:proofsrcc5}
   \end{figure}
   For both $\subrulep{g_3}{g_1}$ and $\disjrulep{g_1}{g_3}$ false,
with the truth value of $\subrulep{g_1}{g_3}$ unknown, a check of
Table \ref{table:rcc5} shows that the only possibilities are
$\rccponame\grpr{g_1}{g_3}$ and $\rccppname\grpr{g_1}{g_3}$, in
agreement with the result from a composition table.
    \par
   Whether RCC is preferable to the framework of this paper depends
entirely upon the type of knowledge to be modelled.  If almost all
relationships between granules are known completely, so that one of
the eight predicates of Table \ref{table:rcc5} applies to each pair
$\grpr{g_1}{g_2}$, then RCC will likely be superior.  However, in the
context of substantial incomplete information, RCC will be overwhelmed
by the need for disjunctive representation, rendering the approach of
this paper the preferred one.
 \end{metalabpara}

 \begin{metalabpara}{discussion}{}
    {Relationship to AFDs}\envlabel{disc:afds}
   In \mycite{DeBra87_mfdbs} and \mycite[Sec.\ 5.2]{ParedaensDGV89}, a
framework for dependencies on relational databases which includes both
functional dependencies (FDs) and \emph{afunctional dependencies}
(AFDs) is developed, with a sound and complete inference system.
Since FDs form the original Armstrong context
\mycite{Armstrong74_ifip}, it is worth asking how this work fits into
the framework developed here.
    \par
   AFDs are not true logical negations of FDs.  The relation $R$
satisfies the AFD $\afd{X}{Y}$ if for \textbf{every} tuple $t \in R$,
there is a second tuple $t' \in R$ with $t[X]=t'[X]$ and $t[Y] \neq
t'[Y]$.  On the other hand, the true logical negation of the FD
$\fd{X}{Y}$ has the property that for \textbf{some} tuple $t \in R$,
there is a second tuple $t' \in R$ with $t[X]=t'[X]$ and $t[Y] \neq
t'[Y]$.
   Thus, despite the utility of AFDs, it is not possible to define the
combined inference system for FDs and AFDs by applying the swap
approach of this paper to constraints on FDs.
 \end{metalabpara}

 \begin{metalabpara}{discussion}{}
    {Other contexts}\envlabel{def:otherctxt}
    It should perhaps be mentioned that the techniques of this section
can also be applied to the inference systems $\qbininfpos{G}$ (see
\envref{sum:APinf}) and $\sqbininfpos{G}$ (see
\envref{def:sqbininfpos}), since their underlying frameworks
(consisting of quasi-models and strong quasi-models, respectively),
are easily shown to be Armstrong contexts.  However, this direction
will not be pursued here, since there is no immediate need for the
result.
 \end{metalabpara}


 \section{Relationship to Other Work}\label{sec:lit}

    xxxxx


 \section{Conclusions and Further Directions}\label{sec:cfd}

    In this paper, two main contributions have been made.

 \baxblkc

   \axitem{\textsc{Single-use proof system
                          for binary granule constraints:}}
    A sound and com-
 \preformat{\linebreak}
 plete proof system for the predicates of subsumption and disjunction
of granules,
 \preformat{\linebreak}
 $\bininfpos{G}$, has been developed.  This result builds
upon the earlier work in \mycite{AtzeniPa88_dke}, adding critical new
features, such as modelling of the bottom granule $\botgrsch{G}$ and
top granule $\topgrsch{G}$, and the requirement that all granules
other than $\botgrsch{G}$ have nonempty semantics.  It furthermore
provides single-use proofs, which are necessary for extension to a
proof system that involves the negations of these predicates as well.

   \axitem{\textsc{Proof system augmentation to negative constraints}}
   A general method has been developed to extend any sound and
complete single-use inference system over an Armstrong context
$\armctxt{C}$ to one which is sound and complete on the assertions in
$\armctxt{C}$ together with their negations.  This methods as then
applied to the inference system $\bininfpos{G}$ to obtain the sound
and complete inference $\bininf{G}$ system on all binary predicates of
an SMAS, both positive and negative.

 \eaxblk

  \noindent
  The immediate further direction is the following.

 \baxblkc

   \axitem{\textsc{Extend to granules with boundary:}}
    In spatial modelling, it is often advantageous or even necessary
to be able to work with granules with boundary.  Work is currently
underway to extend the notion of SMAS of this paper to such granules.
This requires replacing the single disjunction predicate
$\disjrulep{g_1}{g_2}$, with two, the first $\disjruleip{g_1}{g_2}$
for interiors and the second $\disjrulebp{g_1}{g_2}$ for boundaries.
The resulting framework can be shown to be an Armstrong context, with
a sound and complete single-use inference system which extends
$\bininfpos{G}$ naturally.  The methods of Sec.\ 6 of this paper apply
to this new framework as well, providing an inference system for both
positive and negative constraints for binary predicates with boundary.
The modelling power of this framework is equivalent to that of the
four-intersection model \mycite{EgenhoferFr91_ijgis} and RCC8
\mycite{Stell04_amai}.

 \eaxblk


\bibliography{jabbrev,logic,logicdb,dbtheory,math,theorcs,compling,logicpr,knowlrep,expercs,oodb,ai,workflow,gran,aggr,misccs,local}
\bibliographystyle{alpha}


 \end{document}